\documentclass[preprint2,longabstract]{aastex}
\usepackage{longtable,txfonts,color}
\usepackage{enumerate}
\usepackage{afterpage}

\newcommand{\Jykms}{\,Jy\,km~s$^{-1}$}
\newcommand{\Jykmst}{Jy\,km~s$^{-1}$}
\newcommand{\kms}{\,km\,s$^{-1}$}
\newcommand{\kmst}{km\,s$^{-1}$}
\newcommand{\etal}{et~al.~}
\newcommand{\HI}{\mbox{\normalsize H\thinspace\footnotesize I}}
\newcommand{\HII}{\ion{H}{2}}
\newcommand{\sol}{\ifmmode_{\mathord\odot} \else $_{\mathord\odot}$ \fi}
\newcommand{\msun}{M$_{\sol}$}
\newcommand{\B}{$B$}
\newcommand{\R}{$R$}
\newcommand{\II}{$I$}
\newcommand{\J}{$J$}
\newcommand{\HH}{$H$}
\newcommand{\K}{$K_s$}
\newcommand{\cf}{{cf.}\ }

\definecolor{grey}{rgb}{0.5,0.6,0.7}

\def\deg{{^\circ}}

\shorttitle{The Parkes HI Zone of Avoidance Survey}
\shortauthors{Staveley-Smith et al.}

\begin{document}

\title{The Parkes HI Zone of Avoidance Survey}

\author{L. Staveley-Smith }
\affil{International Centre for Radio Astronomy Research, University of Western Australia, Crawley, WA 6009, Australia }
\affil{and ARC Centre of Excellence for All-sky Astrophysics}
\author{R.C. Kraan-Korteweg}
\affil{Astrophysics, Cosmology and Gravity Centre (ACGC), Department of Astronomy, University of Cape Town, Private Bag X3,
Rondebosch 7701, South Africa}
\author{A.C. Schr\"oder }
\affil{South African Astronomical Observatory, PO Box 9, Observatory 7935, Cape Town, South Africa }
\author{P. A. Henning}
\affil{Department of Physics and Astronomy, University of New Mexico,1919 Lomas Blvd. NE, Albuquerque, NM 87131, USA}

\author{B.S. Koribalski}
\affil{CSIRO Astronomy and Space Science, Australia Telescope National Facility, P.O. Box 76, Epping, NSW 1710, Australia}
\author{I.M. Stewart}
\affil{Sterrewacht Leiden, University of Leiden, Niels Bohrweg 2, 2333 CA Leiden, The Netherlands}
\and
\author{G. Heald}
\affil{ASTRON, the Netherlands Institute for Radio Astronomy, Postbus 2, 7990
AA, Dwingeloo, The Netherlands}
\affil{and Kapteyn Astronomical Institute, Postbus 800, 9700 AV, Groningen, The Netherlands}

%



\begin{abstract}


A blind \HI~survey of the extragalactic sky behind the southern Milky Way has been conducted with the multibeam receiver on the 64-m Parkes radio telescope.
The survey covers the Galactic longitude range $212\deg < \ell < 36\deg$ and Galactic latitudes 
\mbox{$|b| < 5\deg$} to an rms sensitivity of 6 mJy per beam per 27\kms\ channel, and yields 883 galaxies to a recessional velocity of 12,000\kms.
The survey covers the sky within the \HI~Parkes All-Sky Survey (HIPASS) area to greater sensitivity, finding lower \HI-mass galaxies at all distances, and probing more completely the large-scale structures at and beyond the distance of the Great Attractor.
Fifty-one percent of the \HI~detections have an optical/near-infrared (NIR) counterpart in the
literature. A further 27\% have new counterparts found in existing, or newly obtained, optical/NIR images. The counterpart rate drops in regions of high foreground stellar crowding and extinction, and for low-\HI~mass objects.
Only 8\% of all counterparts have a previous optical redshift measurement.
The \HI~sources are found independently of Galactic extinction, although the detection rate drops in regions of high Galactic continuum.  
The survey is incomplete below a flux integral of approximately 3.1 Jy\kms\ and mean flux density of approximately 21 mJy, with 75\% and 81\% of galaxies being above these limits, respectively. 
Taking into account dependence on both flux and velocity width, and constructing a scaled dependence on the flux integral limit with velocity width ($w^{0.74}$), completeness limits of 2.8 Jy\kms\ and 17 mJy are determined, with 92\% of sources above these limits.
A notable new galaxy is HIZOA J1353$-$58, a possible companion to the Circinus galaxy. Merging this catalog with the similarly-conducted northern extension (Donley et al.~2005), large-scale structures are delineated, including those within the Puppis and Great Attractor regions, and the Local Void.
Several newly-identified structures are revealed here for the first
time. Three new galaxy concentrations (NW1, NW2 and NW3) are key in
confirming the diagonal crossing of the Great Attractor Wall between
the Norma cluster and the CIZA J1324.7-5736 cluster. Further contributors
to the general mass overdensity in that area are two new clusters (CW1 and
CW2) in the nearer Centaurus Wall, one of which forms part of the striking
$180\degr$ ($100h^{-1}$Mpc) long filament that
dominates the southern sky at velocities of $\sim 3000$\kms, and the
suggestion of a further wall at the Great Attractor distance at slightly
higher longitudes.

\end{abstract}


\section{Introduction}


In addition to serving as a direct probe of neutral hydrogen gas,
observations of the 21-cm line of \HI\ allow the detection of galaxies
through the thickest Galactic obscuration and the presence of the highest
foreground stellar confusion, which mask galaxies at optical/IR wavelengths.
The historical Zone of Avoidance (ZOA) has prevented study of the
distribution of galaxies behind the Milky Way and is most pronounced in
the optical.  Dedicated low-Galactic-latitude optical searches and surveys
of galaxies in the infrared have narrowed the ZOA, but still fail to detect
galaxies where Galactic emission, dust obscuration, and stellar confusion
prevent the recognition of background galaxies.  The near-infrared (NIR) has been
particularly fruitful for finding large numbers of low-latitude galaxies.
However, the most homogeneous NIR wide-angle redshift survey, the 2MASS
Redshift Survey (Huchra \etal 2012), is still not an all-sky survey, as it
retains a gap of $5\deg - 8\deg$ around the Galactic Plane.  While this
does not affect our understanding of galaxy populations, as there is no
reason to think galaxies in the ZOA are in any way different from
high-latitude ones, the lack of information in the ZOA contributes
uncertainty to our understanding of dynamics in the local Universe.
Controversial results persist for the apex and convergence radius for the
CMB dipole from galaxy redshift and peculiar velocity surveys (e.g., Erdogdu
\etal 2006; Watkins \etal 2009; Lavaux \etal 2010; Lavaux and
Hudson~2011; Courtois \etal 2012, Ma \etal 2012; Springob \etal 2014).
While insufficient depth in redshift space of the relevant datasets
(e.g.,~optical, 2MASS, IRAS, galaxy clusters) may be a factor, the
incomplete mapping of the ZOA is also important (e.g., Kraan-Korteweg and
Lahav~2000, Loeb and Narayan~2008).

Beyond the ability to find galaxies in regions of arbitrarily high
extinction and stellar confusion, using the 21-cm line of \HI\ to detect
galaxies has the added benefit of immediate redshift measurement,
eliminating the need that two-dimensional imaging surveys have for
follow-up observations to obtain redshifts.  This blind \HI\ survey technique for finding
hidden galaxies was pioneered by Kerr and Henning (1987), but technology
allowing large areas of the sky to be surveyed at 21-cm became available
only relatively recently, with L-band multibeam receivers installed on the 
Parkes, Arecibo, Jodrell Bank Lovell, and Effelsberg radio telescopes.  

We report here on a
21-cm \HI\ survey of the southern hemisphere ZOA, fully covering the area
$212\deg < \ell < 36\deg$; $|b| < 5\deg$ to a sensitivity of 6\,mJy at velocity resolution of 27\kms\, with
the 64 m Parkes radio telescope.
We refer to this survey as HIZOA-S.  A shallow survey of this area, with
partial data and much lower sensitivity (15\,mJy), was presented by Henning et
al.~(2000), and an intermediate-depth study of the Great Attractor region
was conducted by Juraszek \etal (2000), with rms 20\,mJy at slightly better velocity 
resolution.  Two extensions to the north
($\ell = 36\deg - 52\deg$ and $\ell = 196\deg - 212\deg$) have been studied
at identical sensitivity to HIZOA-S by Donley \etal (2005). We refer to this as
HIZOA-N. A region of the Galactic Bulge, above and below our latitude range
around the Galactic Center, has also been surveyed to higher Galactic
latitude, but has not been fully analysed yet. Preliminary descriptions
appear in Shafi (2008) and Kraan-Korteweg \etal (2008).  The current work
presents the data for the main full-sensitivity southern ZOA survey,
HIZOA-S.  The discussion of large-scale structures (\S7) makes use of the
combined southern and northern data.  We refer to the combined sample as
HIZOA.  The \HI\ Parkes All-Sky Survey (HIPASS; Meyer \etal 2004) also
covers the area, but to 2-3 times lower sensitivity than the current work
and is unable to define detailed structure at Great Attractor distances.

An early fully-sampled, blind, but shallow \HI\ survey (Koribalski \etal 2004) has delineated local large-scale structures at low Galactic latitudes.  
Further, pointed 21-cm observations of partially-obscured galaxies have revealed
large-scale structures in selected regions (Kraan-Korteweg, Henning and
Schr\"o\-der 2002; Schr\"o\-der, Kraan-Korteweg and Henning 2009), but the
current work covers the entire southern ZOA, over 1800 deg$^2$, in a blind, deep,
and complete fashion.  The remainder of the great circle of the ZOA in the
northern hemisphere is now being surveyed by the Arecibo (Henning \etal  
2010, McIntyre \etal 2015) and Effelsberg (Kerp \etal 2011) radio telescopes.  There are
also ongoing efforts to uncover the three-dimensional distribution of
partially-obscured galaxies with pointed 21-cm observations of selected
galaxies visible in the NIR with the Nan\c{c}ay radio telescope (van Driel \etal 2009, Ramatsoku \etal
2014).  
The combination of \HI~and deep NIR imaging in regions of
modest extinction and stellar confusion allows determination of peculiar
velocities via the Tully-Fisher relation, and thus provides information on galaxy
flows.  Such surveys are possible in the ZOA with the advent of
large \HI~surveys, such as the current work, and follow-up NIR imaging of the
counterparts of the HIZOA-galaxies (Williams \etal 2014,
Said \etal in prep.). In the future, all-sky \HI\ surveys, using the
Australian Square Kilometre Array Pathfinder (ASKAP: Johnston \etal 2007; Koribalski 2012),
and the APERture Tile In Focus system on the Westerbork telescope in the
north (APERTIF: Verheijen \etal 2009), will allow comprehensive and
complete mapping of large-scale structures as revealed by neutral hydrogen.
Preparatory work is already underway for such large interferometric ZOA
surveys; for instance the Westerbork Synthesis Radio Telescope (WSRT) has
been used for a deep ZOA mosaic with a field of view close to that of the
forthcoming APERTIF system. Ultimately, the SKA itself will extend
such \HI\ surveys further in redshift.

In \S\ref{obsred}, we present details of the observations and data reduction for the
survey; in \S\ref{sample} we describe the sample compilation and \HI~parametrization.
In \S\ref{HIparam}, \HI~properties of the sample are presented, and multiwavelength
counterparts are described in \S\ref{cmatch}. \S\ref{compl} presents the survey's completeness,
reliability and parameter accuracy.  In \S\ref{s:lss}, new large-scale structures and
their relationships to high-latitude structures are discussed. We present
conclusions in \S\ref{concl}.

\section{Observations and Data Reduction}  \label{obsred}

\subsection{Multibeam Observations}


The observations described here were taken with the 21-cm multibeam receiver
(Staveley-Smith et al.~1996) at the Parkes telescope between 1997 March 22 and
2000 June 8, contemporaneously with the southern component of HIPASS. The observations cover the Galactic
longitude range $212\degr < \ell < 36\degr$ in 23 separate regions, each 8\degr\
wide in longitude, and cover the latitude range $5\degr > b > -5\degr$ with
almost uniform sensitivity. As noted above, later observations (Donley \etal 2005) extended the
longitude range by 16\degr\ in each direction (HIZOA-N). The observational
parameters are identical to those described in Donley \etal (2005), but are
summarised in Table~\ref{t:surveyparams} for completeness. 

Each of the 23 longitude regions (8\degr\ in longitude by 10\degr\ in
latitude) was scanned in raster fashion by the
telescope at a rate of 1\degr\ min$^{-1}$, in a direction of increasing, or decreasing, 
Galactic longitude. Each scan therefore
lasted 8 minutes plus overlap and turn-around time. The central beam
was scanned at constant latitude; however, to minimise bandpass effects, the
receiver was not rotated (`parallactified') during the scan. The other
12 beams tended therefore to deviate slightly from constant latitude. The receiver
rotation was adjusted such that the feed rotation was approximately
15\degr\ from the direction of the scan at the mid-point of each observation
(Staveley-Smith 1997). This ensured almost-Nyquist coverage of the sky even for a single
scan, covering a region $1\fdg7 \times 8\degr$ in size. However, for sensitivity
reasons, each of the 23 regions was raster scanned 425 times, giving a total
integration time of approximately 2100 s for each point of the sky (with the actual value 
dependent on gridding strategy) and an rms noise of 6\,mJy\,beam$^{-1}$. The observations
were conducted at night over a period of three years, and there was considerable redundancy
in the data. Each point in the sky was visited hundreds of times, allowing mitigation of 
the already low levels of radio frequency interference (RFI) in this band. For more details,
see Meyer \etal (2004). The main issue encountered was confusion with Galactic recombination
lines (see Meyer \etal 2004 or Alves \etal 2015) which, being diffuse and associated with radio
continuum emission, was straightforward to recognise in the image domain.

The central observing frequency was 1394.5 MHz with a bandwidth of 64\,MHz and a channel spacing of 
62.5 kHz. This corresponds to a velocity range of $-1,280 < cz < 12,740$\kms, and a channel spacing of 13.2\kms\
at zero redshift. After Hanning smoothing, the velocity resolution was 27\kms. The 
correlator integration time was 5 sec, resulting in negligible smearing along the scan direction.
Two orthogonal linear polarisations were recorded, and combined to Stokes $I$ in the gridding step (see below).
Calibration against the continuum radio sources Hydra A and PKS B1934-638 was usually monitored 
on a daily or weekly basis to check system performance. However, all calibration was referenced to
a continuously firing noise diode inserted in each of the 13 feeds at 45\degr\ to the
transducers. No evidence was found for any time variability in the noise diode amplitude.
Neither is there any measurable gain-elevation effect at the Parkes telescope at these
frequencies.

\begin{deluxetable}{lcl}
\tablecolumns{3}
\tablewidth{0pc}
\tablecaption{HIZOA survey parameters.\label{t:surveyparams}}
\tablehead{
\colhead{Parameter} & \colhead{Value} & \colhead{unit}}
\startdata
Galactic longitude & $212\degr < \ell < 36\degr$  & \\
Galactic latitude & $-5\degr < b < -5\degr$ & \\
Velocity range & $-1,280 < cz < 12,740$ & \kmst \\
Velocity resolution & 27 & \kmst \\
Scan rate & 1.0 & degree min$^{-1}$ \\
Beam size (FWHP) & 15.5 & arcmin \\
Correlator cycle & 5 & s \\
Integration time per beam & 2100 & s\\
Cube rms & 6 & mJy  \\
\enddata
\end{deluxetable}

\subsection{Data Reduction through to Cubes} 


The spectral data was reduced at the telescope in real time using the {\sc
LiveData}
package\footnote{http://www.atnf.csiro.au/computing/software/livedata}.  
{\sc LiveData} buffered the 
incoming data and applied position interpolation so that the correct beam position was assigned
to the mid-point of each correlator spectrum. It then applied a barycentric correction using an FFT shift and
smoothed the data with a Tukey 25\% filter (see Barnes \etal 2001). Bandpass correction
was then applied using reference spectra derived from the median of all spectra taken within 2 min 
($\pm2\degr$ in the scan direction) of each spectrum being corrected, except where a scan boundary
was encountered. The reference spectrum was calculated separately for each beam and polarisation. Up to 
49 spectra, including the spectrum being corrected, were used to form each reference spectrum. 
The median statistic helped to suppress any RFI and any compact
HI emission in the reference spectrum, which would otherwise appear as a spatial sidelobe. 
The bandpass was then corrected, and the flux density calibration applied. 

Automated and manual quality control measures were used to ensure that high data quality was maintained.
Scans containing bad data, or scans containing telescope or correlator errors were always 
re-observed, unless the errors were not recognised until after the completion of all ZOA
observations.

Gridding into sky cubes was performed using {\sc gridzilla}, also written especially for
the large datasets arising from multibeam observations (this data set consists of over 24
million spectra). Cubes were made using a simple
top-hat median filter of all the data for a given spectral channel lying within a radius of
6\arcmin\ of each pixel. Although this technique loses S/N ratio relative to normal least-squares
parametric techniques, it proved extremely effective in removing residual RFI or
variable baseline ripple without extensive manual intervention. As noted in Barnes \etal
(2001), this particular gridding kernel requires an adjustment to the flux scale of 28\%
in order that the flux density of compact sources is preserved. This is applied directly
to the cubes.

The pixel size of the final cubes is 4\arcmin\ in RA and Dec and 13.2\kms\ in velocity 
(at zero redshift). The gridded FWHP beamsize is approximately 15\farcm5, compared with the 
normal Parkes beam at 1400 MHz of 14\farcm3, averaged across the 13 beams. Most of the 
considerable continuum emission was removed using the bandpass correction procedure 
described above. However, residual continuum emission remained, and this was further 
suppressed using the `scaled template method' ({\sc luther}), written by one
of us and described in Barnes \etal 
(2001), where a weighted spectral template derived from the strongest sources in the field 
was scaled in amplitude to fit to the spectrum at each point in the cube. Finally, the
data cubes were Hanning-smoothed to minimise ripple from strong Galactic \HI\ signals. This
results in a final velocity resolution of 27\kms. Spectra from the final cubes are available
for download\footnote{http://www.atnf.csiro.au/research/multibeam/release -- select `Data Source = ZOA'.}.

\section{Sample Compilation and \HI\ Parametrization} \label{sample}


The initial 
HIPASS automated source list (Meyer \etal 2004) required the visual inspection of 33 times
the number of galaxies that appeared in the final catalog. Due to residual
baseline excursions, especially in the presence of strong continuum, the efficiency
of automated algorithms applied to the HIZOA data was even lower. Therefore, we chose to create
the HIZOA catalog by visual searches alone.  Each Hanning-smoothed cube was
independently visually inspected by 2 or 3 authors, over the entire
velocity range from $-1,200$\kms\ to $12,700$\kms, using the visualization
package {\sc karma} (Gooch 1996).  The candidate lists created by each
searcher were given to a single author who served as ``adjudicator'' for
all cubes to produce as uniform a final catalog as possible.  There were no
quantitative {\it a priori} selection criteria, although for a signal to be
accepted as extragalactic \HI\ it had to be at or exceeding the 
5$\sigma$ level in peak flux density, at least marginally extended in velocity, and
cleanly separated from Galactic \HI\ in velocity space.  Galactic gas tends
to be more diffuse than external galaxies at our resolution, which are typically
unresolved by the Parkes beam, although compact clouds certainly exist.
While it is difficult to securely discriminate between High-Velocity Clouds
and nearby dwarf galaxies in \HI , the former are generally spatially more
extended and visibly related to lower-velocity Galactic gas in our cubes.
Generally, the \HI\ sources showed distributions in velocity space
consistent with known \HI\ sources, i.e. either two-horned, flat-topped, or
Gaussian profiles, and are well separated from Galactic gas velocities.

Once the final list of sources was made, the determination of \HI\
parameters was done using the program {\sc mbspect} within the {\sc miriad}
package (Sault \etal 1995).  For each source, zeroth-moment maps were made,
and the centroid of the \HI\ emission was obtained by Gaussian fitting with
either a FWHM equal to the gridded telescope beam for unresolved galaxies,
or a Gaussian of matching width in the case of resolved galaxies.  Using
this fitted position, the weighted sum of the emission along the spectral
dimension of the datacube was calculated, producing the
one-dimensional \HI\ spectrum.  Each spectrum was visually inspected, and a
low-order polynomial was fit to the line-free channels and subtracted, to
remove any slowly-varying spectral baseline.  The total flux due to \HI\ was
then determined from this baseline-subtracted spectrum by integrating
across the channels containing 21-cm emission.  The heliocentric velocity
(in the optical convention, $v = cz$) of each source was determined by
taking the average of the velocity values at the 50\% of peak flux points
on the profile.  Linewidths at the 50\% or 20\% of peak flux levels were
measured ($w_{50}$ and $w_{20}$ respectively), using a width-maximizing
algorithm.  To correct for instrumental broadening due to the coarse
velocity resolution after Hanning smoothing (27\kms), the values of
$w_{50}$ and $w_{20}$ were decreased by 14 and 21\kms , respectively
(Henning \etal\ 2000).  The errors on all values were calculated using the
formalism of Koribalski \etal (2004); they do not take into account
baseline fitting errors.  The errors that depend on linewidth (errors on
heliocentric velocity, $w_{50}$ and $w_{20}$) were calculated using the
observed (uncorrected) linewidths and are therefore somewhat conservative
values.  In the case that the linewidth at 20\% of peak flux was not
robustly measurable due to the signal's being too close to the noise level,
no value is listed in the catalog (Table~\ref{tabdetsex}).  Because the
errors on heliocentric velocity and $w_{50}$ also depend on $w_{20}$, via
measurement of the steepness of the profile edges, the sample average value
of ($w_{20} - w_{50}) = 42$\kms\ was used to calculate these two errors
when $w_{20}$ was not available for a particular source.

\begin{figure*}[ht]
\begin{center}
\includegraphics[bb=40 100 550 800,clip=true,width=0.65\textheight]{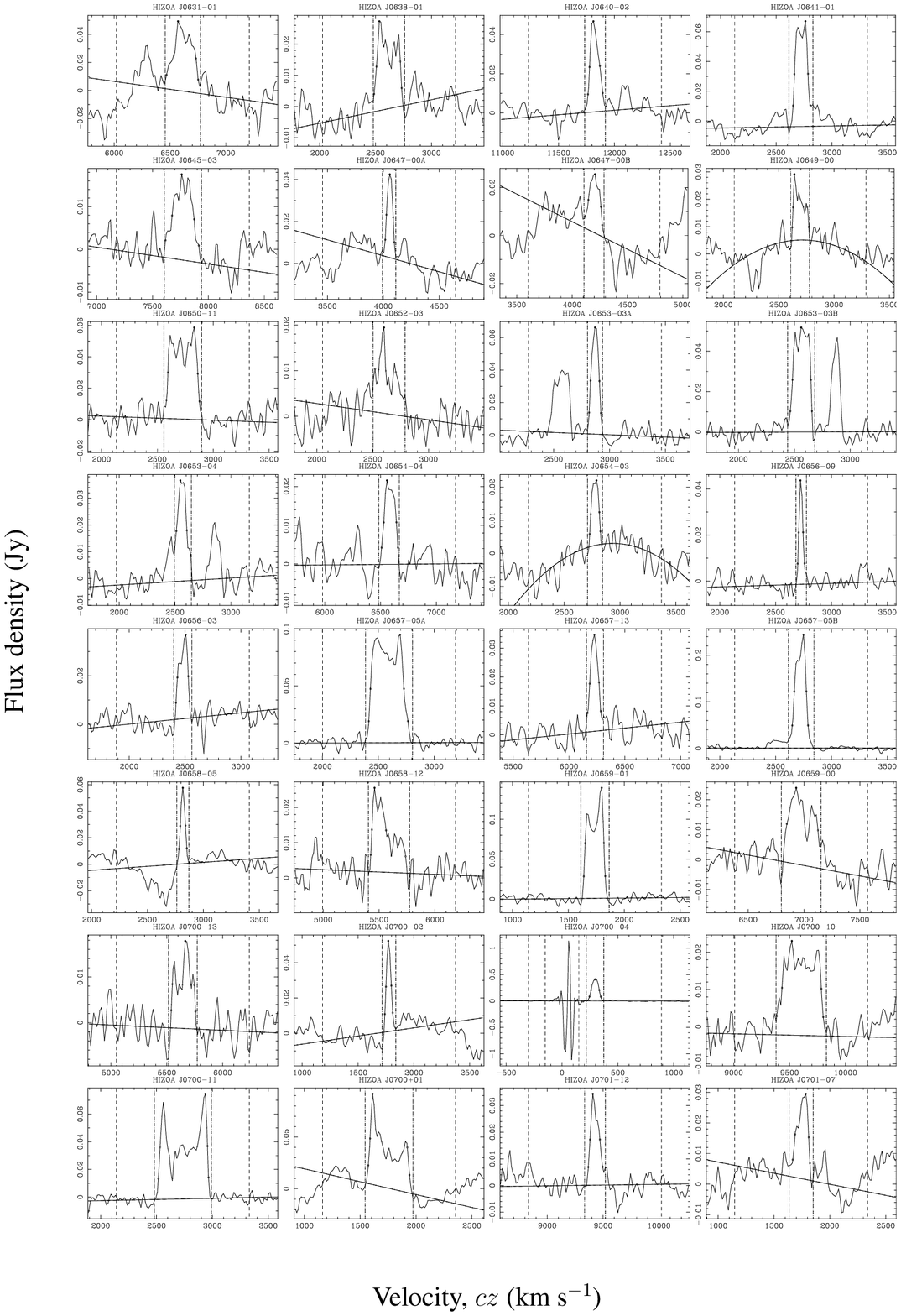}
\end{center}
\caption{Example \HI\ spectra of the newly detected galaxies in the HIZOA-S
survey. Low order baselines (indicated by the solid line) are fitted, excluding the
detections themselves (which are bracketed by the dash-dot vertical lines) and excluding the low and high-velocity 
edges to the left and right of the dashed vertical lines, respectively. 20\% and 50\% profile
markers are visible. The remaining 801 spectra are available in the online
version of the Journal.\label{f:dets}}
\label{exampleplot}
\end{figure*}

%
\begin{deluxetable}{
l@{\extracolsep{0.mm}}    c@{\extracolsep{1.mm}}            r@{\extracolsep{2.mm}}                                   r@{\extracolsep{4.mm}}                         
 r@{\extracolsep{1.mm}}                                     r@{\extracolsep{-1.mm}}                                  r@{\extracolsep{2.mm}}                        
 r@{\extracolsep{1.mm}}                                     r@{\extracolsep{1.mm}}                                   r@{\extracolsep{3.mm}}                       
 r@{\extracolsep{-1.mm}}                                    r@{\extracolsep{-1.mm}}                                  r@{\extracolsep{1.mm}}
 r}
\tabletypesize{\scriptsize}
\tablecolumns{14}
\tablewidth{0pc}
\tablecaption{\HI\ and derived parameters. This table is published in its entirety in the electronic edition. \label{tabdetsex}} 
\tablehead{
HIZOA ID   & f                                                & \multicolumn{1}{c@{\extracolsep{2.mm}}}{RA}            & \multicolumn{1}{c@{\extracolsep{4.mm}}}{Dec}      
 & \multicolumn{1}{c@{\extracolsep{1.mm}}}{$l$}               & \multicolumn{1}{c@{\extracolsep{-1.mm}}}{$b$}          & \multicolumn{1}{r@{\extracolsep{2.mm}}}{$E(B-V)$}   
 & \multicolumn{1}{c@{\extracolsep{1.mm}}}{$v_{\rm hel}$}     & \multicolumn{1}{c@{\extracolsep{1.mm}}}{$w_{50}$}      & \multicolumn{1}{c@{\extracolsep{3.mm}}}{$w_{20}$} 
 & \multicolumn{1}{c@{\extracolsep{-1.mm}}}{Flux}             & \multicolumn{1}{c@{\extracolsep{-1.mm}}}{$v_{\rm LG}$} & \multicolumn{1}{c@{\extracolsep{1.mm}}}{$D$}       
 & \multicolumn{1}{c@{\extracolsep{1.mm}}}{$\log M_{\rm HI}$} \\
                &                                             & \multicolumn{2}{c@{\extracolsep{2.mm}}}{J2000.0}        
 & \multicolumn{1}{c@{\extracolsep{1.mm}}}{[deg]}             & \multicolumn{1}{c@{\extracolsep{-1.mm}}}{[deg]}        &                                                   
 & \multicolumn{1}{c@{\extracolsep{1.mm}}}{[\kmst]}           & \multicolumn{1}{c@{\extracolsep{1.mm}}}{[\kmst]}       & \multicolumn{1}{c@{\extracolsep{3.mm}}}{[\kmst]}  
 & \multicolumn{1}{r@{\extracolsep{-1.mm}}}{[\Jykmst]}        & \multicolumn{1}{c@{\extracolsep{-1.mm}}}{[\kmst]}      & \multicolumn{1}{c@{\extracolsep{1.mm}}}{[Mpc]}   
 & \multicolumn{1}{c@{\extracolsep{1.mm}}}{[\msun]} \\
\multicolumn{1}{c}{(1)}      & \multicolumn{1}{c}{(2)}        & \multicolumn{1}{c@{\extracolsep{2.mm}}}{(3a)}          & \multicolumn{1}{c@{\extracolsep{4.mm}}}{(3b)}      
 & \multicolumn{1}{c@{\extracolsep{1.mm}}}{(4a)}              & \multicolumn{1}{c@{\extracolsep{-1.mm}}}{(4b)}         & \multicolumn{1}{c@{\extracolsep{2.mm}}}{(5)}      
 & \multicolumn{1}{c@{\extracolsep{1.mm}}}{(6)}               & \multicolumn{1}{c@{\extracolsep{1.mm}}}{(7)}           & \multicolumn{1}{c@{\extracolsep{3.mm}}}{(8)}      
 & \multicolumn{1}{c@{\extracolsep{-1.mm}}}{(9)}              & \multicolumn{1}{c@{\extracolsep{-1.mm}}}{(10)}         & \multicolumn{1}{c@{\extracolsep{1.mm}}}{(11)}    
 & \multicolumn{1}{c@{\extracolsep{1.mm}}}{(12)}  
}
\startdata
J0631$-$01   & &  06 31 49.3  & $-01$ 40 25  &  212.19  &  $-5.13$  &    0.77  &   6634$\pm$\phantom{}22  &  184$\pm$\phantom{}44  &  262$\pm$\phantom{}67   &    9.2$\pm$\phantom{00}3.1   &  6475  &   86.3  &  10.21 \\
J0638$-$01   & &  06 38 16.6  & $-01$ 28 56  &  212.75  &  $-3.60$  &    1.09  &   2615$\pm$\phantom{0}8  &  200$\pm$\phantom{}16  &  226$\pm$\phantom{}25   &    4.7$\pm$\phantom{00}1.0   &  2453  &   32.7  &   9.07 \\
J0640$-$02   & &  06 40 53.6  & $-02$ 39 50  &  214.10  &  $-3.56$  &    1.00  &  11824$\pm$\phantom{0}6  &   78$\pm$\phantom{}12  &  111$\pm$\phantom{}18   &    4.3$\pm$\phantom{00}0.8   & 11656  &  155.4  &  10.38 \\
J0641$-$01   & &  06 41 01.9  & $-01$ 41 34  &  213.25  &  $-3.09$  &    0.82  &   2725$\pm$\phantom{0}5  &  108$\pm$\phantom{}11  &  134$\pm$\phantom{}16   &    8.6$\pm$\phantom{00}1.4   &  2561  &   34.1  &   9.37 \\
J0645$-$03   & &  06 45 20.1  & $-03$ 05 58  &  215.00  &  $-2.77$  &    1.14  &   7774$\pm$\phantom{}13  &  183$\pm$\phantom{}25  &  277$\pm$\phantom{}38   &    3.8$\pm$\phantom{00}0.6   &  7602  &  101.4  &   9.96 \\
J0647$-$00A  & &  06 47 03.5  & $-00$ 36 31  &  212.97  &  $-1.25$  &    0.78  &   4058$\pm$\phantom{0}8  &   47$\pm$\phantom{}16  &   60$\pm$\phantom{}23   &    2.4$\pm$\phantom{00}1.0   &  3895  &   51.9  &   9.18 \\
J0647$-$00B  & &  06 47 18.8  & $-00$ 50 33  &  213.21  &  $-1.30$  &    0.78  &   4203$\pm$\phantom{}12  &  100$\pm$\phantom{}24  &  133$\pm$\phantom{}36   &    2.7$\pm$\phantom{00}0.9   &  4039  &   53.8  &   9.26 \\
J0649$-$00   & &  06 49 40.4  & $-00$ 10 36  &  212.89  &  $-0.47$  &    0.78  &   2693$\pm$\phantom{0}6  &  114$\pm$\phantom{}12  &  129$\pm$\phantom{}18   &    2.2$\pm$\phantom{00}0.6   &  2530  &   33.7  &   8.77 \\
J0650$-$11   & &  06 50 13.7  & $-11$ 13 08  &  222.81  &  $-5.35$  &    0.82  &   2728$\pm$\phantom{0}6  &  253$\pm$\phantom{}12  &  275$\pm$\phantom{}19   &   13.1$\pm$\phantom{00}1.9   &  2525  &   33.7  &   9.54 \\
J0652$-$03   & &  06 52 14.0  & $-03$ 40 01  &  216.29  &  $-1.49$  &    1.06  &   2614$\pm$\phantom{}12  &  173$\pm$\phantom{}23  &    \nodata              &    2.5$\pm$\phantom{00}0.7   &  2436  &   32.5  &   8.78 \\
J0653$-$03A  & &  06 53 10.6  & $-03$ 59 57  &  216.69  &  $-1.44$  &    1.23  &   2870$\pm$\phantom{0}4  &   66$\pm$\phantom{0}8  &   84$\pm$\phantom{}12   &    5.1$\pm$\phantom{00}0.8   &  2691  &   35.9  &   9.19 \\
J0653$-$03B  & &  06 53 21.1  & $-03$ 53 32  &  216.61  &  $-1.35$  &    1.17  &   2565$\pm$\phantom{0}5  &  151$\pm$\phantom{}10  &  173$\pm$\phantom{}16   &    8.2$\pm$\phantom{00}1.2   &  2386  &   31.8  &   9.29 \\
J0653$-$04   & &  06 53 52.5  & $-04$ 00 18  &  216.77  &  $-1.28$  &    1.25  &   2559$\pm$\phantom{0}7  &   80$\pm$\phantom{}14  &    \nodata              &    3.8$\pm$\phantom{00}0.8   &  2379  &   31.7  &   8.95 \\
J0654$-$04   & &  06 54 10.0  & $-04$ 44 26  &  217.46  &  $-1.55$  &    1.23  &   6586$\pm$\phantom{0}9  &   99$\pm$\phantom{}19  &  126$\pm$\phantom{}28   &    2.3$\pm$\phantom{00}0.7   &  6403  &   85.4  &   9.60 \\
J0654$-$03   & &  06 54 41.8  & $-03$ 15 38  &  216.21  &  $-0.76$  &    0.76  &   2777$\pm$\phantom{0}7  &   63$\pm$\phantom{}14  &   75$\pm$\phantom{}21   &    1.5$\pm$\phantom{00}0.5   &  2600  &   34.7  &   8.62 \\
J0656$-$09   & &  06 56 10.9  & $-09$ 34 29  &  222.00  &  $-3.31$  &    0.70  &   2726$\pm$\phantom{0}3  &   25$\pm$\phantom{0}7  &   40$\pm$\phantom{}10   &    1.9$\pm$\phantom{00}0.4   &  2525  &   33.7  &   8.69 \\
J0656$-$03   & &  06 56 16.1  & $-03$ 42 29  &  216.78  &  $-0.62$  &    0.99  &   2481$\pm$\phantom{0}5  &   83$\pm$\phantom{}10  &  101$\pm$\phantom{}15   &    3.0$\pm$\phantom{00}0.6   &  2301  &   30.7  &   8.82 \\
J0657$-$05A  & &  06 57 30.0  & $-05$ 11 12  &  218.24  &  $-1.02$  &    1.01  &   2580$\pm$\phantom{0}5  &  286$\pm$\phantom{0}9  &  329$\pm$\phantom{}14   &   25.7$\pm$\phantom{00}2.0   &  2394  &   31.9  &   9.79 \\
J0657$-$13   & &  06 57 46.0  & $-13$ 10 58  &  225.40  &  $-4.59$  &    0.64  &   6227$\pm$\phantom{0}8  &   69$\pm$\phantom{}16  &   97$\pm$\phantom{}24   &    2.8$\pm$\phantom{00}0.8   &  6014  &   80.2  &   9.63 \\
J0657$-$05B  & &  06 57 57.8  & $-05$ 19 26  &  218.41  &  $-0.98$  &    1.09  &   2721$\pm$\phantom{0}3  &   94$\pm$\phantom{0}6  &  120$\pm$\phantom{}10   &   25.4$\pm$\phantom{00}2.6   &  2535  &   33.8  &   9.83 \\
J0658$-$05   & &  06 58 20.7  & $-05$ 58 48  &  219.04  &  $-1.19$  &    1.18  &   2817$\pm$\phantom{0}6  &   32$\pm$\phantom{}11  &   50$\pm$\phantom{}17   &    2.7$\pm$\phantom{00}0.9   &  2628  &   35.0  &   8.90 \\
J0658$-$12   & &  06 58 34.8  & $-12$ 19 52  &  224.73  &  $-4.03$  &    0.58  &   5500$\pm$\phantom{}22  &  113$\pm$\phantom{}43  &  291$\pm$\phantom{}65   &    4.0$\pm$\phantom{00}0.9   &  5289  &   70.5  &   9.67 \\
J0659$-$01   & &  06 59 18.8  & $-01$ 31 14  &  215.18  &  $ 1.06$  &    0.69  &   1734$\pm$\phantom{0}3  &  154$\pm$\phantom{0}6  &  171$\pm$\phantom{0}9   &   18.3$\pm$\phantom{00}1.9   &  1561  &   20.8  &   9.27 \\
J0659$-$00   & &  06 59 51.5  & $-00$ 23 42  &  214.24  &  $ 1.69$  &    0.60  &   6985$\pm$\phantom{}12  &  264$\pm$\phantom{}23  &    \nodata              &    5.6$\pm$\phantom{00}1.3   &  6816  &   90.9  &  10.04 \\
J0700$-$13   & &  07 00 07.2  & $-13$ 54 18  &  226.31  &  $-4.41$  &    0.52  &   5648$\pm$\phantom{0}8  &  183$\pm$\phantom{}15  &  196$\pm$\phantom{}23   &    2.6$\pm$\phantom{00}0.7   &  5432  &   72.4  &   9.51 \\
J0700$-$02   & &  07 00 19.4  & $-02$ 23 44  &  216.08  &  $ 0.88$  &    0.73  &   1774$\pm$\phantom{0}5  &   34$\pm$\phantom{}10  &   50$\pm$\phantom{}14   &    2.5$\pm$\phantom{00}0.7   &  1597  &   21.3  &   8.43 \\
J0700$-$04   & &  07 00 29.5  & $-04$ 12 28  &  217.71  &  $ 0.09$  &    0.87  &    297$\pm$\phantom{0}3  &   60$\pm$\phantom{0}6  &   79$\pm$\phantom{0}9   &   30.6$\pm$\phantom{00}3.8   &   113  &    1.5  &   7.22 \\
J0700$-$10   & &  07 00 51.0  & $-10$ 21 40  &  223.22  &  $-2.64$  &    0.58  &   9608$\pm$\phantom{}12  &  338$\pm$\phantom{}24  &  394$\pm$\phantom{}37   &    7.6$\pm$\phantom{00}1.3   &  9403  &  125.4  &  10.45 \\
J0700$-$11   & &  07 00 58.1  & $-11$ 47 17  &  224.51  &  $-3.26$  &    0.56  &   2744$\pm$\phantom{0}3  &  410$\pm$\phantom{0}7  &  429$\pm$\phantom{}10   &   18.5$\pm$\phantom{00}1.4   &  2534  &   33.8  &   9.70 \\
J0700$+$01   &*&  07 01 04.2  & $+01$ 54 02  &  212.34  &  $ 3.01$  &    0.45  &   1755$\pm$\phantom{0}6  &  319$\pm$\phantom{}12  &  348$\pm$\phantom{}19   &   15.5$\pm$\phantom{00}2.3   &  1595  &   21.3  &   9.22 \\
J0701$-$12   & &  07 01 21.0  & $-12$ 14 31  &  224.96  &  $-3.39$  &    0.68  &   9427$\pm$\phantom{0}6  &   75$\pm$\phantom{}12  &  107$\pm$\phantom{}18   &    2.9$\pm$\phantom{00}0.6   &  9215  &  122.9  &  10.01 \\
J0701$-$07   & &  07 01 58.5  & $-07$ 18 38  &  220.64  &  $-1.00$  &    0.94  &   1750$\pm$\phantom{0}8  &  107$\pm$\phantom{}16  &  133$\pm$\phantom{}24   &    3.1$\pm$\phantom{00}0.8   &  1555  &   20.7  &   8.50 \\
J0702$-$11   & &  07 02 09.6  & $-11$ 34 48  &  224.46  &  $-2.91$  &    1.50  &   2819$\pm$\phantom{0}6  &   49$\pm$\phantom{}12  &   68$\pm$\phantom{}18   &    2.3$\pm$\phantom{00}0.6   &  2609  &   34.8  &   8.81 \\
J0702$-$15   & &  07 02 12.9  & $-15$ 27 57  &  227.93  &  $-4.66$  &    0.51  &   8145$\pm$\phantom{}13  &   43$\pm$\phantom{}26  &   52$\pm$\phantom{}39   &    0.9$\pm$\phantom{00}0.7   &  7923  &  105.6  &   9.36 \\
J0702$-$03A  & &  07 02 15.3  & $-03$ 13 38  &  217.04  &  $ 0.93$  &    1.76  &   6685$\pm$\phantom{}12  &  402$\pm$\phantom{}23  &  417$\pm$\phantom{}35   &    5.2$\pm$\phantom{00}1.5   &  6504  &   86.7  &   9.96 \\
J0702$-$03B  & &  07 02 34.5  & $-03$ 18 30  &  217.15  &  $ 0.97$  &    0.99  &   2579$\pm$\phantom{0}6  &  123$\pm$\phantom{}11  &  136$\pm$\phantom{}17   &    3.7$\pm$\phantom{00}0.8   &  2398  &   32.0  &   8.95 \\
J0702$-$12   & &  07 02 50.3  & $-12$ 19 37  &  225.20  &  $-3.10$  &    0.67  &   9327$\pm$\phantom{0}6  &   79$\pm$\phantom{}12  &   92$\pm$\phantom{}18   &    2.4$\pm$\phantom{00}0.7   &  9114  &  121.5  &   9.93 \\
J0704$-$13   & &  07 04 25.1  & $-13$ 42 07  &  226.60  &  $-3.39$  &    0.65  &   9000$\pm$\phantom{0}9  &   78$\pm$\phantom{}18  &    \nodata              &    3.4$\pm$\phantom{00}0.9   &  8782  &  117.1  &  10.04 \\
J0705$+$02   & &  07 05 37.1  & $+02$ 37 01  &  212.22  &  $ 4.35$  &    0.39  &   1743$\pm$\phantom{0}4  &   38$\pm$\phantom{0}9  &   58$\pm$\phantom{}13   &    6.2$\pm$\phantom{00}1.4   &  1583  &   21.1  &   8.81 \\
J0705$-$12   & &  07 05 39.9  & $-12$ 59 55  &  226.11  &  $-2.80$  &    0.60  &   5455$\pm$\phantom{0}4  &  185$\pm$\phantom{0}9  &  198$\pm$\phantom{}13   &    5.8$\pm$\phantom{00}0.9   &  5239  &   69.9  &   9.83 \\
J0706$-$04   & &  07 06 15.9  & $-04$ 55 23  &  219.00  &  $ 1.04$  &    0.51  &   2481$\pm$\phantom{0}6  &   53$\pm$\phantom{}12  &   71$\pm$\phantom{}18   &    2.4$\pm$\phantom{00}0.7   &  2292  &   30.6  &   8.72 \\
J0706$-$06   & &  07 06 55.2  & $-06$ 24 24  &  220.40  &  $ 0.51$  &    0.71  &   2508$\pm$\phantom{0}4  &   69$\pm$\phantom{0}7  &   83$\pm$\phantom{}11   &    5.3$\pm$\phantom{00}0.9   &  2313  &   30.8  &   9.08 \\
J0707$-$11   & &  07 07 12.3  & $-11$ 30 20  &  224.96  &  $-1.78$  &    0.60  &   1730$\pm$\phantom{0}8  &  162$\pm$\phantom{}16  &  173$\pm$\phantom{}23   &    2.3$\pm$\phantom{00}0.7   &  1518  &   20.2  &   8.34 \\
J0707$-$14   & &  07 07 24.5  & $-14$ 24 55  &  227.57  &  $-3.07$  &    0.61  &   2732$\pm$\phantom{}10  &   32$\pm$\phantom{}20  &   60$\pm$\phantom{}30   &    1.3$\pm$\phantom{00}0.6   &  2511  &   33.5  &   8.53 \\
J0708$-$08   & &  07 08 18.6  & $-08$ 17 01  &  222.22  &  $-0.05$  &    0.64  &   6472$\pm$\phantom{0}7  &  230$\pm$\phantom{}13  &  267$\pm$\phantom{}20   &    5.2$\pm$\phantom{00}0.7   &  6270  &   83.6  &   9.93 \\
J0708$-$01   & &  07 08 40.8  & $-01$ 25 41  &  216.17  &  $ 3.18$  &    0.24  &   7483$\pm$\phantom{0}9  &  186$\pm$\phantom{}17  &    \nodata              &    3.8$\pm$\phantom{00}0.8   &  7306  &   97.4  &   9.93 \\
J0709$-$03   & &  07 09 01.0  & $-03$ 57 37  &  218.46  &  $ 2.09$  &    0.33  &   3663$\pm$\phantom{0}6  &   78$\pm$\phantom{}12  &   90$\pm$\phantom{}18   &    2.0$\pm$\phantom{00}0.6   &  3476  &   46.4  &   9.01 \\
J0709$-$05   & &  07 09 34.2  & $-05$ 24 14  &  219.81  &  $ 1.55$  &    0.40  &   1719$\pm$\phantom{0}5  &  248$\pm$\phantom{}10  &  273$\pm$\phantom{}15   &   14.4$\pm$\phantom{00}1.6   &  1527  &   20.4  &   9.15 \\
J0710$-$07   & &  07 10 51.0  & $-07$ 57 48  &  222.23  &  $ 0.65$  &    0.55  &   2405$\pm$\phantom{0}8  &   92$\pm$\phantom{}16  &  123$\pm$\phantom{}24   &    2.6$\pm$\phantom{00}0.7   &  2203  &   29.4  &   8.72 \\
J0711$-$05   & &  07 11 45.0  & $-05$ 19 26  &  219.99  &  $ 2.07$  &    0.34  &   3657$\pm$\phantom{}12  &   34$\pm$\phantom{}25  &   66$\pm$\phantom{}37   &    2.2$\pm$\phantom{00}1.1   &  3464  &   46.2  &   9.05 \\
J0712$-$09   & &  07 12 48.6  & $-09$ 18 58  &  223.65  &  $ 0.46$  &    0.56  &   2437$\pm$\phantom{}10  &   84$\pm$\phantom{}20  &    \nodata              &    2.9$\pm$\phantom{00}0.9   &  2230  &   29.7  &   8.78 \\
J0713$-$01   & &  07 13 24.2  & $-01$ 31 03  &  216.79  &  $ 4.19$  &    0.15  &   4957$\pm$\phantom{0}5  &   47$\pm$\phantom{0}9  &   63$\pm$\phantom{}14   &    1.7$\pm$\phantom{00}0.4   &  4778  &   63.7  &   9.22 \\
J0713$-$07A  & &  07 13 34.5  & $-07$ 51 10  &  222.44  &  $ 1.30$  &    0.53  &   2471$\pm$\phantom{0}3  &   57$\pm$\phantom{0}7  &   78$\pm$\phantom{}10   &   14.2$\pm$\phantom{00}1.9   &  2268  &   30.2  &   9.48 \\
\enddata
\end{deluxetable}



\subsection{The Catalog}

The \HI\ profiles for the first 32 of the 883 galaxies in the survey are
shown in Figure~\ref{f:dets} (the remainder are available in the online
version of the Journal). Vertical lines indicate the spectral ranges used
for baseline subtraction, and the linear or polynomial fit is shown.  On
each profile, the peak, 50\% and 20\% levels are noted with dots.
Table~\ref{tabdetsex} is an example page of the full catalog (the table is
published in its entirety in the electronic edition of the Astronomical
Journal); it lists \HI\ parameters and derived quantities for the galaxies
in the following columns:

{\it Columns (1) and (2)} - Source name and flag.  HIZOA galaxies in the
Great Attractor region which were first reported by Juraszek \etal (2000)
and in HIZOA-N (Donley \etal 2005) retain their original names, even if the
position measurement has been improved by this survey. Names affected by
those positional changes are indicated with an asterisk in Col.~2. The \HI\
parameters quoted are for the current work.

{\it Columns (3a and 3b)} - Equatorial coordinates (J2000.0) of the fitted
position. 

{\it Columns (4a and 4b)} - Galactic coordinates.

{\it Column (5)} - Reddening $E(B-V)$ as derived from the IRAS/DIRBE
maps (Schlegel \etal 1989) and corrected with a factor of 0.86 as derived
by Schlafly \& Finkbeiner (2011). 

{\it Column (6)} - Heliocentric velocity and error.

{\it Column (7)} - Velocity width at 50\% of peak flux density, corrected
for instrumental broadening, and associated error.

{\it Column (8)} - Velocity width at 20\% of peak flux density, corrected
for instrumental broadening, and associated error.

{\it Column (9)} - \HI\ flux integral and associated error.

{\it Column (10)} - Velocity of the galaxy corrected to the Local Group
frame of reference via: 
$$ v_{\rm LG} = v_{\rm hel} + 300 \sin \ell \cos b $$ 

{\it Column (11)} - Distance to the galaxy in Megaparsec, based on $v_{\rm
LG}$ and H$_0$ = 75\kms Mpc$^{-1}$.

{\it Column (12)} - Logarithm of the \HI\ mass.

\section{\HI\ Properties of the Sample}  \label{HIparam}

\begin{figure}
\centering
\includegraphics[width=0.44\textwidth]{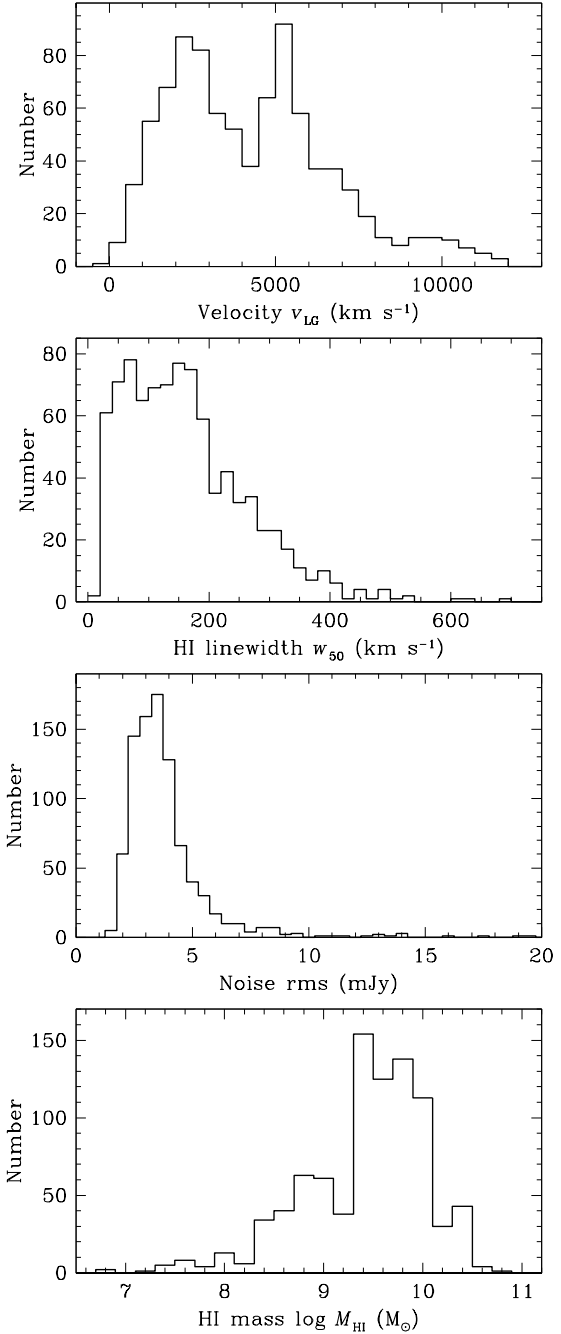}
\caption{HI\ parameters of the 883 galaxies detected in the HIZOA-S
  survey. From top to bottom the histograms display the radial velocity
  $v_{\rm LG}$, the linewidth $w_{50}$, the clipped rms noise at the position of the
  detected galaxy, and the \HI-mass distribution. 
}
\label{hist_hiparam}
\end{figure}
\begin{figure}[th]
\centering
\includegraphics[width=0.44\textwidth]{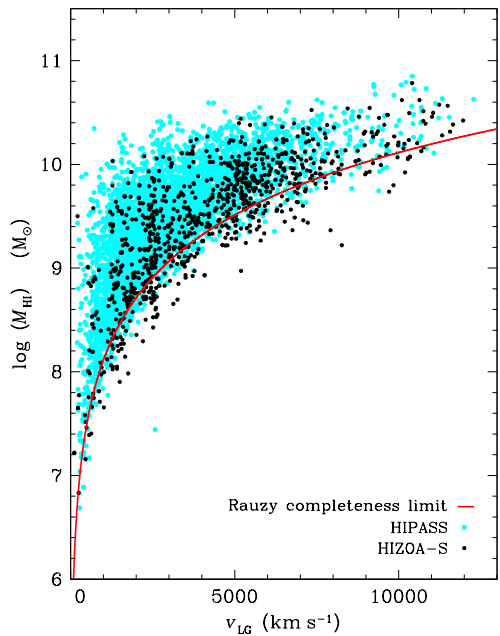}
\caption{\HI\ mass versus velocity in the Local Group frame for the HIZOA-S
(black) and HIPASS samples (light blue; Meyer \etal 2004). The Rauzy
completeness limit for flux integral alone (3.1\Jykms) is shown as a red
curve.
}
\label{distmhiplot}
\end{figure}

An overview of the \HI~properties of the sample of 883 galaxies is shown in
Fig.~\ref{hist_hiparam}.  The top panel shows the distribution of the
galaxies' recessional velocities in the Local Group frame of reference.
The distribution generally reflects the noise-limited sensitivity of the
survey, but also reveals the overdensity of galaxies in the Great Attractor
region, at about 5000\kms.  This feature is apparent despite the averaging
across a variety of large-scale structures in this wide-angle survey,
described in more detail in \S7.  The HIZOA survey was designed to map
large-scale structure at the distance of the Great Attractor more
completely than the HIPASS survey, which did not have the sensitivity
to trace structure beyond about 4000\kms\ (the rms for HIPASS is
$13-22$\,mJy in the Galactic Plane versus an average of 6\,mJy for HIZOA).
This allowed HIZOA to detect galaxies at lower \HI\ masses, and thus probe
lower down the \HI\ mass function at Great Attractor distances
(eg. Fig.~\ref{distmhiplot}).

The next panel shows the distribution of \HI\ line\-widths measured at
half-peak, $w_{50}$, which has a mean value of 163\kms\ and a median of
147\kms, with large, non-Gaussian variation from the smallest value of
17\kms\ to 699\kms\ maximum. Three detections, which we believe are not confused, have $w_{50} > 600$\kms ,
the largest of which (J1416$-$58, $w_{50} = 699$\kms ) is a perfectly edge-on spiral.
The iteratively-clipped rms noise at the
location of each detected source is shown in the next panel.  Due to the clipping
and the limited velocity range over which the measurement is made, this is
normally much lower than the overall cube rms of 6\,mJy. Some
galaxies were found in areas with rms as high as $10 - 20$\,mJy at the edge
of the field or near strong Galactic foreground radiation.  The lowest
panel shows the distribution of the \HI~masses, which ranges from 6.8 to
10.8 in the logarithm, with a mean of 9.5 and a median of 9.6.  The most massive \HI\ object,
HIZOA J0836$-$43, has been the subject of radio interferometry and deep NIR
\clearpage

\noindent 
follow-up observations, showing it to be a high surface brightness, massive
disk galaxy with an extended \HI~disk (Donley \etal 2006; Cluver \etal 
2008, 2010).


The well-known Circinus galaxy lies within the survey area, and is
re-detected here as HIZOA J1413$-$65.  The narrow velocity width,
low-\HI-mass ($\log M_{\rm HI} = 7.24$) detection HIZOA J1353$-$58, less than $8\deg$ from
Circinus, has a recessional velocity that is only about 200\kms\ different from
the large galaxy.  At the low recession velocities of these objects of only a
few hundred \kms, Hubble flow distances are very uncertain, but it seems
likely this newly-detected object is a previously-unknown galaxy which is related to the
Circinus galaxy.  As its velocity corrected to the Local Group frame is
negative, and thus a Hubble distance is undefined in this frame, we adopt a
redshift-independent distance measurement to Circinus, 4.2\,Mpc (Tully \etal
2009), as the distance to both HIZOA J1413$-$65 and HIZOA J1353$-$58.

The Circinus galaxy as well as four other detections (J1509$-$52,
J1514$-$52, J1532$-$56, J1616$-$55) are significantly extended, that is,
resolved with respect to the $15\farcm5$ beam.

Figure~\ref{distmhiplot} shows the distribution of \HI\ mass as a function
of velocity in the Local Group frame for the HIZOA-S and HIPASS
(Meyer \etal 2004) samples. This illustrates the higher sensitivity of the
HIZOA survey over HIPASS, as HIZOA finds lower \HI-mass galaxies at all
distances, but also reflects the smaller area covered, which produced fewer
detections overall, compared to HIPASS.  The completeness limit for HIZOA-S
is also shown in the figure and is fully described in \S6.  As in the top
panel of Fig.~\ref{hist_hiparam}, despite the inclusion of a variety of
large-scale structures probed in this wide-angle survey, clear
overdensities at $\sim\!2500$\kms\ (Hy/Ant and Puppis) and
$\sim\!5000$\kms\ (the Great Attractor region) are evident in the velocity
distribution of the HIZOA galaxies.  These structures are discussed in
detail in \S7.

\section{Multiwavelength Counterparts} \label{cmatch}


To find counterparts for the \HI\ detections, we have (a) searched the
literature through online databases, and (b) searched
images at various wavelengths. 

Our main literature search was done using the NASA/IPAC Extragalactic
Database (NED)\footnote{http://ned.ipac.caltech.edu/}. We submitted a list
of the \HI\ detections through the NED batch service using a $4\farcm5$
search radius (status 2013 August).  For comparison, we also conducted a
search with the SIMBAD\footnote{http://simbad.u-strasbg.fr/simbad/}
Astronomical Database.  There were only a couple of cases in which SIMBAD had a
galaxy listed but NED did not. On the other hand, SIMBAD listed more
(unclassified) IRAS sources than NED, which we looked at (using
their colors as a first-order indicator for galaxies). The searches also
supplied a redshift where available.  Particular attention was paid to galaxies
with optical redshifts since these have negligible positional
uncertainties. 


For the visual search, we downloaded images from the SuperCOSMOS Sky
Surveys\footnote{http://www-wfau.roe.ac.uk/sss/} (\B -band), the Digitized
Sky Surveys
(DSS)\footnote{http://www3.cadc-ccda.hia-iha.nrc-cnrc.gc.ca/en/dss/} (\R -
and \II -band),
2MASS\footnote{http://irsa.ipac.caltech.edu/applications/2MASS/} (\K -band)
as well as UKIDSS\footnote{http://surveys.roe.ac.uk/wsa/} and
VISTA\footnote{http://horus.roe.ac.uk/vsa/} wherever possible (mainly \K
-band but if not available we used the band closest in
wavelength). Finally, we added \K -band images (as well as in \J\  or \HH\ 
when necessary) which were obtained at the Infrared Survey Facility at
Sutherland (IRSF) for the purpose of extracting photometry on all galaxies
within the HIZOA search circles. A first set of this catalog was published by
Williams \etal (2014) and the rest is in preparation (Said \etal in
prep.). 

While the field-of-view of the IRSF images is $8\arcmin\times8\arcmin$, the
downloaded images were extracted as $9\arcmin\times9\arcmin$ or, in some
cases, $10\arcmin\times10\arcmin$. The 2MASS images are only $8\farcm5$
wide and clearly not all of the $4\farcm5$-radius search circle could be
covered in a single image. 2MASS images covering the missing bits of the
search area were added when no suitable candidate could be found at any
other passband. In addition, the DIRBE/IRAS
maps\footnote{http://www.astro.princeton.edu/\~schlegel/dust/}
(Schlegel \etal 1998) were downloaded and viewed together with the images
to take into account the amount and distribution of extinction in the
area. WISE\footnote{http://irsa.ipac.caltech.edu/applications/wise/} images
(Wright \etal 2010) were used to better distinguish between galaxies and
Galactic objects.

Images were viewed with DS9\footnote{http://ds9.si.edu/} and overlaid with
regions indicating (i) the \HI\ position, (ii) the $4\farcm5$ search
radius, and (iii) the positions of the objects found with NED and
SIMBAD. We added positions from HIPASS (Meyer \etal 2004) and the HIZSS
shallow survey catalog (Henning \etal 2000). While displaying the \HI\
parameters, these images were searched simultaneously by eye to detect any
possible galaxy in the field. Most galaxies found in the literature were
visible in at least one passband, and quite often unpublished galaxies
could be found, especially blue late-type galaxies (often not or barely
visible in the NIR) or, on deep NIR images, galaxies at high extinctions or
near the Galactic bulge where 2MASS has difficulty detecting galaxies due
to severe star crowding.


If more than one galaxy was found, the most likely counterpart was selected
based on the \HI\ parameters, the appearance of the galaxy on the images and
the extinction information.  In some cases no reasonable decision could be
made on which of the galaxies is the counterpart, hence, both possible (but
ambiguous) candidates are retained. In other
cases, the chosen candidate was classified as `probable', indicating that
another galaxy could also be considered as the counterpart albeit less
likely. This class was also used when no second candidate was visible but
the appearance of the galaxy was not an ideal match with the \HI\
parameters. Finally, sometimes more than one candidate is considered to
contribute to the profile (i.e., the profile is likely to be `confused').

Some of the counterparts lie beyond the nominal $4\farcm5$ search
radius. They were found when a larger search circle seemed appropriate (\cf
the discussion below).



The counterparts are listed in Table~\ref{tabcrossex} (an example page of
which is shown here), which is published in its entirety in the electronic
edition of the Astronomical Journal. It is divided into three parts,
reflecting the different types or counterparts found as explained
above. Table~\ref{tabcrossex}a lists those HIZOA detections which either
have a single or no counterpart. Table~\ref{tabcrossex}b presents 
detections where more than one galaxy is assumed to contribute to the \HI\
profile (note that where the confusing partner was another of our \HI\
detections we do not list them separately).  Finally,
Table~\ref{tabcrossex}c lists those cases where more than one candidate was
found but, judged by the profile, only {\it one of them} is the likely
counterpart. The columns are as follows:

{\it Column (1)} - Source name as in Table~\ref{tabdetsex}.

{\it Columns (2a and 2b)} - Galactic coordinates of the \HI\ detection.

{\it Column (3)} - Distance to the \HI\ galaxy in Megaparsec, as in
Table~\ref{tabdetsex}.

{\it Column (4)} - Logarithm of the \HI\ mass, as in Table~\ref{tabdetsex}.

{\it Column (5)} - Extinction in the \B -band, converted from $E(B-V)$
given in Table~\ref{tabdetsex} using $R_{\rm B}=4.14$. A star denotes an
extinction value deemed to be uncertain during the search (e.g., due to high
spatial variability).

{\it Column (6)} - Classification of the counterpart; `d' = definite, `p' =
probable, `a' = ambiguous, `c' = confused candidate; `--' = no candidate.

{\it Column (7)} - Flags for counterparts in major catalogs: `I' stands for
IRAS Point Source Catalog (Helou \& Walker 1988), `M' for 2MASX
(Jarrett \etal 2000), `W' for Williams \etal 2014 (the IRSF catalog), `H'
for the HIPASS catalogs (South, Meyer \etal 2004 and North, Wong \etal
2006) or `h' for other HIPASS publications (Kilborn \etal 2002;
Ryan-Weber \etal\ 2002), `S' for the \HI\ Parkes ZOA Shallow Survey (HIZSS,
Henning \etal 2000), and `Z' for earlier HIZOA publications (Juraszek \etal
2000 and Donley \etal 2005).

{\it Column (8)} - Velocity in the literature (\cf NED and
HyperLEDA\footnote{http://leda.univ-lyon1.fr/}): `o' = optical (and, in one
case, `x' for X-ray), `h' = \HI\ (note that Juraszek \etal 2000 and
Henning \etal 2000 are not included here since their \HI\ parameters are
not independent of ours, but see flags in Col.~7).

{\it Column (9)} - Source `l' for name and coordinates: `N' is listed in NED, `S'
is listed in SIMBAD, `c' is coordinates measured on DSS or NIR images (note
that some published coordinates were not centered properly so we give the
measured ones).

{\it Column (10)} - Note in the appendix.

{\it Columns (11a and 11b)} - Equatorial coordinates (J2000.0) of the
counterpart.

{\it Column (12)} - Distance between the \HI\ fitted position and the
counterpart position in arcminutes.

{\it Column (13)} - One name in the literature in this order of preference:
NGC, ESO, RKK/WKK, CGMW, 2MASS, others. 

%
\begin{deluxetable}{
l                                                  r@{\extracolsep{1.mm}}                                       r@{\extracolsep{0.mm}}
 r@{\extracolsep{-1.mm}}                           r@{\extracolsep{1.mm}}                                       r@{\extracolsep{1.mm}}
 c@{\extracolsep{2.mm}}                            c@{\extracolsep{0.mm}}                                       c@{\extracolsep{0.mm}}
 c@{\extracolsep{0.mm}}                            c@{\extracolsep{0.mm}}                                       c@{\extracolsep{0.mm}}
 c@{\extracolsep{2.mm}}                            c@{\extracolsep{-2.mm}}                                      c@{\extracolsep{0.mm}}
 c@{\extracolsep{-2.mm}}                           c@{\extracolsep{1.mm}}                                       c@{\extracolsep{2.mm}}
 l@{\extracolsep{2.mm}}                            l@{\extracolsep{1.mm}}                                       r@{\extracolsep{4.mm}}
 l}
\tabletypesize{\scriptsize}
\tablecolumns{22}
\tablewidth{0pc}
\tablecaption{Crossmatches of the \HI\ detections. This table is published in its entirety in the electronic edition. \label{tabcrossex}}
\tablehead{
HIZOA ID                                           & \multicolumn{1}{c@{\extracolsep{1.mm}}}{$l$}               & \multicolumn{1}{c@{\extracolsep{0.mm}}}{$b$}   
 & \multicolumn{1}{c@{\extracolsep{-1.mm}}}{$D$}   & \multicolumn{1}{c@{\extracolsep{1.mm}}}{$\log M_{\rm HI}$} & \multicolumn{1}{c@{\extracolsep{1.mm}}}{$A_B$}             
 & class                                           & I                                                          & M
 & W                                               & H                                                          & S
 & Z                                               & o                                                          & h 
 & l                                               & c                                                          & Note  
 & \multicolumn{1}{c@{\extracolsep{2.mm}}}{RA}     & \multicolumn{1}{c@{\extracolsep{1.mm}}}{Dec}               & \multicolumn{1}{c@{\extracolsep{4.mm}}}{$d_{\rm sep}$}   
 & \multicolumn{1}{l@{\extracolsep{1.mm}}}{Name}   \\  
                                                   & \multicolumn{1}{c@{\extracolsep{1.mm}}}{[deg]}             & \multicolumn{1}{c@{\extracolsep{0.mm}}}{[deg]} 
 & \multicolumn{1}{c@{\extracolsep{-1.mm}}}{[Mpc]} & \multicolumn{1}{c@{\extracolsep{1.mm}}}{[\msun]}           & \multicolumn{1}{c@{\extracolsep{1.mm}}}{[mag]}           
 &                                                 &                                                            &  
 &                                                 &                                                            &  
 &                                                 &                                                            &   
 &                                                 &                                                            & 
 & \multicolumn{2}{c@{\extracolsep{1.mm}}}{J2000.0}                                                             & \multicolumn{1}{c@{\extracolsep{4.mm}}}{[$\prime$]} 
 &                                                 \\
\multicolumn{1}{c}{(1)}                            & \multicolumn{1}{c@{\extracolsep{1.mm}}}{(2a)}              & \multicolumn{1}{c@{\extracolsep{0.mm}}}{(2b)}   
 & \multicolumn{1}{c@{\extracolsep{-1.mm}}}{(3)}   & \multicolumn{1}{c@{\extracolsep{1.mm}}}{(4)}               & \multicolumn{1}{c@{\extracolsep{1.mm}}}{(5)}  
 & (6)                                             & \multicolumn{6}{c@{\extracolsep{0.mm}}}{(7)}                
                                                   & \multicolumn{2}{c@{\extracolsep{0.mm}}}{(8)}                
 & \multicolumn{2}{c@{\extracolsep{0.mm}}}{(9)}                                                                & (10) 
 & \multicolumn{1}{c@{\extracolsep{2.mm}}}{(11a)}  & \multicolumn{1}{c@{\extracolsep{1.mm}}}{(11b)}             & \multicolumn{1}{c@{\extracolsep{4.mm}}}{(12)}  
 & \multicolumn{1}{l@{\extracolsep{1.mm}}}{(13)} 
}
\startdata
\multicolumn{13}{l}{} \\ 
\multicolumn{13}{l}{(a) HI detections with single cross-matches:} \\
\multicolumn{13}{l}{} \\   
J0631$-$01    &  212.19  &  $-5.13$   &    86.3  &  10.21  &    3.2\phantom{*} &    p   &  -&-&-&H&-&-& -&h&  -&c  &  n  &  06 32 09.5   &          $-$01 36 53    &   6.1  &   --                        \\     
J0638$-$01    &  212.75  &  $-3.60$   &    32.7  &   9.07  &    4.5\phantom{*} &    d   &  -&-&-&-&-&-& -&-&  -&c  &  -  &  06 38 20.3   &          $-$01 28 38    &   1.0  &   --                        \\ 
J0640$-$02    &  214.10  &  $-3.56$   &   155.4  &  10.38  &    4.2\phantom{*} &    d   &  -&M&-&-&-&-& -&-&  N&-  &  -  &  06 40 56.02  &          $-$02 38 32.6  &   1.4  &   2MASX J06405601-0238326   \\  
J0641$-$01    &  213.25  &  $-3.09$   &    34.1  &   9.37  &    3.4\phantom{*} &    d   &  -&-&-&-&-&-& -&-&  -&c  &  -  &  06 40 58.4   &          $-$01 44 03    &   2.6  &   --                        \\ 
J0645$-$03    &  215.00  &  $-2.77$   &   101.4  &   9.96  &    4.7\phantom{*} &    d   &  -&M&-&-&-&-& -&-&  N&-  &  -  &  06 45 23.44  &          $-$03 07 14.3  &   1.5  &   2MASX J06452346-0307141   \\ 
J0647$-$00A   &  212.97  &  $-1.25$   &    51.9  &   9.18  &    3.2\phantom{*} &    -   &  -&-&-&-&-&-& -&-&  -&-  &  n  &  --           &          $ $            &   --   &   --                        \\ 
J0647$-$00B   &  213.21  &  $-1.30$   &    53.8  &   9.26  &    3.2\phantom{*} &    p   &  -&M&-&-&-&-& -&-&  N&-  &  -  &  06 47 13.16  &          $-$00 49 50.1  &   1.6  &   2MASX J06471318-0049501   \\ 
J0649$-$00    &  212.89  &  $-0.47$   &    33.7  &   8.77  &    3.2\phantom{*} &    d   &  -&-&-&-&-&-& -&-&  -&c  &  -  &  06 49 29.7   &          $-$00 09 03    &   3.1  &   --                        \\ 
J0650$-$11    &  222.81  &  $-5.35$   &    33.7  &   9.54  &    3.4\phantom{*} &    d   &  I&M&-&H&-&-& o&h&  N&-  &  -  &  06 50 10.61  &          $-$11 15 13.0  &   2.2  &   CGMW 1-0411               \\ 
J0652$-$03    &  216.29  &  $-1.49$   &    32.5  &   8.78  &    4.4\phantom{*} &    d   &  -&M&-&-&-&-& -&-&  N&-  &  -  &  06 52 00.20  &          $-$03 40 34.2  &   3.5  &   CGMW 1-0424               \\ 
J0653$-$03A   &  216.69  &  $-1.44$   &    35.9  &   9.19  &    5.1\phantom{*} &    p   &  -&-&-&H&-&-& -&h&  -&c  &  n  &  06 53 13.6   &          $-$04 00 04    &   0.7  &   --                        \\ 
J0653$-$03B   &  216.61  &  $-1.35$   &    31.8  &   9.29  &    4.8\phantom{*} &    d   &  -&-&-&H&-&-& -&h&  -&c  &  n  &  06 53 21.8   &          $-$03 52 57    &   0.6  &   --                        \\ 
J0653$-$04    &  216.77  &  $-1.28$   &    31.7  &   8.95  &    5.2\phantom{*} &    -   &  -&-&-&-&-&-& -&-&  -&-  &  n  &  --           &          $ $            &   --   &   --                        \\ 
J0654$-$04    &  217.46  &  $-1.55$   &    85.4  &   9.60  &    5.1\phantom{*} &    d   &  -&-&-&-&-&-& -&-&  -&c  &  -  &  06 53 59.7   &          $-$04 42 16    &   3.3  &   --                        \\ 
J0654$-$03    &  216.21  &  $-0.76$   &    34.7  &   8.62  &    3.2\phantom{*} &    p   &  -&-&-&-&-&-& -&-&  -&c  &  -  &  06 54 39.0   &          $-$03 16 23    &   1.0  &   --                        \\ 
J0656$-$09    &  222.00  &  $-3.31$   &    33.7  &   8.69  &    2.9\phantom{*} &    d   &  -&-&-&-&-&-& -&-&  -&c  &  -  &  06 56 09.2   &          $-$09 36 31    &   2.1  &   --                        \\ 
J0656$-$03    &  216.78  &  $-0.62$   &    30.7  &   8.82  &    4.1\phantom{*} &    p   &  -&-&-&H&-&-& -&h&  -&c  &  -  &  06 56 23.0   &          $-$03 45 26    &   3.4  &   --                        \\ 
J0657$-$05A   &  218.24  &  $-1.02$   &    31.9  &   9.79  &    4.2\phantom{*} &    d   &  I&M&-&H&-&-& -&h&  N&-  &  n  &  06 57 21.50  &          $-$05 08 59.6  &   3.1  &   CGMW 1-0464               \\ 
J0657$-$13    &  225.40  &  $-4.59$   &    80.2  &   9.63  &    2.7\phantom{*} &    d   &  -&M&-&-&-&-& -&-&  N&-  &  -  &  06 57 40.63  &          $-$13 11 15.2  &   1.3  &   2MASX J06574062-1311150   \\ 
J0657$-$05B   &  218.41  &  $-0.98$   &    33.8  &   9.83  &    4.5\phantom{*} &    d   &  I&M&W&H&S&-& o&h&  N&-  &  n  &  06 58 02.90  &          $-$05 20 41.0  &   1.8  &   CGMW 1-0470               \\ 
J0658$-$05    &  219.04  &  $-1.19$   &    35.0  &   8.90  &    4.9\phantom{*} &    -   &  -&-&-&H&-&-& -&h&  -&-  &  -  &  --           &          $ $            &   --   &   --                        \\ 
J0658$-$12    &  224.73  &  $-4.03$   &    70.5  &   9.67  &    2.4\phantom{*} &    d   &  -&M&-&-&-&-& -&-&  N&-  &  -  &  06 58 30.63  &          $-$12 22 00.8  &   2.4  &   CGMW 1-0472               \\ 
J0659$-$01    &  215.18  &  $ 1.06$   &    20.8  &   9.27  &    2.8\phantom{*} &    d   &  -&-&-&H&S&-& -&h&  N&c  &  -  &  06 59 20.4   &          $-$01 31 31    &   0.5  &   CGMW 1-0476               \\ 
J0659$-$00    &  214.24  &  $ 1.69$   &    90.9  &  10.04  &    2.5\phantom{*} &    d   &  I&M&-&-&-&-& o&-&  N&-  &  -  &  06 59 53.72  &          $-$00 25 27.5  &   1.8  &   CGMW 1-0479               \\ 
J0700$-$13    &  226.31  &  $-4.41$   &    72.4  &   9.51  &    2.2\phantom{*} &    d   &  -&-&W&-&-&-& -&-&  -&c  &  -  &  07 00 02.4   &          $-$13 52 05    &   2.5  &   --                        \\ 
J0700$-$02    &  216.08  &  $ 0.88$   &    21.3  &   8.43  &    3.0\phantom{*} &    d   &  -&-&-&H&-&-& -&h&  -&c  &  -  &  07 00 31.8   &          $-$02 22 45    &   3.3  &   --                        \\ 
J0700$-$10    &  223.22  &  $-2.64$   &   125.4  &  10.45  &    2.4\phantom{*} &    p   &  -&M&-&-&-&-& -&-&  N&-  &  -  &  07 00 34.38  &          $-$10 20 15.1  &   4.3  &   2MASX J07003437-1020151   \\ 
J0700$-$11    &  224.51  &  $-3.26$   &    33.8  &   9.70  &    2.3\phantom{*} &    d   &  -&M&-&H&-&-& -&h&  N&-  &  -  &  07 00 56.14  &          $-$11 47 34.3  &   0.6  &   CGMW 1-0488               \\ 
J0700$+$01    &  212.34  &  $ 3.01$   &    21.3  &   9.22  &    1.8\phantom{*} &    d   &  I&M&-&H&-&Z& o&h&  N&-  &  -  &  07 01 03.30  &\phantom{$-$}01 54 40.6  &   0.7  &   UGC03630                  \\ 
J0701$-$12    &  224.96  &  $-3.39$   &   122.9  &  10.01  &    2.8\phantom{*} &    d   &  -&M&-&-&-&-& -&h&  N&-  &  -  &  07 01 25.75  &          $-$12 15 22.0  &   1.4  &   CGMW 1-0491               \\ 
J0701$-$07    &  220.64  &  $-1.00$   &    20.7  &   8.50  &    3.9\phantom{*} &    d   &  -&-&-&-&-&-& -&-&  -&c  &  -  &  07 01 47.2   &          $-$07 19 36    &   3.0  &   --                        \\ 
J0702$-$11    &  224.46  &  $-2.91$   &    34.8  &   8.81  &    6.2\phantom{*} &    -   &  -&-&-&-&-&-& -&-&  -&-  &  -  &  --           &          $ $            &   --   &   --                        \\ 
J0702$-$15    &  227.93  &  $-4.66$   &   105.6  &   9.36  &    2.1\phantom{*} &    d   &  -&-&-&-&-&-& -&-&  -&c  &  -  &  07 02 18.1   &          $-$15 26 47    &   1.7  &   --                        \\ 
J0702$-$03A   &  217.04  &  $ 0.93$   &    86.7  &   9.96  &    7.3*           &    d   &  -&M&-&-&-&-& -&-&  N&-  &  -  &  07 02 15.33  &          $-$03 13 46.6  &   0.1  &   CGMW 1-0497               \\ 
J0702$-$03B   &  217.15  &  $ 0.97$   &    32.0  &   8.95  &    4.1\phantom{*} &    -   &  -&-&-&H&-&-& -&h&  -&-  &  -  &  --           &          $ $            &   --   &   --                        \\ 
J0702$-$12    &  225.20  &  $-3.10$   &   121.5  &   9.93  &    2.8\phantom{*} &    d   &  -&-&-&-&-&-& -&-&  -&c  &  -  &  07 02 52.9   &          $-$12 20 16    &   0.9  &   --                        \\ 
J0704$-$13    &  226.60  &  $-3.39$   &   117.1  &  10.04  &    2.7\phantom{*} &    d   &  -&M&-&-&-&-& -&-&  N&-  &  n  &  07 04 25.31  &          $-$13 46 26.0  &   4.3  &   CGMW 1-0523               \\ 
J0705$+$02    &  212.22  &  $ 4.35$   &    21.1  &   8.81  &    1.6\phantom{*} &    d   &  -&-&-&H&-&Z& -&h&  N&c  &  -  &  07 05 38.5   &\phantom{$-$}02 37 18    &   0.4  &   [H92] 16                  \\ 
J0705$-$12    &  226.11  &  $-2.80$   &    69.9  &   9.83  &    2.5\phantom{*} &    d   &  -&-&-&H&-&-& -&h&  -&c  &  -  &  07 05 41.8   &          $-$13 00 30    &   0.8  &   --                        \\ 
J0706$-$04    &  219.00  &  $ 1.04$   &    30.6  &   8.72  &    2.1\phantom{*} &    d   &  -&-&-&-&-&-& -&-&  -&c  &  -  &  07 06 14.0   &          $-$04 57 09    &   1.8  &   --                        \\ 
J0706$-$06    &  220.40  &  $ 0.51$   &    30.8  &   9.08  &    2.9\phantom{*} &    d   &  -&-&-&-&-&-& -&-&  -&c  &  -  &  07 06 58.0:  &          $-$06 25 10:   &   1.0  &   --                        \\ 
J0707$-$11    &  224.96  &  $-1.78$   &    20.2  &   8.34  &    2.5\phantom{*} &    d   &  -&M&-&-&-&-& -&-&  N&-  &  -  &  07 07 16.68  &          $-$11 30 27.8  &   1.1  &   2MASX J07071668-1130281   \\ 
J0707$-$14    &  227.57  &  $-3.07$   &    33.5  &   8.53  &    2.5\phantom{*} &    -   &  -&-&-&-&-&-& -&-&  -&-  &  -  &  --           &          $ $            &   --   &   --                        \\ 
J0708$-$08    &  222.22  &  $-0.05$   &    83.6  &   9.93  &    2.7\phantom{*} &    d   &  -&-&-&-&-&-& -&-&  -&c  &  -  &  07 08 25.8   &          $-$08 17 01    &   1.8  &   --                        \\ 
J0708$-$01    &  216.17  &  $ 3.18$   &    97.4  &   9.93  &    1.0\phantom{*} &    d   &  -&-&-&-&-&-& -&-&  N&-  &  -  &  07 08 36.83  &          $-$01 26 30.8  &   1.3  &   2MASX J07083681-0126306   \\ 
J0709$-$03    &  218.46  &  $ 2.09$   &    46.4  &   9.01  &    1.4\phantom{*} &    d   &  -&-&-&-&-&-& -&-&  S&c  &  n  &  07 08 57.8   &          $-$03 58 39    &   1.3  &   DSH J0708.9-0358          \\ 
J0709$-$05    &  219.81  &  $ 1.55$   &    20.4  &   9.15  &    1.6\phantom{*} &    d   &  I&M&-&H&S&-& o&h&  N&-  &  -  &  07 09 34.59  &          $-$05 25 40.5  &   1.4  &   CGMW 1-0575               \\ 
J0710$-$07    &  222.23  &  $ 0.65$   &    29.4  &   8.72  &    2.3\phantom{*} &    p   &  -&-&-&-&-&-& -&-&  -&c  &  -  &  07 10 57.9   &          $-$07 54 31    &   3.7  &   --                        \\ 
J0711$-$05    &  219.99  &  $ 2.07$   &    46.2  &   9.05  &    1.4\phantom{*} &    d   &  -&-&-&-&-&-& -&-&  -&c  &  -  &  07 11 54.0   &          $-$05 18 13    &   2.5  &   --                        \\ 
J0712$-$09    &  223.65  &  $ 0.46$   &    29.7  &   8.78  &    2.3\phantom{*} &    d   &  I&M&W&-&-&-& -&h&  N&-  &  -  &  07 12 48.61  &          $-$09 18 22.6  &   0.6  &   CGMW 1-0617               \\ 
\enddata
\end{deluxetable}


\subsection{Results of counterpart search}

Of the 883 \HI\ detections, we found cross matches for 688 (78\%). At least
18 (2\%) \HI\ detections have more than one counterpart, that is, more than
one galaxy contributes to the \HI\ profile, see Table~\ref{tabcrossex}b
(not counted here are detections that are confused but could be separated
into their \HI\ components which are all listed in
Table~\ref{tabdetsex}). For another 17 (2\%) \HI\ detections, there is no
unambiguous counterpart. Instead we list two possible candidates each in
Table~\ref{tabcrossex}c (note that these profiles do not show any
indication of confusion).

A total of 295 (33\%) detections have been detected previously in \HI ,
where 256 (28\%) have been detected with HIPASS and 110 (12\%) are also
listed in the HIZSS catalog.


Of the 708 cross matches found (from Table~\ref{tabcrossex}c we count
arbitrarily only the first entry each), 328 (46\%) are listed in NED or
SIMBAD ($N_{\rm S}=3$), an additional 138 (19\%) are detected by
Williams \etal (2014) on deep NIR images, and 240 (34\%) have no optical/IR
counterparts. There are 128 (18\%) IRAS counterparts (with a further 8 (1\%)
uncertain identifications), 248 (35\%) are listed in 2MASS, and 58 (8\%)
have an optical (that is, independent) velocity.



\begin{figure}[t] 
\centering
\includegraphics[width=0.45\textwidth]{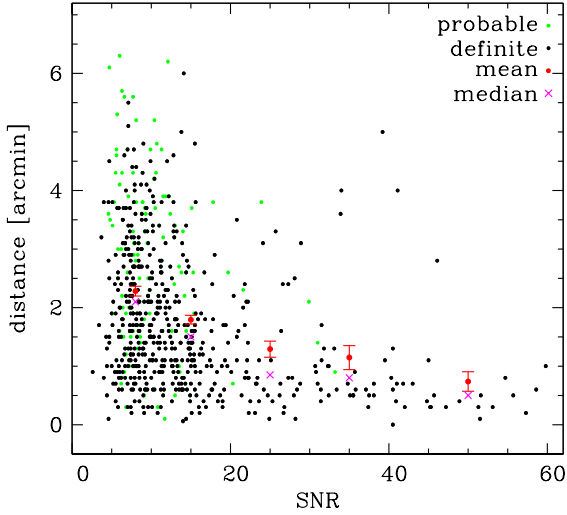}
\caption{Separation between the \HI\ position and the optical/NIR counterpart
position as a function of signal-to-noise ratio. The mean and error as well
as the median of the definite candidates (black dots) are indicated with red
dots and magenta crosses, respectively, for the SNR bins $6-10$, $10-20$,
$20-30$, $30-40$, and $40-60$; not included are the probable candidates
(green dots).  The beam size is $15\farcm5$ and the pixel size is $4\farcm0$.
}
\label{distsnrplot}
\end{figure}

Of the 558 single cross matches (that is, matches from
Table~\ref{tabcrossex}a only and {\it excluding} probable matches), the
median distance between the \HI\ position and the actual position is
$1\farcm4$, while 95\% of the cross matches have a distance
$<3\farcm8$. The dependence of the mean and median distance on
signal-to-noise ratio (SNR) is shown in Fig.~\ref{distsnrplot}. Aside from
low SNR, unusually large distances are probably due to 
offset emission or confusion (two or more galaxies contributing
to the emission) (\cf Table~\ref{tabcrossex}b).
%


For example, in Fig.~\ref{distsnrplot}, two of the five objects with
SNR~$>\!30$ and distance~$>\!2'$ have $|b|>5\degr$ and two are extended. In
one case (J0817$-$29B with $v=1654$\kms , $w_{50}=158$\kms , SNR$=41$, dist$=4\arcmin$) there is a very
bright, nearby \HI\ detection 20\arcmin\ away (J0817$-$30 with $v=1662$\kms ,
$w_{50}=186$\kms ).



Koribalski \etal (2004) find that 95\% of their optical cross matches have a
positional offset $<3\farcm2$. If we use a cut in peak flux similar to
theirs ($S_p = 0.116$\,Jy), our offset for 95\% of our cross matches is
$<3\farcm7$. Reasons for the difference can be our smaller sample size for
this cut ($N=57$) and the fact that Koribalski \etal use the nearest match
in HyperLEDA, while we make an informed decision which of the close-by
galaxies is the counterpart. In fact, we find that for 13\% of our \HI\
detections there is a galaxy closer than the cross match.

\subsection{Properties of the counterpart sample }

\begin{figure}[tb]
\centering
\includegraphics[width=0.45\textwidth]{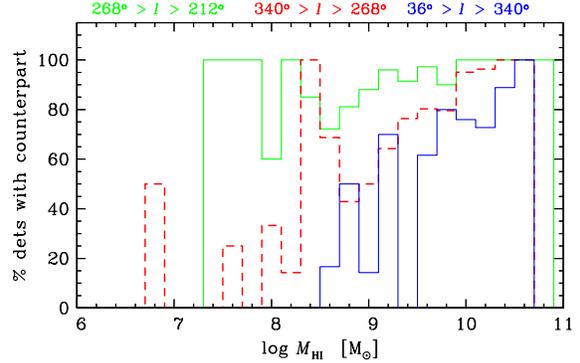}
\caption{\HI\ mass of HIZOA-S detections with counterparts identified,
for different Galactic longitude as indicated.
}
\label{histmhifig}
\end{figure}

With a counterpart found for 78\% of the \HI\ detections, we can
investigate whether there are any systematics that affect the finding of
counterparts. Obviously there is a dependence on extinction, although we
find that the identification of counterparts is fairly independent of
extinction up to $A_{\rm B}\simeq\!13^{\rm m}$ due to the available deep
NIR imaging (VISTA, UKIDSS, IRSF).

Galaxies with high \HI\ mass and broad linewidths are more easily
recovered than (late-type) dwarf galaxies, although this also depends on
the extinction.  Figure~\ref{histmhifig} shows histograms of the \HI\ mass
of cross-identified \HI\ detections dependent on the location in the ZOA
(Puppis area $212\degr < l < 268\degr$ in green; GA area $268\degr < l <
340\degr$ in dashed red; Local Void area $340\degr < l < 36\degr$ in
blue). In the Puppis area, far from the Galactic Bulge, we find almost all
HIZOA detections have cross matches. However, the detection rate decreases
with \HI\ mass for galaxies found in the Local Void / Galactic Bulge area
where stellar crowding and extinction severely affects the
cross-identification in the optical/NIR.

\section{Completeness, Accuracy and Reliability} \label{compl}


\subsection{Completeness}

\begin{figure}[th]
\centering
\includegraphics[width=0.45\textwidth]{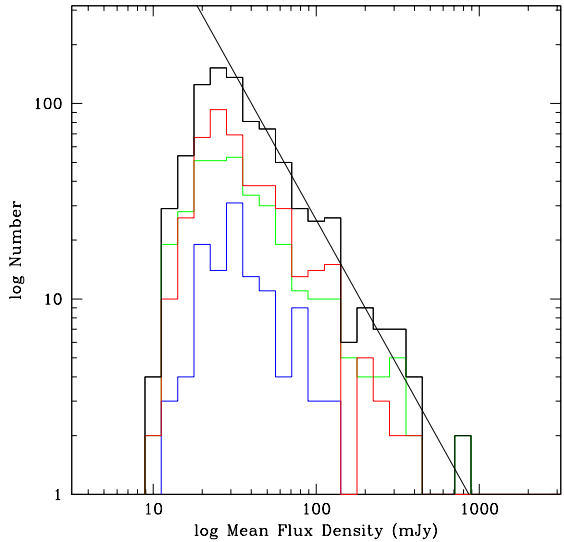}
\caption{Log of the mean \HI\ flux density $S$, (flux integral divided by the linewidth
$w_{50}$) for all HIZOA-S galaxies that lie within the survey limits at which
full sensitivity is reached ($|b| < 5\degr$). The black line corresponds to a slope
of $-3/2$ that would be expected for a homogeneous distribution. The colored histograms
subdivide the survey area into three equal-sized intervals of $\Delta \ell =60\degr$ (similar to Figs.~\ref{histmhifig}). 
}
\label{hist_meanflux}
\end{figure}

Previous analysis of the data in the otherwise identical northern
extension of this survey, HIZOA-N (Donley \etal 2005), has concluded
that the completeness limit lies at a mean flux density of
22\,mJy. That is, galaxies with profiles whose mean flux density, $S$,
is greater than 22\,mJy are generally detected with high
completeness. Galaxies below this threshold can be detected, but with
increasingly poor completeness. The main exceptions to this limit, as
also noted by Donley \etal, are in regions of high rms at the edge of
the field of view and, more importantly, towards bright radio
continuum regions in the Galactic Plane. Bright Galactic continuum not
only raises the receiver temperature and lowers sensitivity, but can
also give rise to non-flat spectral baselines, hampering
detectability. A histogram of the mean flux density, $S$, for 
HIZOA-S constrained by $212\degr \le \ell \le 36\degr$ and $|b| \le
5\degr$ is shown in Fig.~\ref{hist_meanflux}, as well as three subsets
of different Galactic longitude ranges of width $\Delta \ell =
60\degr$. In agreement with Donley et al., incompleteness in mean flux
density is obvious below 30\,mJy. Incompleteness seems to drop off
more quickly in the region represented by the blue histogram which
contains the Local Void. This is probably due to a
combination of the higher continuum in this region (see below) and
large-scale structure.

\begin{figure*}[t]
\centering
\includegraphics[width=1.0\textwidth]{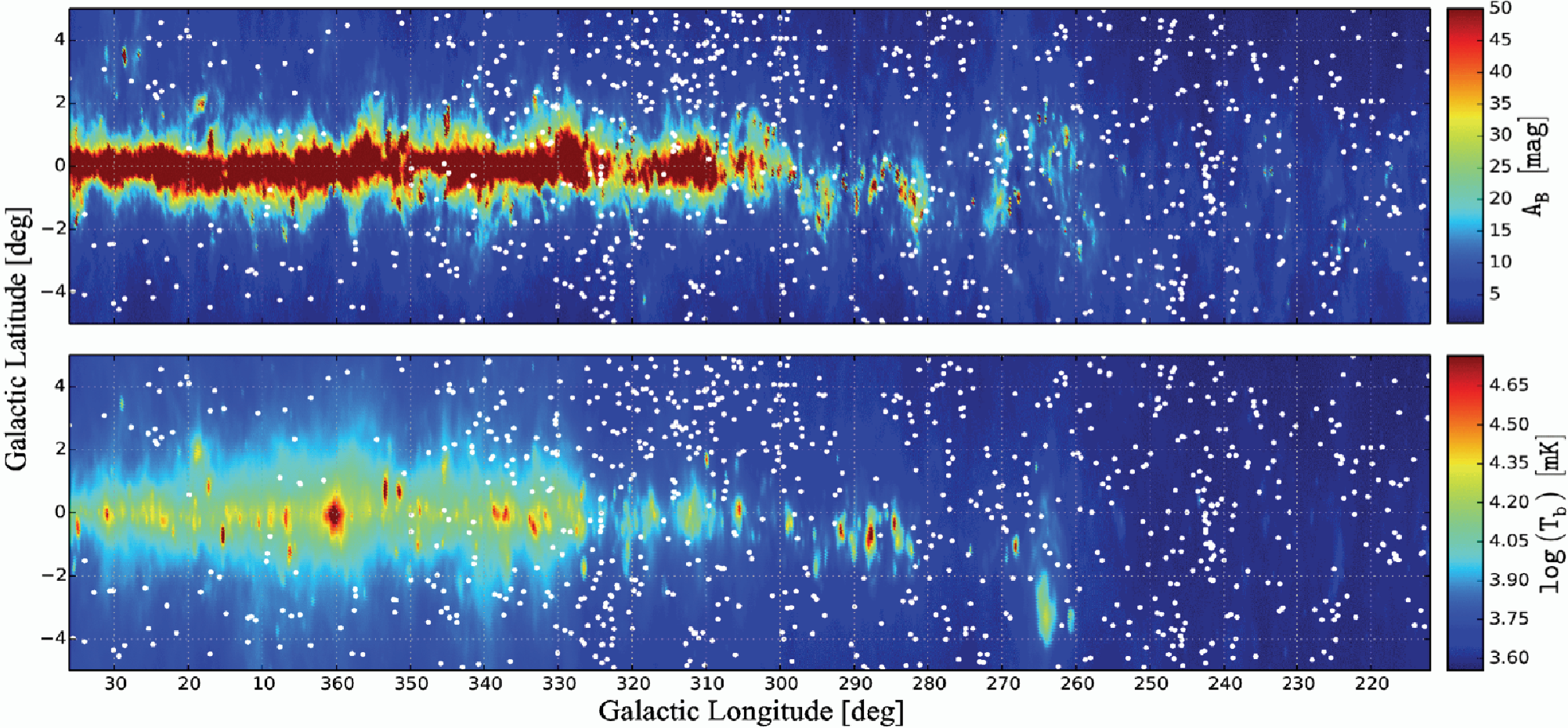}
\caption{Distribution in Galactic coordinates of the HIZOA-S galaxies
  (white dots) superimposed on the re-calibrated DIRBE dust extinction maps
  (Schlafly \& Finkbeiner 2011), top panel, and the Galactic background
  continuum maps (Calabretta \etal 2014), bottom panel. Note that galaxies
  are detected at some of the highest dust column density values, while
  this is not true for the highest continuum levels (see also
  Fig.~\ref{det_rate_dep}).
}
\label{dust_cont}
\end{figure*}

\begin{figure}
\centering
\includegraphics[width=0.45\textwidth]{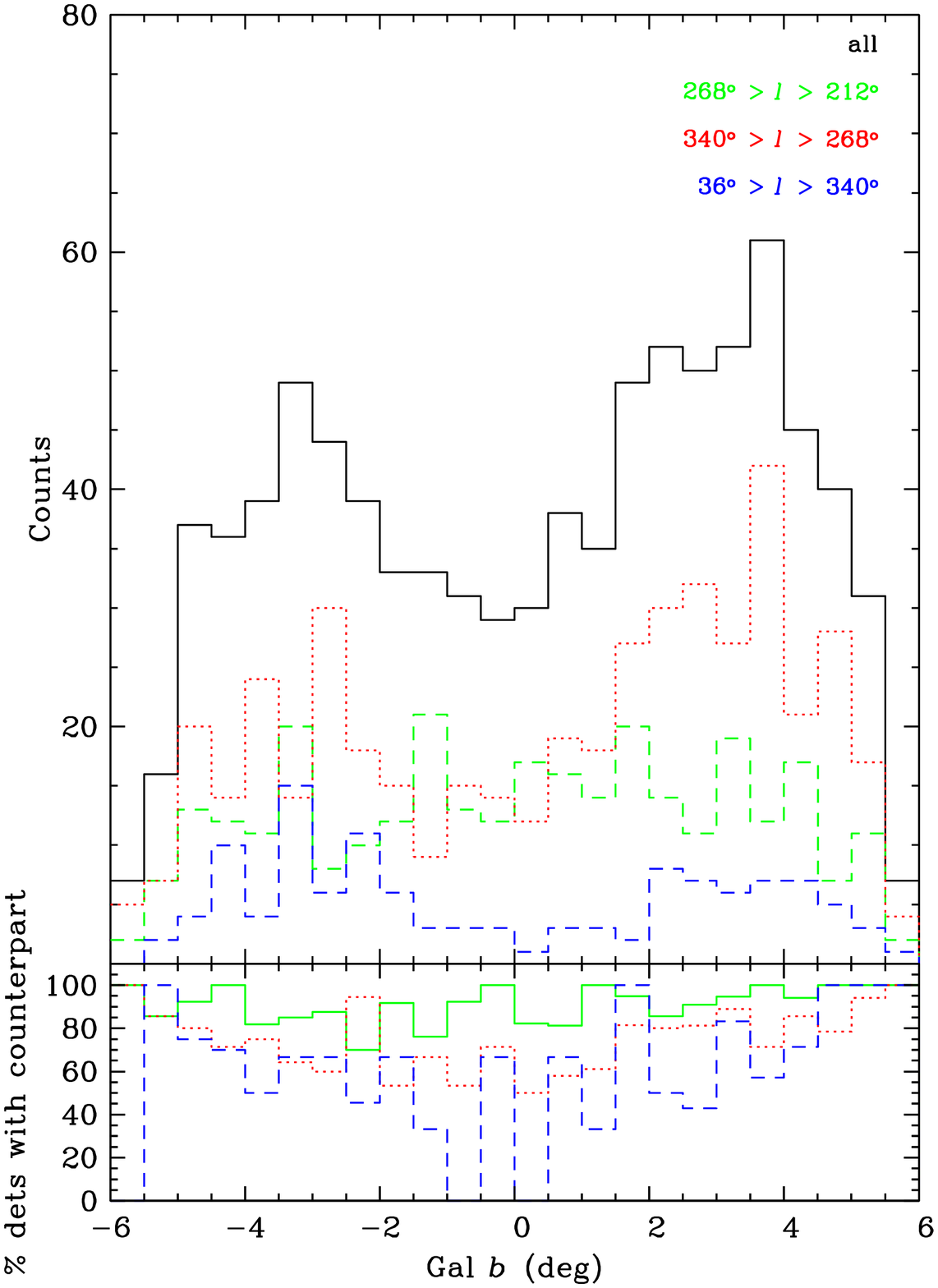}
\caption{Number of detections plotted as a function of Galactic
latitude for three different longitude ranges as indicated at the top right, and
for the combined sample (solid black line). The counterpart identification rate
is plotted in the lower panel.\label{f:hist_galb_gall}}
\end{figure}

\begin{figure*}
\centering
\includegraphics[width=0.70\textwidth]{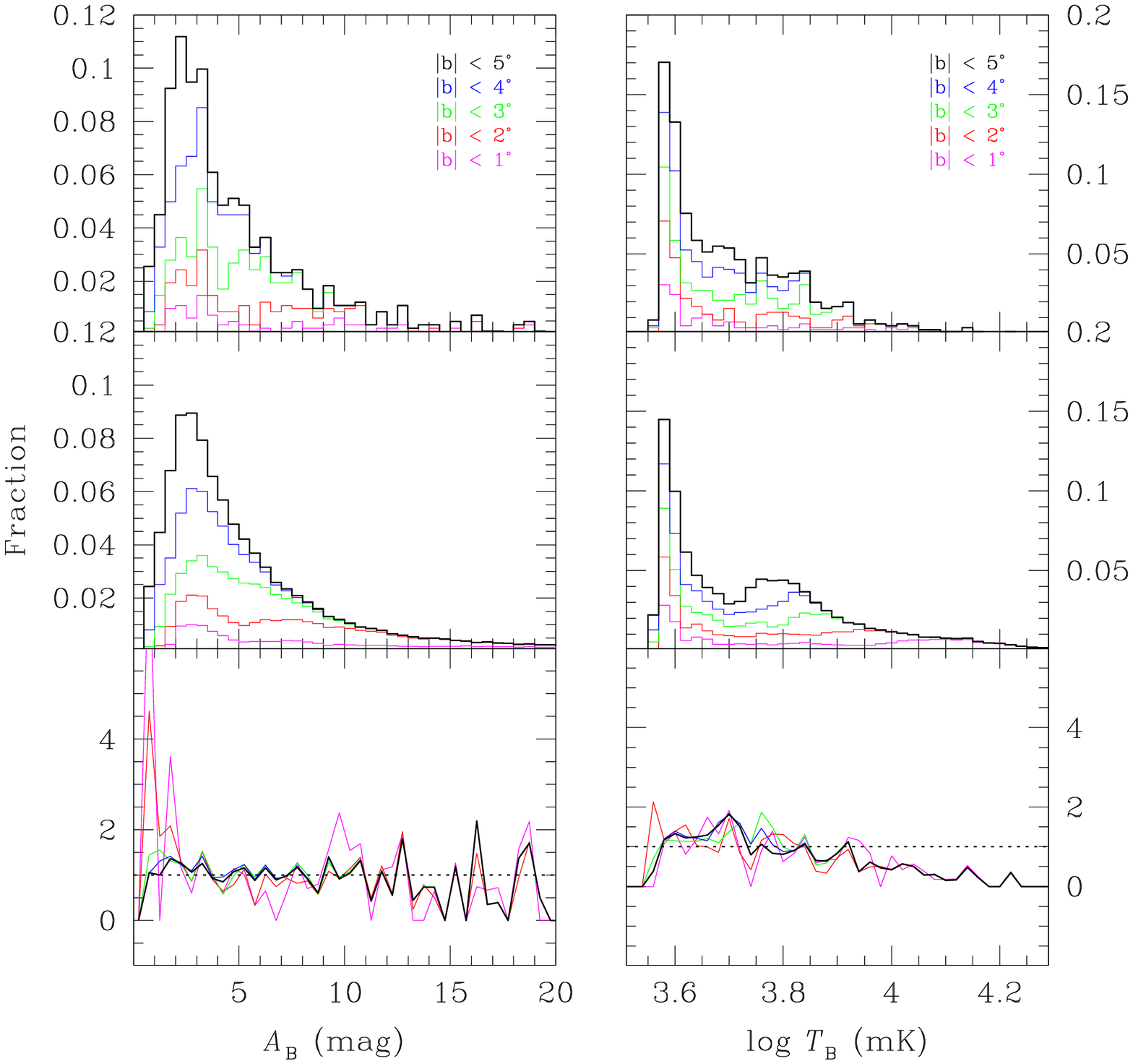}
\caption{Detection rate as a function of foreground extinction (left panels)
and continuum background (right panels). Colors represent different latitude
cuts: black, blue, green, red, magenta for $|b| <  5,4,3,2,1\degr$
respectively. {\it Top panels:} distribution of $A_B$ and $\log T_{\rm B}$ 
levels at the position of HIZOA-S galaxies as a fraction of the total number
within the respective latitude limits.  The continuum shows a much sharper
drop-off  compared to $A_B$ (note the different scales).  {\it Middle
panels:} $A_B$ and $\log T_{\rm B}$ over the surveyed area calculated from a
grid of cells of $0\fdg1 \times 0\fdg1$. Note the smoothness of the
distributions and the similarity to the distribution in the top panel,
apart from a bump around $\log T_{\rm B} \sim 3.7-3.9$. This is due to the
Galactic Bulge and disappears completely when the region is limited to
$\ell < 325\degr$. {\it Bottom panels:} the ratio of $A_B$ and $\log T_{B}$
at the position of HIZOA-S galaxies to the overall survey values (a division
of the top and middle histograms), which should be flat if there is no
dependence. 
}
\label{det_rate_dep}
\end{figure*}

The 883 galaxies detected in HIZOA-S are overlaid on an image of the
Galactic continuum background made from the same multibeam data by
Calabretta, Staveley-Smith \& Barnes (2014) in
Fig.~\ref{dust_cont}, bottom panel. The anti-correlation between Galactic continuum
and galaxy detection is noticeable for $|b|<1\degr$, although note
that the Local Void (\S\ref{s:lss}) results in reduced galaxy
density at all latitudes for $\ell>350\degr$. Figure~\ref{dust_cont}
also shows the correlation with dust extinction (top panel). Although a broadly
similar anti-correlation at $|b|<1\degr$ is evident, a detailed
comparison shows that galaxies can be detected at all optical
extinctions (including $A_B>50$ mag) as long as the Galactic
foreground temperature is $T_B<20$\,K. More quantitatively, the
detection rate as a function of Galactic latitude is plotted in
Fig.~\ref{f:hist_galb_gall}. A deficit in detections is apparent
outside the nominal latitude range $\pm5\degr$ and at
$|b|<1.5\degr$. The latter deficit is much more striking near the
Galactic Center. The relative surface density of galaxies found at
increasing continuum temperature and optical extinction is shown in
Fig.~\ref{det_rate_dep}. The anti-correlation with continuum is much
tighter than for extinction, with $\sim100$\% detectability for
$T_B<7$\,K, decreasing to 50\% at 10\,K and to almost zero at $T_B>16$
K, although the latter represents only 1.9\% of the area surveyed. As
with HIZOA-N, the reason for the low detectability in regions of bright 
continuum is mainly the increased baseline ripple and higher system 
temperature. Very little of the Galactic Plane is optically thick at
1.4 GHz.

The significant numbers of galaxies detected in this survey allow a
more detailed characterisation of completeness than made by Donley \etal
(2005). We use the technique of Rauzy (2001), which is a modified 
$V/V_{\rm max}$ test. This has previously been used for \HI\ surveys
alongside false-source injection techniques by Zwaan \etal (2004)
for HIPASS and Hoppmann \etal (2015) for the Arecibo Ultra Deep
Survey (AUDS), and has been shown to be a robust alternative. Its main
advantage is that, compared with source counts (e.g.,
Fig.~\ref{hist_meanflux}), it is insensitive to large-scale structure,
which is considerable in the southern ZOA.  Disadvantages are: (a)
it does not allow the characterisation of a `soft' rolloff in
completeness; and (b) the method assumes no substantial variation in
the shape of the \HI\ mass function with position, e.g., as a function
of environment. Evolutionary effects (Rauzy 2001) and bright limits
(Johnston, Teodoro \& Hendry 2007) can be incorporated, but are
unnecessary for this analysis.

\begin{deluxetable}{llcclc}
\tablecolumns{6}
\tablewidth{0pc}
\tablecaption{Rauzy completeness limits for flux integral, mean flux density and 
scaled fluxes.
$Tc=-3$ and $Tc=-2$ correspond to the 99.3\% and 97.7\% confidence bounds for
the completeness limits, respectively; $f$ is the fraction of galaxies above the completeness
limit. Completeness limits refer to objects in the main survey region ($212\degr < \ell < 36\degr$,  
$5\degr > b > -5\degr$) and Galactic foreground brightness $T_B<7$\,K.\label{t:rauzy}}
\tablehead{
\colhead{}    &  \multicolumn{2}{c}{$T_c=-3$} &   \colhead{}   &
\multicolumn{2}{c}{$T_c=-2$} \\
\cline{2-3} \cline{5-6} \\
\colhead{Parameter} & \colhead{Limit}   & \colhead{$f$}    &
\colhead{}    & \colhead{Limit}   & \colhead{$f$}}
\startdata
Flux integral, $F$              & 3.1\Jykms          & 0.75 & &  3.3\Jykms          & 0.72   \\
Mean flux density, $S$          & 21\phantom{.}\,mJy & 0.81 & &  22\phantom{.}\,mJy & 0.74 \\
$F (w^u_{50}/160$\kms$)^{0.74}$ & 2.8\Jykms          & 0.92 & &  2.9\Jykms          & 0.90 \\
$S (w^u_{50}/160$\kms$)^{-0.26}$& 17\phantom{.}\,mJy & 0.91 & &  18\phantom{.}\,mJy & 0.90 \\
\enddata
\end{deluxetable} 

\begin{figure}
\centering
\includegraphics[width=0.45\textwidth]{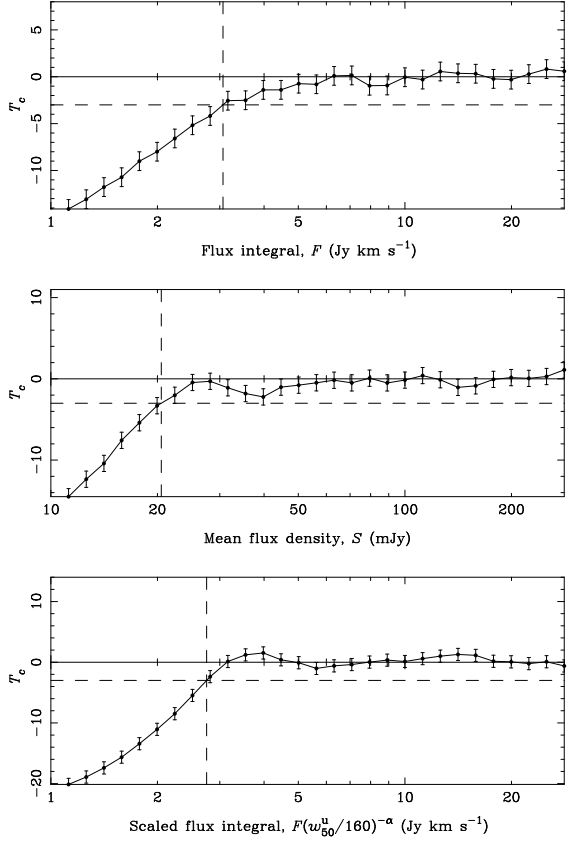}
\caption{Rauzy completeness statistic $T_c$ as a function of faint
  cutoff limit for: (top panel) flux integral, (middle panel) mean flux
  density and (bottom panel) scaled flux integral. $T_c$ values less
  than zero indicate incompleteness, with $T_c=-3$ corresponding to
  the lower 99.3\% confidence bound for completeness. The best
  description of completeness for this survey is given by the scaled
  flux integral with $\alpha=0.74$.\label{f:rauzy}}
\end{figure}

\begin{figure}
\centering
\includegraphics[width=0.45\textwidth]{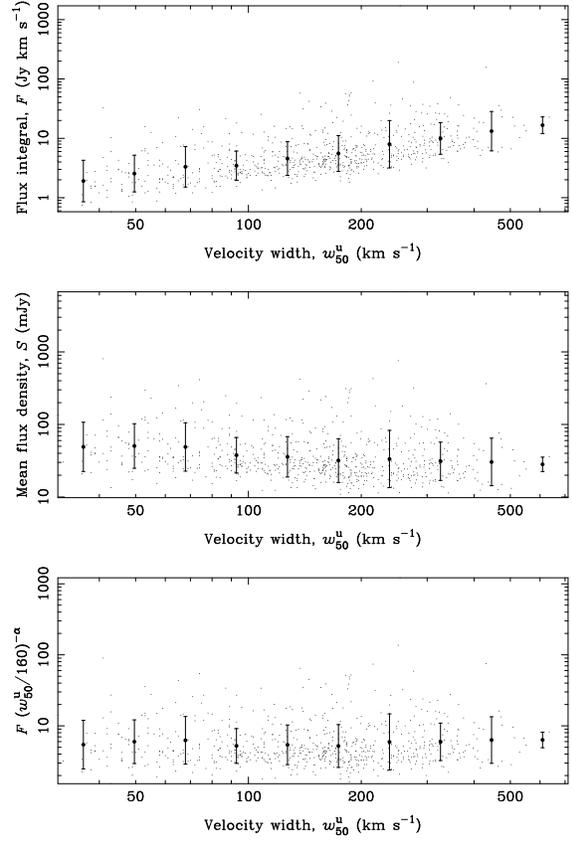}
\caption{Measured velocity width (before resolution correction),
  $w^u_{50}$, plotted against (top panel) flux integral, (middle
  panel) mean flux density, and (bottom panel) scaled flux integral
  for all survey galaxies. Mean logarithmic values in 10
  equally-spaced intervals are shown, with error bars indicating the
  rms dispersion.\label{f:widths}}
\end{figure}

Results of the Rauzy test are presented in Table~\ref{t:rauzy} and
Fig.~\ref{f:rauzy}.  For regions within the main survey area
($212\degr < \ell < 36\degr$, $5\degr > b > -5\degr$) and $T_B<7$\,K,
the $T_c=-3$ completeness limits are $F_{\circ} = 3.1$\Jykms\ for
flux integral and $S_{\circ} = 21$\,mJy for mean flux density. The
Rauzy test indicates that there is only 0.6\% probability that the
actual completeness limits are fainter than this. The mean flux
density limit (defined here as the ratio of flux integral, $F$, to the
velocity width prior to resolution correction, $w^u_{50}$) is
satisfactorily similar to the value of 22\,mJy deduced by Donley \etal
(2005). However, Fig.~\ref{f:rauzy} shows that the flux-integral 
completeness starts to reduce below 100\% well above the formal
completeness limit. To a lesser extent, the mean flux density
completeness also rolls off. The reason for this is that, as shown in
previous blind \HI\ surveys, completeness is a function of both flux
and velocity width. This is illustrated in Fig.~\ref{f:widths} where
galaxies of a given flux integral are clearly easier to detect at
fainter limits when they have a narrow linewidth, $w$.  Conversely,
galaxies of a given mean flux density are easier to detect when they
have a large linewidth.

The radiometer equation suggests that the flux integral selection
limit should scale as $w^{\alpha}$, with $\alpha=0.5$. Conversely, the
flux density selection limit should scale approximately as
$w^{\alpha-1}$. However, the well-known effect of baseline ripple
makes high velocity width galaxies harder to detect in practice.
At Parkes, the primary ripple wavelength corresponds to 1,200 \kms.
Figure~\ref{f:widths} shows that a more appropriate index is
$\alpha=0.74$, where the selection limit is characterised for flux
integral by $F_{\circ} (w^u_{50}/160$\kms$)^{\alpha}$ and for mean
flux density by $S_{\circ} (W^u_{50}/160$\kms$)^{\alpha-1}$. With
this scaling, much tighter selection limits of $F_{\circ} = 2.8$\Jykms\ and
$S_{\circ} =17$\,mJy are obtained. At a velocity width of 50\kms, this
corresponds to flux integral and flux density limits of 1.2\Jykms\ and
24\,mJy, respectively. At a velocity width of 350\kms,  
this corresponds to flux integral and flux density limits of 5.0\Jykms\ and
14\,mJy, respectively. Thus defined, 92\% of the sample is 
brighter than the hybrid selection limit, compared with only 75\% and
81\% for pure flux and flux density-limited samples, respectively.
Additional dependencies, such as with profile shape (Zwaan \etal
2004), are small for the hybrid limit.


\subsection{Comparison with other \HI\ catalogs} \label{complit}


Two subsamples of the data presented here had been previously analysed:
Henning \etal (2000) published 110 bright \HI\ detections based on 8\% of
the full integration time and Juraszek \etal (2000) published 42
\HI\ detections in the GA region based on 16\% of the full integration
time.  All of these detections were recovered.  The only \HI\ detection
with a large positional offset compared to the previous publications
($d>10\arcmin$) is J1616-55 which is an extended, multi-component object
described in detail in Staveley-Smith \etal (1998) (\cf\ note in
Appendix~\ref{notes}).


As is normal between adjacent cubes, there is a $1\degr$-overlap around
$l=36\degr$ and $l=196\degr$ between this catalog and its
extension to the north, HIZOA-N (Donley \etal 2005). The two catalogs have
three galaxies in common. A further detection, found by us at $l=36\fdg06$,
is technically part of HIZOA-N but was missed there.


We compared our detections with HIPASS, which is an independent survey of
the southern hemisphere (Meyer \etal 2004; hereafter HIPASS-South), and its
extension to the north (Wong \etal 2006; hereafter HIPASS-North) 
at 20\% of our integration time. There are 251 detections in common, only
one of which is listed in HIPASS-North (J0705+02). One HIPASS detection is
confused (J1000-58) and was resolved into two detections in our deeper
survey. An additional two detections each are recorded in publications
based on older versions of the HIPASS catalog, namely Kilborn \etal (2002)
and Ryan-Weber \etal (2002).

\begin{deluxetable}{lrrcll}
\tablecolumns{6}
\tablewidth{0pc}
\tablecaption{HIPASS detections not in the HIZOA-S catalog. Values in bold
  typeface represent objects outside the nominal HIZOA range in Galactic
  latitude, low-SNR HIPASS detections, or low HIPASS quality. \label{hipmisstab} }
\tablehead{
\colhead{HIPASS} & \colhead{Gal $b$} & \colhead{SNR} & \colhead{$q^a$} & \colhead{Note} & \colhead{Comment}
}
\startdata
J0817$-$45 & {\bf -5.59} &      9.2  &     1  &      real & below the acceptance limit (noisy area at the edge)    \\  
J0844$-$38 &       2.45  & {\bf 4.6} &{\bf 2} &  not real &                                                        \\  
J0857$-$41 &       2.55  & {\bf 4.5} &{\bf 2} &  not real &                                                        \\  
J1026$-$51 & {\bf  5.09} &      6.2  &     1  &      real & below the acceptance limit (bad baseline at the edge)  \\  
J1232$-$68 & {\bf -5.38} &      5.4  &     1  &      real & below the acceptance limit (noisy area at the edge)    \\  
J1412$-$65 &      -4.09  &    132.4  &     1  &  not real & part of Circinus                                       \\  
J1418$-$63 &      -2.19  & {\bf 3.3} &{\bf 2} &  not real &                                                        \\  
J1439$-$54 & {\bf  5.28} &      6.0  &     1  &      real & below the acceptance limit (faintly visible at edge)   \\  
J1440$-$53 & {\bf  5.58} &      7.8  &     1  &      real & below the acceptance limit (visible at edge)           \\  
J1444$-$53 & {\bf  5.38} & {\bf 4.5} &{\bf 2} &  not real &                                                        \\  
J1516$-$58 &      -0.50  & {\bf 3.5} &{\bf 2} &  not real &                                                        \\  
J1600$-$52 &       0.48  & {\bf 4.2} &{\bf 2} &  not real &                                                        \\  
J1624$-$47 &       1.42  & {\bf 4.0} &     1  &  not real &                                                        \\  
J1655$-$49 &      -3.99  & {\bf 4.4} &{\bf 2} &  not real &                                                        \\  
J1642$-$37 & {\bf  5.66} &      6.3  &     1  &      real & below the acceptance limit (visible at edge)           \\  
J1708$-$37 &       1.35  &      8.0  &     1  &  not real & likely RFI                                             \\  
J1740$-$37 &      -3.80  &      7.9  &     1  &  not real & likely RFI                                             \\  
J1713$-$33 &       2.94  &      9.2  &     1  &  not real & likely RFI                                             \\  
J1718$-$27 & {\bf  5.62} &      6.4  &     1  &      real & below the acceptance limit (faintly visible at edge)   \\  
J1743$-$21 &       4.09  &      5.9  &     1  &      real & below the acceptance limit                             \\  
J1823$-$16 &      -1.44  &      5.4  &     1  &  not real & likely RFI                                             \\  
J1828$-$09 &       0.82  & {\bf 4.0} &     1  &  not real &                                                        \\  
J1817$-$04 & {\bf  5.75} & {\bf 4.6} &     1  &  not real &                                                        \\  
\enddata
{\newline \footnotesize $^a$HIPASS quality flag: $q=1$ = `real', $q=2$ = `have concerns'. }
\end{deluxetable}

There are a further 23 galaxies (all in HIPASS-South) that were not
detected in HIZOA-S, see Table~\ref{hipmisstab}. On closer inspection, we
find 8 detections likely to be real, of which four were detected in the
visual searches but lie below our limit for inclusion (usually due to
locally high rms), while the other four
are located near the edge of a cube and were thus missed due to the higher
noise (though they were also below our acceptance limit).

Of the 15 HIPASS detections considered not to be real, seven were labeled
in the HIPASS catalog as `have concerns' (quality flag $q=2$), all of which
have low SNR ($<5.0$); they could not be confirmed in our cubes. Three
further detections with $q=1$ have similarly low SNR and could not be
confirmed by us either. One HIPASS detection (J1412$-$65) forms part of
Circinus (our J1413$-$65 detection). The remaining four detections seem to be
caused by RFI: they are narrow peaks with high SNR in the HIPASS spectra,
but nothing is evident in the HIZOA cubes.

\begin{figure}[t]
\centering
\includegraphics[width=0.47\textwidth]{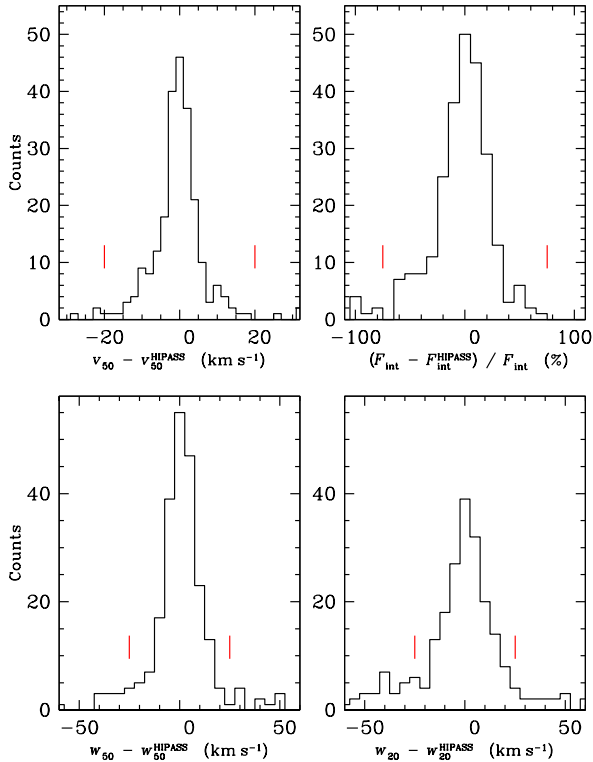}
\caption{Histograms of differences between HIZOA-S and HIPASS parameters:
  $v_{50}$ (top left), Flux (difference percentage, top right), $w_{50}$
  (bottom left) and $w_{20}$ (bottom right). The histograms are truncated,
  where 18 are outside the plotted range for $v_{50}$ (out to $|\Delta
  v_{50}|\simeq 130$\kms), one for flux integral (at -177\%), 17 for
  $w_{50}$ (out to $|\Delta
  w_{50}|\simeq 380$\kms) and 14 for $w_{20}$ (out to $|\Delta
  w_{20}|\simeq 250$\kms). The vertical bars indicate the cuts 
  used for the statistics given in Table~\ref{comphiptab}.
}
\label{hipcompplot}
\end{figure}

In Fig.~\ref{hipcompplot} we compare the parameters of the 254 detections in
common (excluding J1000$-$58). For HIPASS we use their width-maximised $v_{50}$,
$w_{50}$ and $w_{20}$. Note that not all objects have an entry for these parameters,
while our catalog also misses some of the $w_{20}$ parameters. 
Outliers are generally due to two reasons: (i) confused or lopsided
profiles (all extreme cases fall into this category) and (ii)
noisy profiles or profiles affected by a poor baseline.
Table~\ref{comphiptab} gives the mean and standard deviations of the
`core' of the histograms shown in Fig.~\ref{hipcompplot} (indicated with
red vertical bars). In summary, we see no statistically significant
systematic effects. 

\begin{deluxetable}{lllll}
\tablecolumns{5}
\tablewidth{0pc}
\tablecaption{Comparison of HIPASS and HIZOA-S parameters. \label{comphiptab}}
\tablehead{
\colhead{Parameter} & \colhead{Cut} & \colhead{$N$} & \colhead{Mean} & \colhead{Std dev} 
}
\startdata
$v_{50}$           & 20\kms & 227 & $-0.3\pm0.4$\kms            & \phantom{0}5.6\kms  \\
Flux integral, $F$ & 75\%   & 246 & $-0.6\pm1.4$\%              & 23.8\% \\
$w_{50}$           & 25\kms & 214 & $\phantom{-}0.2\pm0.6$\kms  & \phantom{0}8.8\kms  \\
$w_{20}$           & 25\kms & 179 & $\phantom{-}0.9\pm0.7$\kms  & \phantom{0}9.9\kms  \\
\enddata
\end{deluxetable}


We have also compared our detections with the deeper, pointed Parkes
survey of southern ZOA galaxies by Kraan-Korteweg \etal (2002) and Schr\"oder
\etal (2009) which has a typical rms
of 
$2-6$\,mJy (hereafter referred to as PKS sample). We have 39 detections
in common. In six cases the ID given in the PKS sample seems to be
mis-judged (J1222$-$57, J1337$-$58B (confused profile), J1405$-$65, J1550$-$58,
J1553$-$61, J1612$-$56; see Appendix~\ref{notes} for details), which is mostly
due to off-beam detections in the high galaxy density area in the GA
region.  The 23 detections in the PKS sample not recovered by us are
usually below the HIZOA sensitivity limit or lie near the edge of our
survey area and are thus lost in the noise ($N=20$). Three PKS detections
were detected in our visual searches but were just under the acceptance
limit.
%

A comparison of parameters gives similar values of mean and scatter as for
HIPASS though with larger uncertainties due to the smaller number of
samples. Outliers are due to confused profiles ($N=4$) or low SNR/baseline
variations ($N=2$).


\begin{figure*}[tb]
\centering
\includegraphics[width=0.9\textwidth]{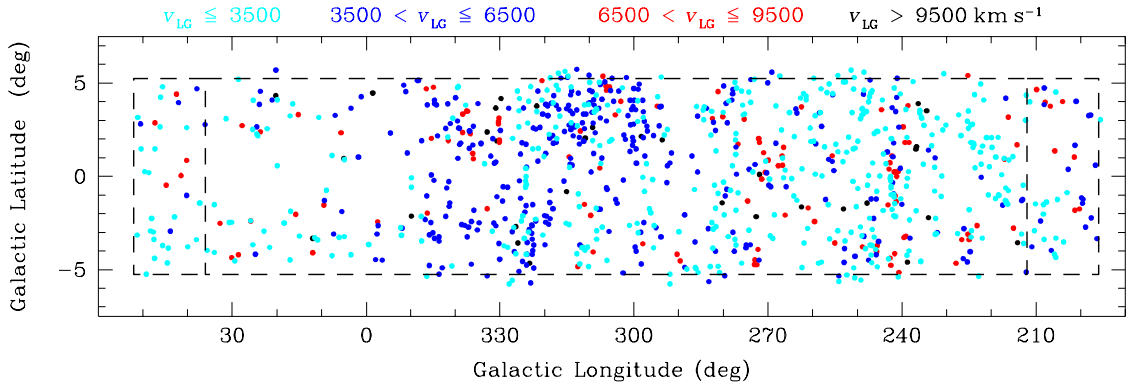}
\caption{Distribution in Galactic coordinates of the 957 \HI-detected
  galaxies in the merged HIZOA catalog: HIZOA-S ($36\degr > \ell > 212\degr, |b| < 5\degr$) 
  and HIZOA-N  ($196\degr <  \ell < 212\degr$ and  $36\degr < \ell < 52\degr$) respectively. 
  The survey areas are indicated by the dashed lines.The dots are color-coded as a function
  of velocity. Note the predominance of galaxies in the general GA region
  near $\ell \sim 312\deg$ and the overall variation of large-scale
  structure as a function of longitude (see also Fig.~\ref{vhist_reg}).
}
\label{mb}
\end{figure*}

We have also used HyperLEDA to extract
all galaxies with \HI\ velocities within our survey region. Next to the
HIPASS and PKS detections discussed above, there are 39 \HI\ detections not
detected by us.  Of these, 31 are too faint or too close to the survey
limits to be detectable by us. Four of these detections were found in the visual
searches of the HIZOA cubes, but were under the acceptance limit. Two detections seem to be
off-beam detections of one of our detections: (i) J0806$-$27 was attributed
to PGC022808 by Matthews \& Gallagher (1996); and (ii) J0749$-$26B was correctly attributed to CGMW~2-1330 
by Matthews \etal (1995), but with the wrong coordinates (that is, it was cross-matched with PGC100722 in
HyperLEDA).  Two bright detections (with $F>50$\Jykms ) could not be
confirmed by us: ESO494-013 (reported by Kraan-Korteweg \& Huchtmeier
(1992) and which we detected as J0802$-$22 at a different velocity) and
PGC2815809 (Huchtmeier \etal 2001); they are both detections with the
Effelsberg Radio Telescope and possibly RFI.
%


\subsection{Reliability}

Within the boundaries of the full-sensitivity survey and the source
detection adjudication process (\S\ref{sample}), the sample is expected to
be nearly 100\% reliable (cf.\ Donley \etal 2005).  The high reliability
was ensured by a reasonably high cut-off in signal-to-noise ratio. No
marginal detections were included in the catalog, though follow-up
observations are planned to confirm some of these.

We have conducted two checks on the reliability: 

(i) Of 56 detections in the overlap regions between adjacent
cubes, only two were not detected in both cubes: J1847+04 belongs
nominally to the northern extension, and J0818$-$33 is faint and was not
recognised in one cube due to large baseline variations near the edge of
the cube. 

(ii) We have cross-checked our detections with the non-detections in the
deeper (but pointed) Parkes survey of optically-selected galaxies in the
southern ZOA by Kraan-Korteweg \etal (2002) and Schr\"oder \etal (2009) and
found no false positives (with the caveat that a pointed survey has a
different selection function than a blind survey).




\section{Large-Scale Structure}
\label{s:lss}

In this section we investigate the large-scale distribution of the galaxies
detected in this survey and discuss newly identified structures in the context of
known large-scale structures in the immediate vicinity. For the latter we
use publicly available data archives like HyperLEDA (Paturel \etal 2003)
and, for some discussion, the 2MASS Redshift Survey (2MRS; Huchra \etal
2012). We combine our survey with the northern extension (HIZOA-N; Donley
\etal 2005). That survey was observed with the same telescope using an
identical strategy and analysed in exactly the same manner by the same
team. The structures identified in HIZOA-N (two $16\degr \times 10\degr$
fields) can be better appreciated by combining with the current HIZOA-S
catalog. The combined HIZOA survey data cover the Galactic longitude range
$52\degr > \ell > 196\degr$ ($\Delta \ell = 216\degr$) for the latitude
range of $|b| < 5\degr$ (see Fig.~\ref{mb}).

Duplicates from overlap regions in the two surveys were eliminated. As
discussed in \S\ref{complit}, three galaxies were published in Donley 
\etal (2005) but nominally belong in the current HIZOA-S catalog, while there
was one galaxy identified in the HIZOA-S cubes that technically belongs in
HIZOA-N but was not listed there. This means that the total number of
galaxies in the merged list sums to $N=957$ rather than $N=960$ ($883 +
77$).

For the large-scale structure discussion, all HIZOA galaxies will be
used. Note though that 62 galaxies lie just beyond the nominal latitude
limit ($|b| > 5\degr$) and a further one outside the longitude limit. The
total number in the full-sensitivity survey area, away from the edges, is
therefore $N=894$ galaxies which translates to an average density of 0.41
galaxies per square degree for the nominal survey area of 2160 deg$^2$.

\subsection{2D and redshift distributions of the HIZOA galaxies}

Figure~\ref{mb} displays the distribution around the Galactic Plane of the
957 galaxies detected in the HIZOA survey. The survey areas are marked by
the dashed line. The 63 galaxies that lie just beyond the nominal survey
borders are easily identifiable. The resulting galaxy distribution shows
remarkable substructures. These stand out even when ignoring the
color-coding of the dots that symbolises different redshift ranges (cyan:
$500-3500$\kms; blue: $3500-6500$\kms; red: $6500-9500$\kms; black:
$9500-12500$\kms).  The right-hand part of the plot shows a fairly smooth
distribution of galaxies with an inkling of a filamentary feature crossing
the Plane vertically in Puppis at about $\ell \sim 245\degr$. This part has
an average detection rate (0.44 deg$^{-2}$), similar to the mean of the
survey. The middle part shows a clear density enhancement (0.6 deg$^{-2}$)
which is associated with the Great Attractor (GA), whereas the left part,
centered around the Galactic Bulge, has few galaxies particularly at the
lowest latitudes. The density of the detections is about half of that in
the Puppis region, and a third of that in the GA region. This is mostly due
to the dominance of the Local Void (LV; Tully \& Fisher 1987;
Kraan-Korteweg \etal 2005) rather than a bias due to its location behind
the Galactic Bulge. As discussed in \S\ref{compl}, there is no dependence of the
detection rate on foreground extinction levels, and only a minor reduction
where the brightness temperature is elevated ($T_B \ga 7$ K). Only a small
part of the survey area, generally limited to $|b| \la 1\fdg0$, has a
higher detection threshold (cf.\ Figs.~\ref{dust_cont}
and~\ref{det_rate_dep}). 

The variation of the detection rate is not the only systematic apparent in
Fig.~\ref{mb}. The mean redshifts also show a striking difference as a
function of longitude. The general Puppis area is dominated by nearby
structures (cyan). The area centered on the GA is dominated by higher
velocities (dark blue), particularly for longitudes of ($290\degr -
340\degr$). The few galaxies uncovered at $340\degr \la \ell \la 52\degr$
are mainly nearby but with some at higher redshift (red and black),
suggestive of being at the far side of the LV.

\begin{figure}[t!!!]
\centering
\includegraphics[width=0.47\textwidth]{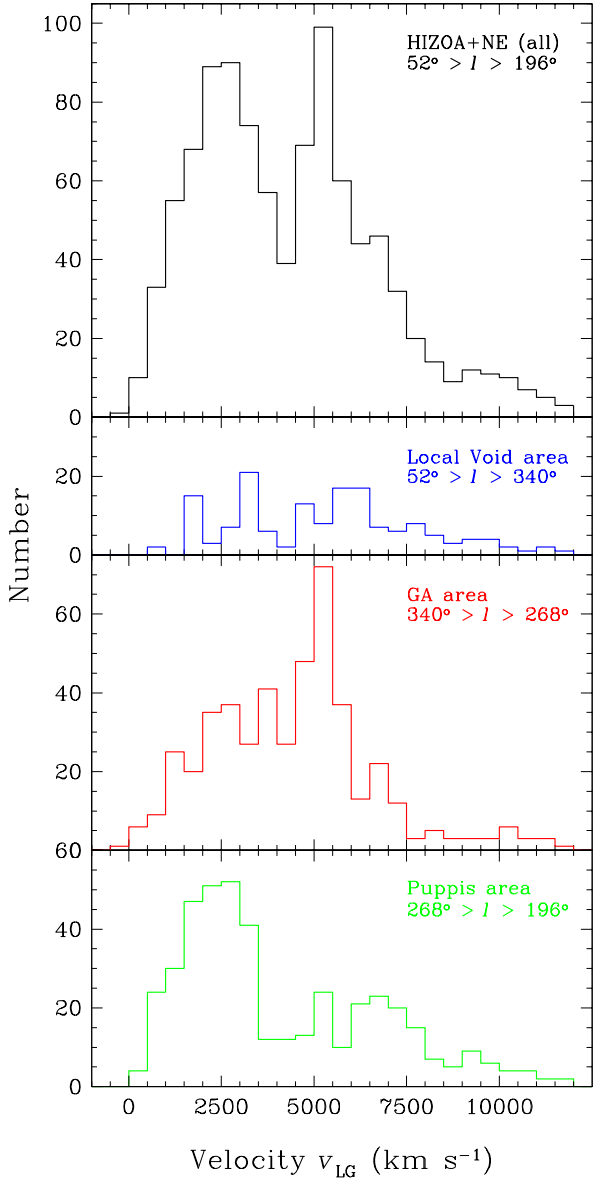}
\caption{Velocity histogram ($v_{\rm LG}$) of the galaxies detected in the
  combined HIZOA survey ($52\degr > \ell > 196\degr$). The top panel shows
  all galaxies ($N=957$); the subsequent panels are subdivided in three
  regions of equal width ($\Delta \ell = 72\degr$). Each one describes and
  encompasses vastly different structures, namely the Local Void (LV;
  $N=154$), the Great Attractor (GA; $N=462$) and the Puppis ($N=341$)
  region.
}
\label{vhist_reg}
\end{figure}

This can be better appreciated in Fig.~\ref{vhist_reg} which shows the
overall velocity histogram (top panel), together with three velocity
histograms subdivided in longitude ranges of width $\Delta \ell = 72\degr$
each. The histogram of the whole survey displays a fairly steep rise up to
velocities of about 3000\kms\ followed by a gradual drop-off towards the
highest velocities which is consistent with expectations from the loss of
sensitivity at higher redshifts. A distinct peak around 5000\kms\ 
superimposed on the gradual drop-off is due to the GA overdensity
(see 3rd panel).  Note that galaxies are found all the way out to the
velocity limit of the survey, and thus probe the local volume considerably
deeper than HIPASS (Meyer \etal 2004).

\begin{figure*}[t]
\centering
\includegraphics[width=0.8\textwidth]{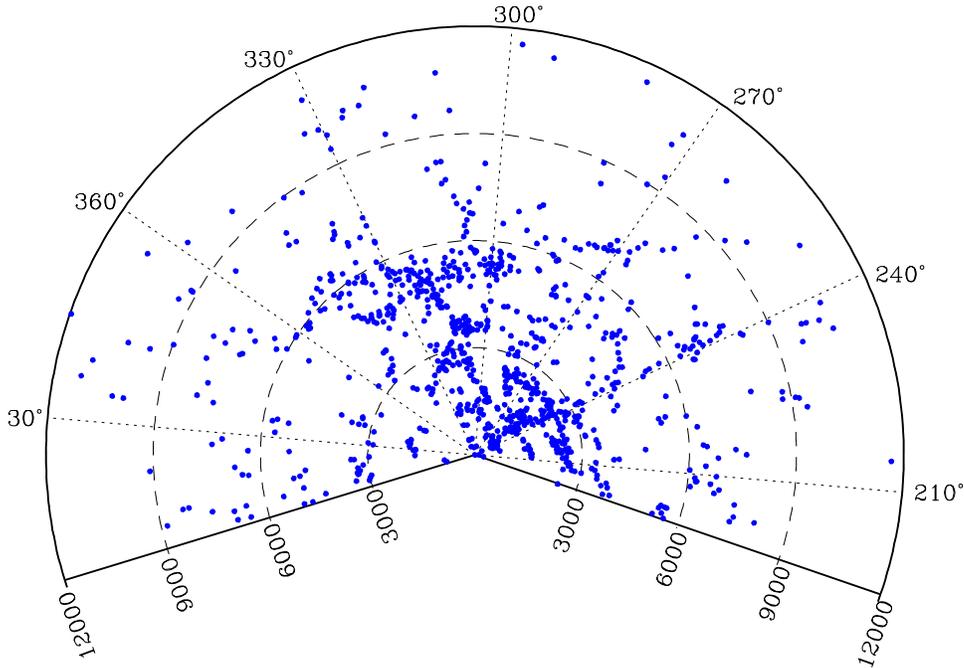}
\caption{A redshift wedge plot for $v_{LG} < 12\,000$\kms\ for the
  latitude range $|b| < 5\fdg25$ for the \HI-detected galaxies in the HIZOA
  survey. Note the wall-like structure of the GA, the Norma supercluster,
  at about $\sim 4800$\kms\ that stretches in the range $345\degr \ga \ell
  \ga 290\degr$. 
  }
\label{wedge}
\end{figure*}

\begin{figure*}[t!]
\centering
\includegraphics[width=0.8\textwidth]{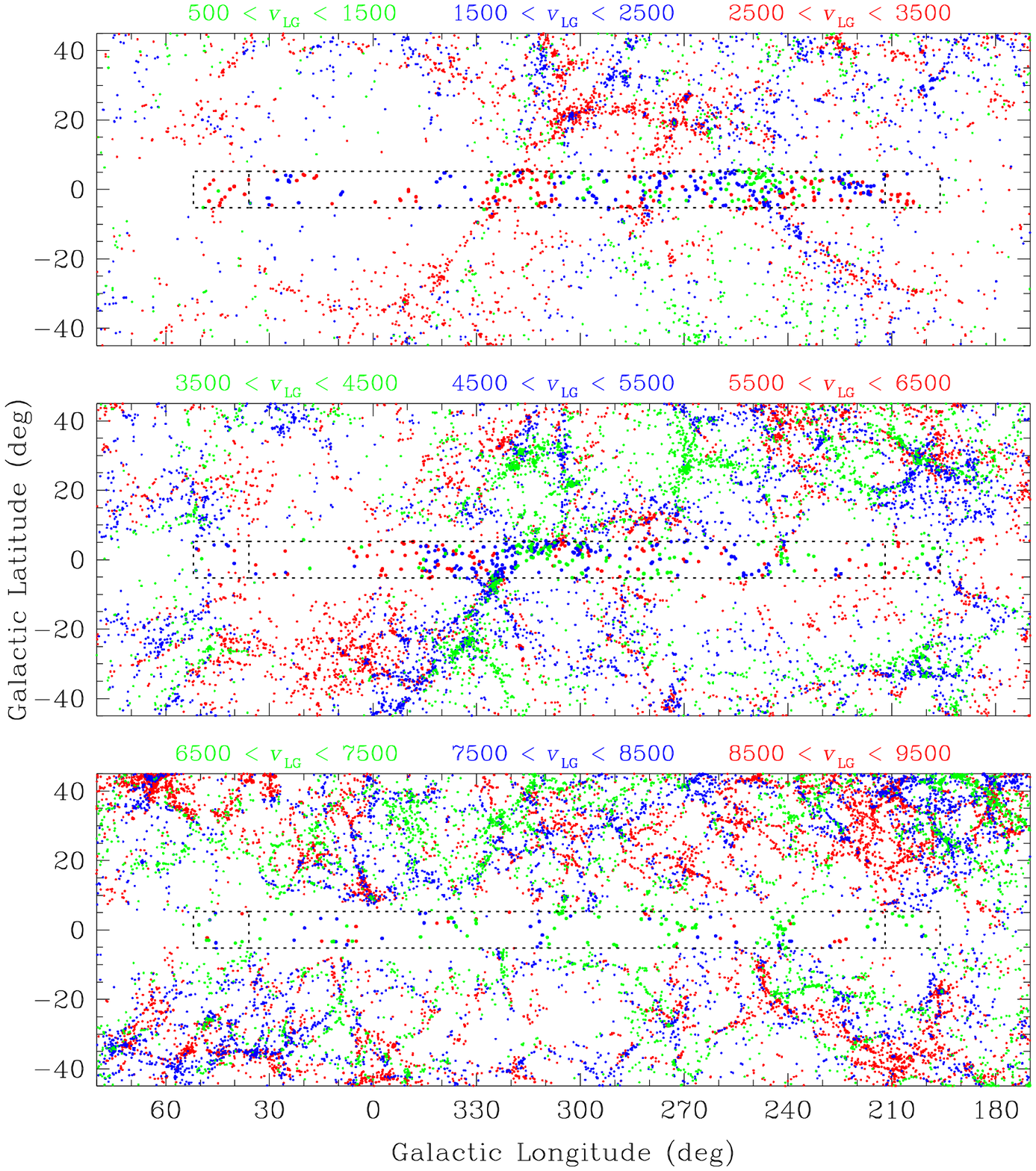}
\caption{Three sky projections in Galactic coordinates in redshift
  intervals of $\Delta v = 3000$\kms\ within $500 < v_{LG} < 9500$\kms. 
  The survey areas are outlined.  The most distant slice ($9500 <
  v_{LG} <12000$\kms) is not presented due to the scarcity of detections
  at those redshifts ($N=36$); but see Fig.~\ref{wedge}).  The HIZOA
  detections (larger dots) are combined with redshifts available in LEDA up
  to latitudes of $|b| < 45\degr$ to investigate the connectivity of newly
  revealed large-scale structures with known structures. Within the
  individual panels, the color-coding demarcates narrower velocity
  intervals ($\Delta v = 1000$\kms).
}
\label{onsky}
\end{figure*}

In the second panel, the LV clearly dominates. There are very few
galaxies up to about 4500\kms, and only one below 1500\kms. A broad
peak ranging from about $4500-6500$\kms\ seems to demarcate the right-hand side of the boundary
of the LV ($\ell \sim 345\degr$;  see Fig.~\ref{wedge}) as well as the far side of
the LV (see also Figs.~\ref{wedge} and~\ref{onsky}). A more detailed
discussion on the extent of the LV can be found in Kraan-Korteweg \etal 
(2008), including preliminary data from an extension of the deep Parkes
multibeam \HI\ survey to higher latitudes around the Galactic Bulge. At
higher velocities the numbers drop down to extremely low levels again,
indicative of a further underdense region behind the LV, which probably is
quite extended on the sky, because it is visible also in the third panel.

The third panel with its strong peak at about 5000\kms\ is clearly
dominated by the Norma supercluster (Woudt \& Kraan-Korteweg 2000; Woudt 
\etal 2004; Radburn \etal 2006). At lower velocities the numbers are also
elevated. This can be explained by the nearer filament that crosses the
plane at about 3000\kms\ and links up with the Centaurus clusters at
higher latitudes. The number counts at higher redshifts are particularly
low, and the Norma overdensity seems well separated from other
structures. This is not unexpected for a cosmic web-like Universe where
underdense region are surrounded by wall-like structures.

The bottom panel is dominated by low-velocity galaxies ($v_{\rm LG} \la
3000$\kms). They form part of the quite distinct and much larger filament
that crosses the Plane in Puppis. It will be discussed in more detail in
Fig.~\ref{onsky} which merges the new detections with known features beyond
the ZOA. The slight overdensity around 7500\kms\ is real and seems
suggestive of a more distant filament, one that has already been
highlighted as such in early ZOA \HI\ work by Chamaraux \& Masnou (2004).

\subsection{New large-scale structures and their connectivity} 

In the following we give a qualitative description of new structures
detected in HIZOA, and how they link and contribute to known structures
based on the finalised \HI\ data set. A few results have been previously
presented based on preliminary catalogs (e.g., Kraan-Korteweg, 2005,
Kraan-Korteweg \etal 2005; Henning \etal 2005). A more quantitative
analysis will be given in a subsequent paper; also included will be: (a)
the further extension of the Parkes ZOA surveys to higher latitudes around
the Galactic Bulge ($\pm10\degr$ for the longitude range $36\degr > \ell >
332\degr$, reaching up to higher positive latitudes ($+15\degr$) for
$20\degr > \ell > 348\degr$), where both the deep optical and near-infrared
2MASS surveys fail at identifying galaxies due to stellar crowding; and (b)
near-infrared counterparts of all HIZOA galaxies based on a systematic deep
near-infrared ($J H K_S$) follow-up imaging observations (Williams \etal
2014, Said \etal in prep.).

We investigate the large-scale structures by examining the redshift cone
defined by the HIZOA galaxies (Fig.~\ref{wedge}), and we will refer to
on-sky distributions for various redshift intervals at the same time
(Fig.~\ref{onsky}).

Figure~\ref{wedge} presents a redshift wedge with the HIZOA galaxies. The
width of the wedge corresponds to the HIZOA survey width and includes all
galaxies detected up to $|b| \le 5\fdg25$, the most opaque part of
the Milky Way.  It traces the structures along its full longitude range out
to the velocity limit of $v_{LG} = 12\,000$\kms. Hardly any of these galaxies were
known before the HIZOA survey, apart from a handful of galaxies in the Puppis
area where the dust column density is particularly low. The high efficiency
of tracing large-scale structures with systematic \HI\ surveys without
hindrance by the foreground ``pollution'' of our Milky Way -- which biases
most other multi-wavelength surveys (e.g., Kraan-Korteweg \& Lahav 2000) --
is clearly demonstrated.

\begin{figure*}[t]
\centering
\includegraphics[width=0.59\textwidth]{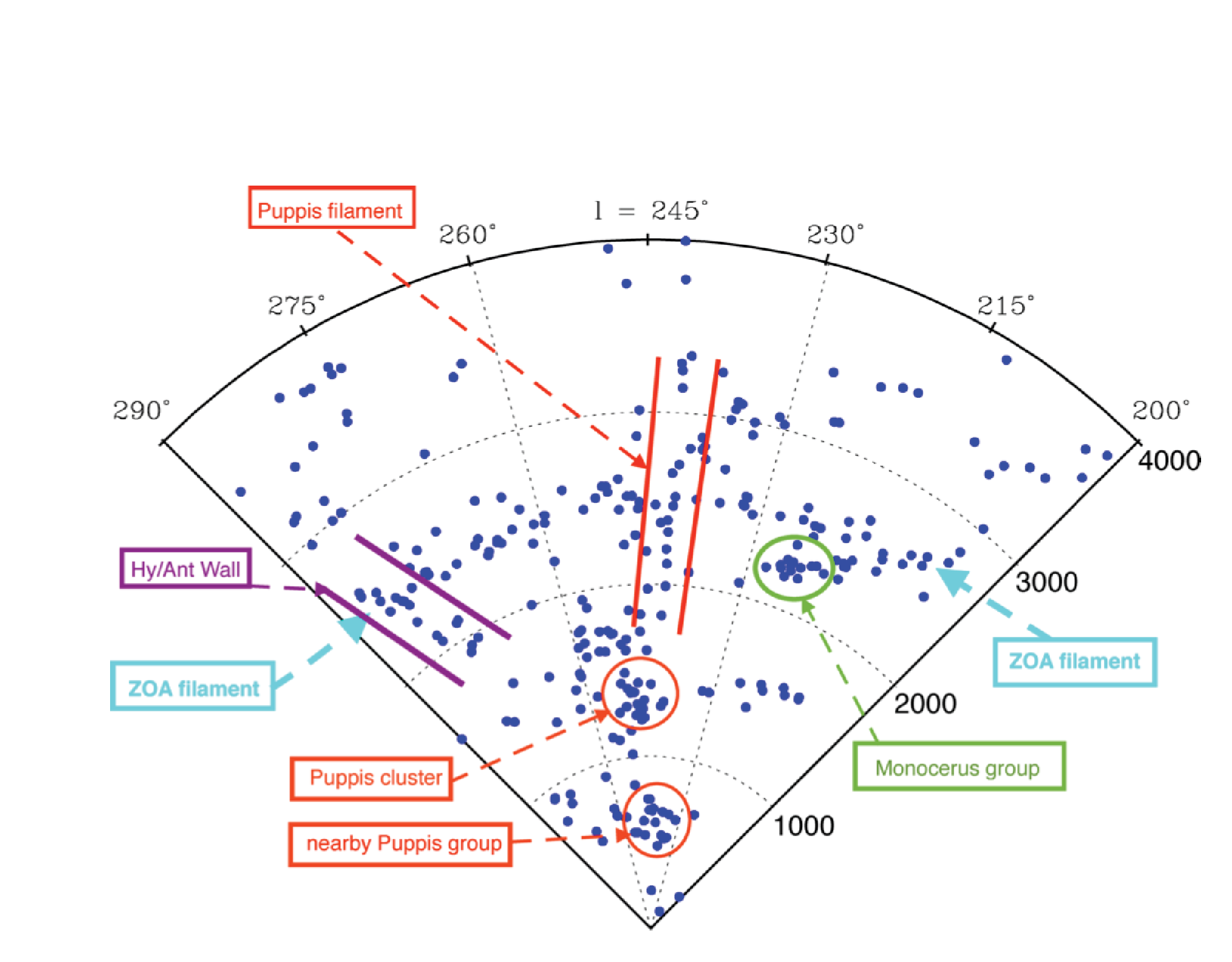}
\caption{Zoomed-in wedge focusing on the crowded Puppis area 
  ($290\degr > \ell > 200\degr$; $|b| < 5\fdg25$; $v_{LG} < 4000$\kms). The most
  prominent features mentioned in the text are marked. A very narrow
  filament or wall (marked `ZOA filament') can be traced at about $\sim
  2500 \pm 100$\kms\ over the full longitude range of the wedge.
}
\label{Puppis}
\end{figure*}

Figure~\ref{onsky} shows on-sky distributions in Galactic coordinates
centered on the southern Galactic Plane.  The \HI-detected galaxies are plotted together
with galaxies surrounding the ZOA using redshifts extracted from
HyperLEDA\footnote{http://leda.univ-lyon1.fr/} (Paturel \etal 2003) for the
Galactic latitude range of $|b| < 45\degr$ to allow the interpretation of
the newly revealed features in context with known large-scale structures. It
should be emphasised that the merged samples are very differently
selected. The HyperLEDA sample contains redshifts purely based on their
availability in the literature and thus does not provide homogenous
coverage. It also contains both radio and optically determined
redshifts. Restricting the data to \HI\ redshifts only results in an
extremely shallow coverage -- except for the Arecibo declination range --
showing again the urgent need for the forthcoming SKA Pathfinder
\HI\ surveys. Another comparison sample could have been the 2MRS (Huchra
\etal\ 2012) which is uniform and all-sky down to latitudes of $|b| >
5\degr$. However, the \HI\ and NIR selected surveys favor vastly different
populations, gas-rich bluish versus early-type red galaxies, which leads to vastly different selection functions. The HIZOA galaxies are densest at low velocities because HIZOA
is so sensitive to gas-rich nearby dwarfs which 2MRS does not easily
detect, whereas the 2MRS numbers steeply rise with redshift and the balance
shifts rapidly.
%
%
We hence prefer to use the HyperLEDA data sample to display the
connectivity of structure across the Milky Way, because it overall it provides
the deepest data set, hence the optimal source to delineate structure:
voids remain empty if real, and filaments will only appear more pronounced
when sampled more deeply.

The panels in Fig.~\ref{onsky} show three contiguous shells in
velocity space, each of width of 3000\kms. Colors indicate finer
redshift intervals within each panel. This figure also demonstrates
quite clearly how well the inner $\pm5\degr$ of the ZOA has been
filled by the HIZOA survey ($N=408, N=368$ and $N=134$ redshifts from
top to bottom respectively). The last shell out to the velocity limit
of the survey is not plotted here. Due to the low sensitivity in that
redshift interval the detection rate is too low ($N = 36$; see also
Fig.~\ref{vhist_reg}) to add much insight into larger structures;
their distribution can be inspected in the redshift slice
(Fig.~\ref{wedge}). 

As before, we will focus on the three areas separately that are dominated by
quite different large-scale structures, i.e., Puppis, GA and LV.

\subsubsection{The Puppis region}

The righthand side of Fig.~\ref{wedge} is quite crowded, particularly
at the low velocity range in the Puppis region ($\ell \sim 245\degr$).
For a better visualisation of these quite nearby structures we show a
zoom-in version of the redshift slice in Fig.~\ref{Puppis}. A nearby
group and a slightly more distant small cluster at 700 and 1400\kms,
respectively, dubbed the Puppis 1 group and the Puppis 2 cluster,
stand out. Both were found already in 1992 by Kraan-Korteweg \&
Huchtmeier through \HI\ follow-up observations of optically visible
galaxies in this area of low foreground extinction. Due to their
proximity, both of these galaxy concentrations were also identified
and described in the shallow survey (Henning \etal 2000) and in the
\HI\ Bright Galaxy Catalog (Koribalski \etal 2004). At slightly
higher redshifts we note the Puppis filament. There is a third
concentration (not marked by a circle) at slightly higher longitudes
and redshifts ($\ell \sim 255\degr, v \sim 1700$\kms), but it does not
stand out as a group in three dimensions. Between 2000 and 3000\kms\
we see a surprisingly narrow filament (henceforth referred to as the
ZOA filament) that can be traced along the full longitude range of the
Puppis wedge. It is well separated from other structures apart from
where this filament crosses the Puppis filament ($\ell \sim 245\degr$)
and the Hydra Wall ($\ell \sim 278\degr$). The Monocerus group ($\ell
\sim 220\degr$; marked as green oval) was identified in Donley \etal (2005). They suggested that the group might form part of the very narrow ZOA filament (see
Fig.~\ref{onsky}). A closer look at the Puppis cone presented here (Fig.~\ref{Puppis}) does, however, not entirely support this because they are
distinct in redshift space.

The Puppis filament itself is a highly interesting feature; it is part of a
very extended structure and can be traced over most of the southern sky
(see top panel of Fig.~\ref{onsky}). From far below the Galactic Plane
($\ell,b,v) \sim (210\degr,-25\degr, 2800$\kms), it extends towards the
ZOA around Puppis ($\ell \sim 245\degr$) crossing at slightly
lower redshifts. From there it continues and connects with Antlia
($273\degr, 19\degr, 2800$\kms).  It continues across Antlia towards the
Centaurus cluster ($302\degr, 21\degr, 3400$\kms) and then folds back
crossing the Milky Way once more -- this second crossing lies just in front
of the Great Wall or Norma Wall crossing discussed in the next section. The
filament traces a near-perfect sinewave-like structure in this projection,
while in an equatorial projection it appears more circular. Its
three-dimensional shape seems to be to be a long, straight
filament. Interesting though is the length over which the filament is
contiguous across the sky ($\sim180\degr$), certainly in comparison to its
narrow width. At a mean redshift of $\sim 3000$\kms\, this translates to
about $\sim\!100 h^{-1}$Mpc. Although interspersed with numerous galaxy
groups, the filament is only about ($\sim\!5 h^{-1}$Mpc) wide, very
different compared to the very broad, foam-like Norma Wall structure.

Various other filaments seem to intersect with the Puppis filament. The
most prominent is a filamentary connection emanating from the Hydra cluster
($270\degr, 26\degr, 3500$\kms) towards Antlia (where it crosses the
Puppis filament), from where it moves downwards in Fig.~\ref{onsky},
crossing the Plane at about $\ell \sim 280\degr$. This continuation to
latitudes of about $-10\degr$ was surmised early on by Kraan-Korteweg
(1989) and Kraan-Korteweg \etal (1995) and now is substantiated with the
present data set. The signature of the crossing is prominent, too, in
Fig.~\ref{Puppis}. It is conceivable that the Monocerus group ($\ell
\sim 220\degr$) lines up with Antlia as well, but the sparsity of data in
the Galactic latitude range between $+5\degr$ and $+10\degr$ precludes such
a confirmation.

At higher redshifts (Fig.~\ref{wedge}) we see a further indication of a
possible cluster (Puppis 3) at about ($\ell,v) \sim (242\degr, 7000$\kms). 
It is very prominent in the middle and bottom panels of the sky
distributions, and appears to be embedded in a filamentary structure within
the survey area. This overdensity is also picked up by Chamaraux \& Masnou
(2004) in their analysis of the Puppis Wall. They claim the existence of a
large void between the Puppis Wall and this galaxy agglomeration. However,
this is not evident from the figures presented here which show numerous
galaxies in between these redshift ranges (see
Fig.\ref{wedge}). Furthermore, the clumping or filament at about 7000\kms\ 
seems to be located in the middle of a large underdense region of an
extent of nearly $40\degr \times 40\degr$.

\subsubsection{The Great Attractor region}

\begin{figure*}[t]
\centering
\includegraphics[width=0.59\textwidth]{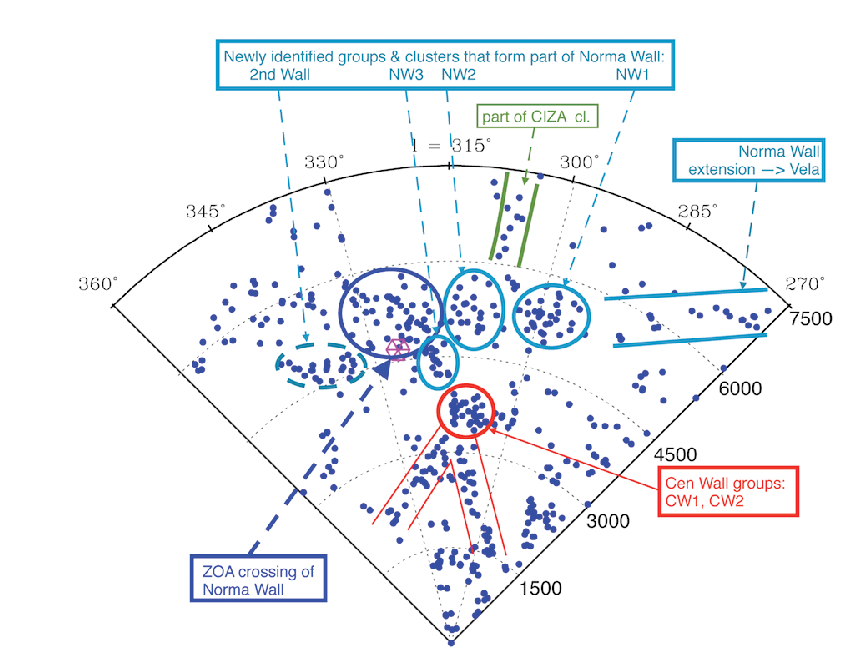}
\caption{Zoomed-in wedge focused on the Norma Wall 
  ($360\degr > \ell > 270\degr$; $|b| < 5\fdg25$; $v_{LG} < 7500$\kms) showing HIZOA
  galaxies. Note the wall-like structure of the GA, the Norma supercluster,
  at about $\sim 4800$\kms\ that stretches from about $340\degr \ga \ell
  \ga 290\degr$. The magenta hexagon marks the center of the Norma cluster
  which is located just below the HIZOA survey area $(\ell,b,v) = (325\degr,
  -7\degr, 4871$\kms).
}
\label{NormaWall}
\end{figure*}

Overall, the large-scale structures revealed by the \HI\ detections in
Fig.~\ref{wedge} are clearly dominated by the GA, also
referred to as the Norma Wall or the Norma supercluster (Kraan-Korteweg 
\etal\ 1994; Woudt \etal 1999, 2004; Radburn-Smith \etal 2006; Jarrett
\etal 2007; Kraan-Korteweg \etal 2011). The Norma Wall is centered on the
Norma cluster, ACO 3627 (Abell, Corwin \& Olowin 1989), which has been
identified as the most massive cluster in the GA region (Kraan-Korteweg
\etal 1996; Woudt \etal 2008). However, the Norma cluster itself ($\ell,b,v) =
(325\fdg3, -7\fdg2, 4871$\kms) lies just outside the
boundaries of our HIZOA survey. The Norma Wall crosses the ZOA diagonally
-- from the Norma Cluster to the CIZA and Cen-Crux clusters on the opposite
side of the ZOA (discussed below). The HIZOA survey traces some of its
previously partly, or fully, obscured components in more detail, and has
uncovered further galaxy concentrations that form part of the Norma
supercluster, adding to the general overdensity in this part of the
sky. A zoom-in redshift cone of the GA region is presented in
Fig.~\ref{NormaWall}.

In this plot, the Norma Wall seems to stretch from $360\degr$ to $290\degr$
, lying always just below the 6000\kms\ circle. Indeed, Woudt \etal (2004,
2008) have shown that both the Norma cluster and the wall in which it is
embedded are well separated in space from other structures. At the lower
longitudes a weak extension towards Vela ($\ell \sim 270\degr$) can be seen
-- marked as Norma Wall extension. The latter is more pronounced at higher
latitudes (see middle panel of Fig.~\ref{onsky}).

This Norma Wall is made up of various agglomerations. We will discuss them one
by one, from left to right, respectively, high to low longitudes. Note that
the galaxies around 6000\kms\ and longitudes of $\sim 340\degr - 360\degr$
are not part of the Norma Wall. They might form part of the outer boundary
of the LV though. The agglomeration around $340\degr$, denoted as
the $2^{\rm nd}$ Wall, is identified for the first time. It is marked with
dashed lines because it is unclear whether this galaxy concentration
extends above latitudes of $|b| > 5\degr$, due to the overall high
extinction and star density this close to the Galactic Bulge. The on-sky
distribution seems to hint at a further wall parallel to the Centaurus Wall
but at higher longitudes (see middle panel of Fig.\ref{onsky}). The
Galactic Bulge extension data, which will provide further \HI\ detections
up to higher latitudes around $(\ell,v) \sim (340\degr,4500$\kms), might
shed further light on this. Nevertheless, this galaxy concentration is a
further contributor to the mass overdensity in this part of the sky.

\paragraph{ZOA crossing of the Norma Wall:} 

The largest clump (marked in dark blue) contains galaxies that start at
the outer fringes of the Norma cluster, A3627 at ($\ell,b,v) = (325\degr,
-7\degr, 4871$\kms) (Woudt \etal 2008), and extends diagonally across the
ZOA (see also middle panel of Fig.~\ref{onsky}). A small cluster, marked as
NW3, has been identified for the first time within the Norma Wall. NW3 lies
deep in the ZOA at ($\ell,b,v) = (319\degr, -2\degr, 4500$\kms). At higher
positive latitudes, the Norma Wall then encompasses two further
condensations, NW2 and NW1 ($\ell \sim 307\degr$ and $300\degr$), which may
be groups (or small clusters). None of these were previously known. Both
are located at $b \sim 4\degr$ and quite distinct in Fig.~\ref{onsky}
(middle panel). In between these two groups, a short Finger of God is
evident at slightly higher redshifts ($\ell,b,v) = (307\degr, +5\degr,
5700$\kms). This is due to galaxies in the outskirts of
Centaurus-Crux/CIZA J\,1324.7$-$5736 cluster, discovered independently by
(i) Fairall, Woudt \& Kraan-Korteweg (1998), Woudt \& Kraan-Korteweg
(2001), and Woudt \etal (2004) from deep optical galaxy searches and
systematic redshift follow-ups, and (ii) Ebeling \etal (2002; 2005) and
Mullis \etal (2005) in their search for X-ray clusters in the Zone of
Avoidance. This cluster is the second most massive cluster within the Norma
Wall, and has about $50 - 70$\% of the mass of the Norma cluster
(Radburn-Smith \etal 2006). From the CIZA cluster, the Norma Wall
continues to higher latitudes and velocities (its lower latitude part is
still visible in the wedge) where it connects to the cluster in Vela, Abell
S0639 at ($\ell,b,v) = (280\degr, +11\degr, 6500$\kms) (Stein 1997).

\paragraph{Centaurus Wall groups:}

Apart from new groups and clusters that form part of the GA overdensity,
other significant new structures are identified at lower redshifts. Notable
is the cluster CW2 centered at about ($316\degr, +3\degr, 3800$\kms) which
is massive enough to display a small Finger of God. To the right of this (within the same
red circle), a possible group of galaxies CW1 is visible at about
($309\degr, 3\fdg5, 3400$\kms).  They lie in front of the GA 
Wall and form part of the Centaurus Wall, a much narrower wall that lines
up with the Centaurus cluster ($302\degr, 21\degr, 3400$\kms). The clumps
form part of the sine-wave feature (top panel of Fig.~\ref{onsky}). Two
nearby filaments are also marked. They appear to point to the newly
identified groups/clusters (marked by red lines), i.e., the one that
approaches the cluster CW2 from the left at lower velocities is part of
the sine-wave ZOA crossing visible in the on-sky projection at those
longitudes ($315\degr-330\degr$). The other filament pointing towards the
CW1 and CW2 clumps from the right (lower longitudes) is more difficult to
discern in the on-sky distribution because it is cut into two parts by
redshift division between the top and middle panel.
 
\paragraph{The PKS\,1343$-$601 cluster:}

What is not very prominent in the HIZOA detections is the cluster around
the strong radio source PKS\,1343$-$601, also known as Cen B. Because the
presence of a strong radio source often is indicative of a massive cluster,
there was previous speculation (Kraan-Korteweg \& Woudt 1999) whether a
further dominant cluster that forms part of the Norma supercluster could
have escaped detection due to its location at very low latitudes ($b =
1\fdg7$, and high extinction ($A_B = 10\fm8$; Schlafly \& Finkbeiner
2011). However, various deep near-infrared studies of the cluster (Nagayama
\etal 2004, Schr\"oder \etal 2007), as well as a careful investigation of
its X-ray-flux by Ebeling \etal (2005) do not support this hypothesis. The
PKS\,1343$-$601 cluster seems to be an intermediate size cluster, and forms
part of the GA overdensity. We find some galaxies in that
general area and velocity, but not many. The strong radio continuum
emission of Cen B (79 Jy at 1410 MHz; Wright \& Otrupcek 1990) suppresses
galaxy detection in its immediate vicinity, but does not explain the
absence of galaxies in our survey. It is possible that CW1, at only slightly higher
latitude, is connected to or is part of this cluster.

Next to the Norma cluster, and the Centaurus-Crux cluster no further X-ray
emission has been found that could point to the existence of a further
massive cluster within the Norma Wall. However X-rays are subject to
photoelectric absorption by Galactic gas and confusion with Galactic
sources, which limits detection close to the Galactic Plane. Nevertheless,
apart from the Norma and Centaurus Wall groups and clusters (NW1, NW2, NW3,
CW1 and CW2), there is no evidence from the current \HI\ survey, or
previous surveys at other wavelengths, for further massive concentrations
of galaxies within the Norma Wall.

\subsubsection{The Local Void region}

The left third of the HIZOA redshift slice (Fig.~\ref{wedge}) is
dominated by the LV. Both Figs.~\ref{wedge} \& \ref{onsky} indicate
that the LV under-density appears to extend in the range of about $45\degr > \ell
> 330\degr$ for radial velocities $v_{LG} \la 6000$\kms, consistent
with Tully's early definition (Tully \& Fisher 1987). The left-hand
boundary had been uncovered in HIZOA-N (Donley \etal2005). The on-sky
projections suggest that the LV is quite symmetric around the
Galactic Plane reaching up to latitudes of about $\pm 40\degr$ on
either side. It is thus quite a formidable structure in the nearby
Universe. However, both the wedge and certainly the on-sky projection
indicate that the LV is not nearly as empty as the previously available 
data suggest.  The reason is that a large fraction of the LV is
located behind the Galactic Bulge (GB) where dust extinction and star
confusion reach much higher latitudes and the optical and NIR ZOA
are substantially wider. The current HIZOA data alone do not reach
high enough Galactic latitudes to bridge the gap. It is for this
reason that the above mentioned GB-extension of HIZOA was
launched. The GB-extension will improve the knowledge of the borders
of the LV and Sagittarius Void ($350\degr, +0\degr, 4500$\kms; Fairall
1998), as well as the Ophiuchus cluster ($0\fdg5, +9\fdg3, 9000$\kms)
studied optically and spectroscopically by Hasegawa \etal (2000), and
Wakamatsu \etal (2005).

The galaxies detected in the LV predominantly seem to be low-mass
(gas-rich) dwarf galaxies. This was also confirmed in a preliminary
analysis of the GB extension (Kraan-Korteweg \etal 2008) which reveals
some quite fine filaments, sparsely populated with low-mass galaxies, that
permeate the LV. They probably are similar structures to what has been
dubbed tendrils by Alpaslan \etal (2014) in the analysis of large-scale
structures in GAMA. The LV population will be studied in more detail in a
forthcoming paper.

\section{Conclusions} \label{concl}


We have presented data from a deep \HI\ survey (HIZOA-S) with the multibeam
receiver on the Parkes telescope in the southern `Zone of Avoidance'. The
region covered ($36\degr > \ell > 212\degr$; $|b|<5\degr$) includes the highest
obscuration for optical/IR observations and, near the Galactic Center, the
highest stellar crowding. Of the 883 HIZOA-S galaxies detected, only 8\% have
existing optical redshifts. Nevertheless, on careful inspection of new and existing
images and data sources, 78\% of the detections have a counterpart visible
in either the optical or the near-IR wavebands. A third of these have not
been previously cataloged. The median distance between the \HI\ position
and the counterpart is $1\farcm4$ (95\% of counterparts have $d<3\farcm8$),
while the FWHP beam size is $15\farcm5$. For detected galaxies with
low \HI\ masses which are in areas of high extinction, we have mostly been
unable to find counterparts.
  
The survey (with an average rms of 6\,mJy) is 2--3 times deeper than HIPASS
(Meyer \etal 2004). To suppress ringing from strong Galactic \HI\ and
recombination line emission, spectra have been Hanning-smoothed, so the
velocity resolution is 50\% poorer (27\kms). For the 251 brighter galaxies
in common, the derived \HI\ parameters (velocity, width and flux) for
HIPASS and HIZOA-S are in excellent agreement.

\begin{deluxetable}{lcccl}
\tablecolumns{5}
\tablewidth{0pc}
\tablecaption{New large-scale structures detected in the HIZOA data.\label{newlss}}
\tablehead{
\colhead{Name}     &   \colhead{$v_{\rm LG}$}   &   \colhead{$\ell$}   & \colhead{$b$}   &  \colhead{Description} \\
\colhead{}         &   \colhead{\kms}           &                      &                 & 
}
\startdata
Pup 1 group                       & $\sim 700$                &  $\sim 245\degr$                   & $\sim +2.5\degr$                & loose group \\
Pup 2 cluster                     &  $\sim 1400$             &  $\sim 248\degr$                   & $\sim +4 \degr$                  & dense group \\
Pup 3 cluster                     &  $\sim 6800$             &  $\sim 242\degr$                   & $\sim 0\degr $                     & cluster \\
Monoceros group              &  $\sim 2200$             &  $\sim 215\degr - 225\degr$  & $\sim -3\degr - +4\degr$ & elongated loose group\\ 
ZOA filament                      & $2500 \pm 150$        & $\sim 210\degr - 285\degr$  & $\sim \pm 4 \degr$            & lengthy filament with small dispersion\\
NW1                                  &  $\sim  5600$             & $\sim 298\degr$                     & $\sim +4\degr$                       & NW galaxy concentration\\
NW2                                  &  $\sim 5200$              & $\sim 307\degr $                    & $\sim +4\degr$               & NW galaxy concentration\\
NW3                                  & $\sim 4500$               & $\sim 319\degr $                    & $\sim -2\degr$                & new NW cluster \\
$2^{\rm nd}$ Wall                &  $\sim 4800$              & $335\degr - 345\degr$           &  $\sim \pm 4\degr$         & possible new wall \\
CW1                                  &  $\sim 3400$               & $\sim 309\degr$                     & $\sim +3.5\degr$                & new CW loose group \\
CW2                                  &  $\sim 3800$               & $\sim 316\degr$                     & $\sim +3\degr$                & new CW cluster \\
\enddata
\end{deluxetable}

The completeness as a function of Galactic background, flux and velocity
width for the HIZOA-S catalog has been characterised in a manner that is
independent of assumptions of homogeneity. This allows HIZOA-S to be used for
quantitative study of the ZOA density field. The flux completeness limit
for the main survey region (excluding edges) and at Galactic continuum temperatures
$<7$\,K, is 2.8\Jykms\ at the mean sample velocity width of 160\kms. This
corresponds to a mean flux density limit of 17\,mJy. The flux integral and
flux density completeness limits scale as $w^{0.74}$ and $w^{-0.26}$,
respectively; i.e., galaxies with narrow widths can be found at lower flux
integrals and higher mean flux densities. There is some incompleteness at
higher fluxes, but we also find 9\% of HIZOA galaxies are detected at lower
fluxes.

Several interesting new objects have been noted in HIZOA data and already
published, including the nearby extended objects HIZOA J1514$-$52, J1532$-$56
and J1616$-$55 (Staveley-Smith \etal 1998), the new nearby galaxy HIZSS 003
(Henning \etal 2000, Massey, Henning \& Kraan-Korteweg 2003, Begum \etal
2005), the fast rotator J1416$-$58 (Juraszek \etal 2000) and the massive
J0836$$-$$43 galaxy (Donley \etal 2006; Cluver \etal 2008, 2010). Notably no
new nearby, massive galaxies similar to Circinus have been discovered, and
there is no further room in the ZOA for such an object to be
hidden within the areas mapped so far. However, the new catalog does contain 
a new galaxy, HIZOA J1353$-$58, which seems to be a possible companion (and 
the only candidate found to date) for the Circinus galaxy.
 
The main new results identified in this paper are the groups and clusters
discussed in \S\ref{s:lss}. For clarity, these are summarized in
Table~\ref{newlss}. The HIZOA survey has clearly proved to be a highly
successful approach in uncovering the large-scale structures in the ZOA
most of which were previously unknown. This is particularly
striking in the redshift cone in Fig.~\ref{wedge}, which reveals
large-scale structure within the deepest obscuration layer of the Milky
Way, reminiscent of the redshift cone of the Coma cluster in the so-called
Great Wall (de Lapparent \etal 1986). The main difference is that the main
Norma cluster lies just beyond the borders of the HIZOA survey.  The
combination with known structures adjacent to the ZOA (Fig.~\ref{onsky})
shows how many of the in \HI-detected features link up. HIZOA has resulted
in an improved census of the GA region and an improved
understanding of the extent of the Norma supercluster. The diagonal
crossing of the supercluster is well traced by the new data and has led to
the identification of further galaxy concentrations that form part of the
wall-like structure, such as the NW3 cluster deep in the plane, and the NW2
and NW1 complexes that connect to the Cen-Crux clusters that lie just at
the border of this survey. Hence, the Norma Wall is now seen to extend from
Pavo II to the Norma cluster, the NW3 cluster, the NW2, NW1, and the CIZA
clusters to the Vela cluster. Also remarkable are the two nearer
clusters that lie within the Centaurus Wall, that have filaments pointing
towards them, very similar to the legs and body of the Great Wall
`stickman'. We have also newly identified a potential second crossing at higher
latitudes. Although not part of the Norma supercluster despite its
redshift, this galaxy concentration also adds to the overall mass of the GA
overdensity. 
    
In the Puppis region, some of the unveiled clusters and groups have been
seen in earlier \HI-observations. However, what is confirmed is the earlier
suspected Hy/Ant Wall, which can now be traced across the ZOA, as well as a
full mapping of the Puppis filament, that forms part of the major sine-wave
structure seen in the top panel of Fig.~\ref{onsky}. A new discovery is the
Monoceros group, and a filamentary feature at mostly constant
redshift that runs within the ZOA for a longitude range of nearly 80
degrees.

HIZOA has also confirmed the large extent of the Local Void. Nevertheless,
quite a few nearby galaxies have been found behind the Galactic Bulge,
indicating that the void is not quite as empty as previously suggested.

These results bode well for the future with the forthcoming \HI\ surveys
that are planned with the SKA Pathfinders ASKAP (Wallaby) and
WSRT/APERTIF. These surveys will together cover the whole sky and provide
for the first time a deeper ($z = 0-0.26$) and well-resolved complete
census of the large-scale structures in the sky, {\sl inclusive} of the
Milky Way.

\acknowledgments

The Parkes radio telescope is part of the Australia Telescope National
Facility which is funded by the Commonwealth of Australia for operation as
a National Facility managed by CSIRO. We would like to thank the staff at
the Parkes Observatory for all their support, D. Barnes, M.R. Calabretta, R.F. Haynes,
A.J. Green, S. Juraszek, M. Kesteven, S. Mader,
R.M. Price, and E.M. Sadler for their assistance during the course of the
survey, and K. Said for early access to the IRSF
imaging data of the HIZOA galaxies as well as his help with some of the
data reduction. 

Parts of this research were conducted by the Australian Research Council
Centre of Excellence for All-sky Astrophysics (CAASTRO), through project
number CE110001020. RKK and ACS acknowledge the research support they
received from the South African National Research Foundation. PAH thanks
the NSF for support for the early stages of this work through the NSF
Faculty Early Career Development (CAREER) Program award AST 95-02268.

This research has made use of: the NASA/IPAC Extragalactic Database (NED)
which is operated by the Jet Propulsion Laboratory, California Institute of
Technology, under contract with the National Aeronautics and Space
Administration; the Digitized Sky Surveys were produced at the Space
Telescope Science Institute under U.S. Government grant NAG W-2166; the
Sloan Digital Sky Survey which is managed by the Astrophysical Research
Consortium for the Participating Institutions; the HIPASS data archive,
provided by the ATNF under the auspices of the Multibeam Survey Working
Group; the SIMBAD database, operated at CDS, Strasbourg, France and of the
SuperCOSMOS Sky Surveys, the WFCAM and VISTA Science Archives, operated at
the Royal Observatory of Edinburgh (WFAU); the HyperLEDA database
(http://leda.univ-lyon1.fr); data products from the Two Micron All Sky
Survey, which is a joint project of the University of Massachusetts and the
Infrared Processing and Analysis Center, funded by the National Aeronautics
and Space Administration and the National Science Foundation; the NASA/
IPAC Infrared Science Archive, which is operated by the Jet Propulsion
Laboratory, California Institute of Technology, under contract with the
National Aeronautics and Space
Administration.

\appendix

\section{Notes to individual detections}  \label{notes}



{\bf J0631$-$01:} The distance between the \HI\ position and the counterpart
is rather large ($d_{\rm sep}=6\farcm1$), but the \HI\ detection is located near the
corner of the cube (l,b = $212\fdg2,-5\fdg1$) where the sensitivity is
lower. On the other hand, the HIPASS position is much closer to the
counterpart ($d_{\rm sep}=2\farcm3$).

{\bf J0647$-$00A:} There is a possible galaxy visible in the NIR next to a
group of stars at (RA,Dec)\,=\,(06:47:06.0, \mbox{-00:}35:27), with a
distance of $1\farcm2$ to the \HI\ position.

{\bf J0653$-$03A, J0653$-$03B, and J0653$-$04:} The spectra of all three
detections show emission from the other galaxies as well. 

{\bf J0657$-$05A and J0657$-$05B:} Both spectra show emission of the other
galaxy as well.

{\bf J0700$-$04:} Begum \etal (2005) observed this
galaxy with the VLA and found that the \HI\ emission comes from two dwarf
galaxies. Both are faintly visible on the deep NIR images.


{\bf J0704$-$13:} Underneath the narrow \HI\ profile there appears to be a
broader profile (under the detection threshold) at a more northern
declination, as can be made out in the data cube. The obvious cross match
for the narrow profile is the nearly face-on galaxy 2MASX
J07042532-1346257, while further north (at about equal distance from the
fitted \HI\ position) lies a galaxy pair, 2MASX J07041909-1338217 and an
unpublished galaxy at (RA,Dec)\,=\,(07:04:18.7,\mbox{-13}:37:47), which is
likely the counterpart of the broader profile.

{\bf J0709$-$03:} This galaxy is not in NED but was found in Simbad.

{\bf J0717$-$22B and J0717$-$22A:} These two detections are close together, and
the images show a close galaxy pair that is likely interacting (it is
bright in IRAS as well as WISE). There is a (single) optical velocity for
the IRAS source ($2750\pm70$\kms, Visvanathan \& Yamada 1996) that agrees
well with the main \HI\ detection (J0717$-$22A). There are also three smaller
galaxies in the area, two of which are likely background, but one (2MASX
J07174038-2224406) is a possible candidate, too. The \HI\ detection
J0717$-$22A comes most likely either from the more edge-on component of the
galaxy pair (CGMW 2-0730) or from both (the profile is lopsided).
The narrow profile J0717$-$22B could come from the less
inclined component of the galaxy pair (CGMW
2-0731), or it could come from a hidden late-type galaxy 
($A_{\rm B} = 3\fm8$). It is less likely coming from 2MASX
J07174038-2224406 which seems nearly edge-on (though at these extinction
levels faint outer spiral arms could be invisible).

{\bf J0718$-$09:} There are two LSB galaxies in the field; the closer and
more extended one is given as the cross match. It is possible, though, that
the other galaxy at (RA,Dec)\,=\,(07:18:14.5,-09:03:03) may contribute to the
signal (cf.\ J0700$-$04). 

{\bf J0725$-$24A:} The profile shows a high-velocity shoulder. A close
inspection of the data cube indicates a fainter \HI\ emission at smaller RA
with the obvious candidate 2MASX J07245535-2430057.

{\bf J0732$-$16:} This seems to be a star forming region with an \HII\ region
and star cluster. There is a 2MASS detection nearby (2MASX
J07315300-1658237) which looks like a small, nearly edge-on galaxy. However,
the extinction in this area is large and spatially highly variable which
means the galaxy could be also be a nearly face-on barred spiral. Due to
the large spatial variation of the extinction a hidden galaxy is also
possible.

{\bf J0733$-$28:} The profile shows a high-velocity shoulder which seems to
be caused by the inclined galaxy at (RA,Dec)\,=\,(07:33:23.2,-28:44:01).


{\bf 0738$-$26:} A galaxy pair is the obvious match for the \HI\ detection. 
Both galaxies seem to be of similar (late) type, with CGMW 2-1056 being
more edge-on and CGMW 2-1059 being less inclined. It is likely that both
have \HI\ and possibly contribute to the \HI\ detection.


{\bf J0742$-$20A:} 
There are at least five galaxies of similar size visible which could form a
group of galaxies. The counterpart is a large spiral that is the most
prominent in the $B$-band. It is also the reddest on WISE images. 

{\bf J0743$-$32:} There are two 2MASS galaxies in the area but both are 
of earlier morphological type than the cross-match and hence less likely
candidates. Nevertheless, they cannot be fully excluded as candidates. 


{\bf J0744$-$26 and J0744$-$25:} 
These two \HI\ detections are only $2\farcm5$ apart. There are five
galaxies visible, two of which are possible candidates. 
2MASX J07442148-2602486 is the larger candidate and a late-type spiral. It
agrees well with the profile of J0744$-$26, while the other galaxy, 2MASX
J07442386-2600016, seems to be an early-type spiral with a smaller diameter which
agrees better with J0744$-$25. Paturel \etal (2003) observed 2MASX
J07442386-2600016 and found a detection at $v=4686$\kms\ though with a
very low SNR of 2.7.
%

{\bf J0745$-$18, J0746$-$18:} Both spectra show emission of the other galaxy as
well.


{\bf J0747$-$21:} 2MASX J07471872-2138080 is confirmed through an optical
velocity of $v=6988\pm40$\kms\ (Visvanathan \& Yamada 1996). A nearby early
spiral could be a companion.

{\bf J0748$-$25A and J0748$-$25B:} These two \HI\ detections are $4\farcm1$
apart (and at different velocities). There are six possible candidates in
the area.  2MASXJ07483666-2510211 was detected by Paturel \etal (2003) at
$v=6809$\kms, while Kraan-Korteweg \etal (in prep.) have observed
2MASXJ07483252-2516431 at $v=6804$\kms. Both used the NRT (Nan\c{c}ay
Radio Telescope) and have measured a similar flux ($\sim\!4.5$\Jykms and
$\sim\!5.1$\Jykms, respectively), while our detection has a higher
flux. We therefore conclude that 2MASXJ07483452-2513191, which is closer to
our \HI\ position, is the counterpart of J0748$-$25B.

Due to the two NRT observations, we can exclude four of the six galaxies as
the candidate for J0748$-$25A. 2MASX J07480561-2513459 has an optical
velocity of $v=7980$\kms\ (Acker \etal 1991) and is an IRAS galaxy. 
It is the most likely candidate for J0748$-$25A.

{\bf J0748$-$26, J0749$-$26A and J0749$-$26B:} These three \HI\ detections are
close together and the profiles are confused. While J0748$-$26 and J0749$-$26B
have obvious candidates, Miriad failed to fit the position of J0749$-$26A and
the distance to the cross match could be larger. There are two candidates,
ESO 493-G017 and 2MASX J07490476-2611061. ESO 493-G017 is an Sab galaxy and
was observed by Chamaraux \etal (1999) with the NRT who find a lower peak
flux ($\sim\!60$\,mJy). 2MASX J07490476-2611061 is therefore the more
likely counterpart. 

{\bf J0751$-$26}: A possible candidate looks small and seems to be a more
distant galaxy. However, an LSB halo cannot be excluded (WISE images show
considerable (foreground) emission in the W3 and W4 bands and are therefore
not conclusive).

{\bf J0753$-$22:} The candidate is a galaxy triple (IRAS source) and, if they
are companions, the \HI\ signal could come from more than one of them. Note
that the \HI\ detection in the literature is associated with CGMW 2-1471
which is the smallest galaxy in the triplet. There are more galaxies in
the area which may indicate a galaxy group.

{\bf J0755$-$23:} At least five candidates are visible in the area. The
largest and brightest, 2MASX J07551159-2304175, is an inclined early spiral
and a likely candidate for this \HI\ mass ($\log M_{\rm HI}=9.8$). Two
other galaxies, CGMW 2-1571 and a new galaxy at (RA,Dec)\,=\,
(07:55:34.1,-23:04:33), are also large but bluer and of later morphological
type. This is possibly a group of galaxies.

{\bf J0756$-$26:} The profile is confused. The HIPASS profile, with a
position $d_{\rm sep}=5\arcmin$ further south, extends only to $v\sim 6700$\kms,
while our profile extends to $v\sim 6800$\kms. Using the HIPASS webpage,
the position of our candidate gives a more similar profile to ours. The
HIZOA cube shows a peak at (RA,Dec)\,$\simeq$\,(07:56:47,-26:16:27) which
was not identified as a separate detection during the search. We conclude
that the confusing partner is 2MASX J07563846-2615018 far in the
south. Huchtmeier \etal (2001) observed this galaxy
with the Effelsberg radio telescope at an \HI\ velocity of 241\kms. The
HIZOA data cube does not show anything at this velocity and with a peak
flux of 67\,mJy and a width of $w_{50}=26$\kms.
%

{\bf J0756$-$31:} There are two equally-sized galaxies close together,
possibly a pair. One seems edge-on, the other has a small inclination. The
profile is more consistent with the less inclined galaxy.

{\bf J0802$-$22:} The counterpart, ESO 494-G013, was also observed with the
Effelsberg radio telescope by Kraan-Korteweg \& Huchtmeier (1992), and
detected at $v=1983$\kms. We could not confirm this detection at the
given peak flux of $\sim 0.15$\,Jy.
%

{\bf J0805$-$27 and J0806$-$27:} These two \HI\ detections are close together
and the profiles cannot be fully separated, that is, some of the \HI\
parameters are uncertain.

{\bf J0807$-$25:} Kraan-Korteweg \& Huchtmeier (1992) detected this galaxy
with the Effelsberg radio telescope at $v=1019$\kms, while our detection is
at $v=7679$\kms. The HIPASS
archive\footnote{http://www.atnf.csiro.au/research/multibeam/release/} does
not show any (bright) detection at this position and velocity.
%

{\bf J0808$-$35:} There are two candidates of similar size: 2MASX
J08083663-3558328 is of a later spiral type and more inclined than 2MASX
J08080066-3558243. The \HI\ detection lies between these two galaxies, and
from a closer inspection of the data cube it is likely that both galaxies
contribute to the signal. Kraan-Korteweg \etal (in prep.), using the NRT,
detected 2MASX J08080066-3558243 without the broader base of our
detection (which is therefore likely due to 2MASX J08083663-3558328).


{\bf J0816$-$27B:} A diffuse emission visible in the \B -band at
(RA,Dec)\,=\,(08:17:07.0,-27:45:15) at a distance of $3\farcm0$ to
the \HI\ position is the likely counterpart if confirmed as a galaxy.

{\bf J0817$-$27:} The \HI\ profile seems to be either disturbed or
confused. The counterpart is NGC 2559 which appears also disturbed in the
optical (Corwin \etal 1985). There is an LSB dwarf close by, at
(RA,Dec)\,=\,(08:17:24.4,-27:28:04), which may be a companion. At about
18\arcmin\ north and south are two other \HI\ detections (J0816$-$27A and
J0816$-$27B) with similar velocities.

{\bf J0817$-$29A:} This cross match is listed in Simbad but not in NED. 


{\bf J0822$-$36:} There is a possible galaxy visible in the NIR at
(RA,Dec)\,=\,(08:22:50.4,-36:36:41), with a distance of $0\farcm5$ to
the \HI\ position, which, if real, is a likely counterpart.

{\bf J0824$-$41:} There are two similar looking galaxies, possibly a galaxy
pair. 2MASX J08240445-4144221 is more inclined, while 2MASX
J08235945-4143111 seems to show a bar and the start of a spiral arm; both
match the profile well enough.  The profile has a low SNR of 5, and it is
not possible to tell whether one or both galaxies contribute to the signal.

{\bf J0827$-$35:} Paturel \etal (2003) have observed IRAS 08254-3512 which is
$5\farcm4$ north of our detection. Their detection is similar to ours with
a slightly lower peak flux. Nothing is visible at the IRAS position, and
the counterpart is more likely an invisible LSB between these two positions
(which would agree with the narrow single-peak profile and low \HI -mass
($\log M_{\rm HI}=8.0$).

{\bf J0827$-$36:} There are two galaxies close together, probably a galaxy
pair. 2MASX J08271103-3629389 is edge-on and seems to match the profile
better; it also has an optical velocity of $9512\pm40$\kms\
(Visvanathan \& Yamada 1996). The pair was detected by IRAS as well. 

{\bf J0834$-$37:} The cross match is a nearly face-on spiral galaxy. There is
a diffuse LSB galaxy visible in the \B -band close by, at
(RA,Dec)\,=\,(08:34:16.9,-37:32:36), which may contribute to the signal.

{\bf J0851$-$44:} There are two possible counterparts visible in the NIR, one
edge-on and one a little inclined. The profile seems to match the galaxy with
less inclination better. 

{\bf J0857$-$39 and J0858$-$39:} Both spectra show emission of the other galaxy
as well.

{\bf J0858$-$45A:} There are two possible counterparts visible in the NIR,
possibly a galaxy pair. 2MASX J08580941-4548126 shows spiral arms, either
disturbed or with a companion close by, while 2MASX J08581399-4550547 is a
very inclined early-type spiral. The profile matches the former slightly
better, and a possible contribution from both galaxies cannot be excluded.

{\bf J0902$-$53:} There is a small edge-on galaxy in the field. Even taking
the extinction into account ($A_{\rm B}=2\fm7$), the galaxy seems too small
for this distance ($v=7122$\kms) and \HI\ mass ($\log M_{\rm HI}=9.8$)
but cannot be fully excluded.

{\bf J0916$-$54A and J0916$-$54B:} Both spectra show emission of the other
galaxy as well.

{\bf J0927$-$48:} The high-velocity shoulder in the profile is likely caused
by J0929$-$48. 

{\bf J0932$-$51:} The cross match is an interacting galaxy pair (IRAS
source). Both are spirals and it is not possible to decide where the \HI\
comes from.


{\bf J0939$-$56:} There is a possible galaxy visible in the NIR at
(RA,Dec)\,=\,(09:39:36.3,-56:53:05) with a distance of $0\farcm8$ to
the \HI\ position, which, if real, matches the profile well.

{\bf J0949$-$47A:} The cross match (ESO 213-G002) is a peculiar S0 galaxy
which seems to have either a companion or a jet to the south. Further south
($d_{\rm sep}=3\farcm6$) is RKK 1666 which is likely to contribute: the data cube shows a faint, more
narrow profile at (RA,Dec)\,$\simeq$\,(09:49:24,-47:59:48).


{\bf J0949$-$56:} The distance between the \HI\ position and the position of
the cross match, $d_{\rm sep}=11\farcm8$, is for a SNR of 77 rather large. A closer inspection of the
data cube revealed a fainter \HI\ detection close-by ($d_{\rm
sep}=1\farcm6$) with an approximate
position of (RA,Dec)\,$\simeq$\,(09:48:18.6,-56:29:49) and $v\simeq 1641$\kms\
which appears as a low-velocity shoulder in the profile of J0949$-$56. The
obvious candidate here is 2MASX J09482319-5627370.

{\bf J0950$-$49:} There are two different optical velocities for this galaxy:
$v=12391\pm500$\kms\ (Kraan-Korteweg \etal 1995) is uncertain and seems high
for a galaxy of this appearance; $v=4043\pm82$\kms\ (Saunders \etal 2000)
is about $3\sigma$ from our \HI\ velocity and agrees better with the
optical appearance of the galaxy.

{\bf J0950$-$52:} A possible candidate visible at all wavelengths is close to
a star which makes it difficult to determine its appearance, but it seems
too small for this velocity ($v=2244$\kms). The star and the extinction
($A_{\rm B}=4\fm3$) could hide a larger halo, but a hidden LSB galaxy is a
more likely cross match. The galaxy is a possible IRAS source (IRAS
09482-5209; though it could be also the bright star to the north).

{\bf J0952$-$61:} No galaxy is visible at the position of RKK\,1733. It is
only $d_{\rm sep}=0\farcm8$ from the 2MASS position which is {\em not} listed by
Kraan-Korteweg (2000) though it is clearly visible in the \B -band. It seems
possible that both IDs indicate the same galaxy.

{\bf J0952$-$55A:} The profile is highly lopsided. The cross match is large
and diffuse with a small bulge in the NIR. It is not possible to
distinguish whether it may be disturbed or may have a small companion.

{\bf J0953$-$49:} There is a large LSB galaxy visible in the \B -band as well
as a smaller inclined galaxy (RKK 1732). The LSB galaxy matches the profile
better.

{\bf J1000$-$58A and J1000$-$58B:} The two counterparts are only $1\farcm5$
apart and the two profiles cannot be fully separated.  

{\bf J1004$-$58:} The profile shows a prominent horn which, according to the
data cube, could be caused by two close-by double horn profiles separated
in both RA and Dec (10:04:34,-58:31:10 and 10:04:04,-58:39:15,
respectively). There are two NIR-bright candidates in these positions,
which are likely counterparts. This is supported by that fact that the
distances between \HI\ position and the two counterparts ($d_{\rm sep}=4\farcm8$ and
$d_{\rm sep}=4\farcm2$) are large for an SNR of 36, \cf Fig.~\ref{distsnrplot}.
%
%

%

{\bf J1024$-$52:} The candidate seems to be a galaxy pair (possibly
interacting since it is an IRAS source). 2MASX J10240866-5159040 is
slightly rounder and of earlier type than RKK 2457, and it is not quite
clear whether only one of them or both contribute to the signal.

{\bf J1058$-$66:} This is a confused profile. the HIPASS detection J1059$-$66
lies further south and shows a broad double-horn profile clearly caused by
ESO 093-G003 with an optical measurement of $v=1470\pm270$\kms\
(Fairall 1980). Our \HI\ detection could not be separated from this
because the ESO galaxy lies beyond the edge of the cube and was therefore
not properly detected.



{\bf J1149$-$64:} There is a large spiral with a smaller companion visible in
the NIR. It is likely that the companion contributes to the profile.

{\bf J1214$-$58:} There are two galaxies in the field, possibly
companions. WKK\,0788 is edge-on and seems to match the profile slightly
better than the later type WKK\,0782. There is no indication of confusion.

{\bf J1222$-$57:} Schr\"oder \etal (2009) have detected this galaxy in the OFF
beam and associated it with WKK\,0969 (which lies $d_{\rm sep}=10\farcm2$ to the
Southeast of our detection) as the only known galaxy in the area, while at
$d_{\rm sep}=6\farcm2$ we find a galaxy that is not listed in the literature as the
more likely counterpart.

{\bf J1234$-$61:} There are two galaxies visible, possibly a galaxy pair. They
have similar appearance except that the galaxy at
(RA,Dec)\,=\,(12:33:47.8,-61:29:57) appears more face-on. The profile appears
to be confused and it is likely that both galaxies are detected.

{\bf J1304$-$58:} The distance of $5\farcm6$ to the cross match is large but
the \HI\ detection is near the edge of the cube and the signal-to-noise
ratio is low. The galaxy is well visible in the \B -band. The other
candidates in the field do not agree as well with the profile.


{\bf J1310$-$57:} There are two bright galaxies (E and S) in the field and
several smaller ones, possibly a group of galaxies.

{\bf J1314$-$58:} This is a strong signal with a lopsided double-horn.  There
are two galaxies in the field, WKK 2020 is more edge-on and a later spiral
than 2MASX J13150989-5856116 which has an optical velocity of
$v=2355\pm50$\kms\ (Fairall \etal 1998). They are likely to form a pair
with both contributing to the signal. This is supported by the fact that
the distance to each of the galaxies ($d_{\rm sep}=2\farcm5$ and $d_{\rm sep}=3\farcm0$,
respectively) is unusually high for the SNR of 96 of the \HI\ detection:
the median SNR of cross-matches with distances between $2\farcm0$ and
$3\farcm0$ is 9.8 with a sigma of 7.2, hence this is a $12\sigma$
deviation. In addition, the \HI\ position lies just between these two
galaxies.
%


{\bf J1329$-$61:} There is a possible galaxy visible in the NIR at
(RA,Dec)\,=\,(13:30:01.0,-61:51:44) with a distance of $1\farcm2$ to
the \HI\ position, which matches the profile well.




{\bf J1337$-$58B:} There are three spiral galaxies in the field, two of which
are of similar size and possibly form a pair. 2MASX J13372458-5852216 is
fairly edge-on and of earlier morphological type than 2MASX
J13373282-5854136 which has an optical velocity of $v=3709\pm70$\kms\
(Visvanathan \& Yamada 1996). It seems likely that both galaxies contribute to the
profile (cf. the profile given in Schr\"oder \etal (2009) for
WKK\,2390\,=\,2MASX J13372458-5852216).


{\bf J1344$-$65:} The counterpart, WKK 2503, was also observed and detected
by Schr\"oder \etal (2009). Their profile, with a higher velocity resolution,
is clearly confused, that is, the peak at $2750-2950$\kms\ seems to
belong to second galaxy which is neither visible in the optical nor in the
NIR.



{\bf J1353$-$58:} This \HI\ detection seems to be a companion of the Circinus
galaxy (J1413$-$65). We have adopted the same distance to calculate the \HI\
mass. 

{\bf J1405$-$65:} There are two LSB galaxies in the field. The fainter and
rounder seems to fit the profile better. The cube shows a possible
faint \HI\ detection under the detection limit at $v\simeq\!3400$\kms\ and
with a larger linewidth which could be the other, more edge-on, galaxy in
the field (WKK\,2924).  WKK\,2924 was also observed by Schr\"oder \etal
2009 where both detections were noted. While the detection at $3410$ was
attributed to WKK\,2924 (as we do), the detection at $2864$\kms\ was
attributed to WKK\,2938 which lies to the south at $d_{\rm sep}=7\farcm2$. However,
the position and flux of our \HI\ detection rule out WKK\,2938 as a
counterpart.

{\bf J1406$-$57:} There is a galaxy pair in the field. Their morphological
types seem to be similar, and 2MASX J14061018-5752098 is edge-on while
2MASX J14062739-5751425 is less inclined. It is likely that both contribute
to the profile (the data cube seems to support a separation in RA).

{\bf J1413$-$65:} This is the well-known Circinus galaxy. It is resolved
with respect to the Parkes beam. 



{\bf J1416$-$58:} This galaxy has the largest linewidth: $w_{50} = 699\pm12$\kms
(and $w_{20} = 730\pm18$\kms). The NIR images show a perfect edge-on galaxy
with a clear bulge. 



{\bf J1424$-$60:} There is a possible galaxy visible in the NIR at
(RA,Dec)\,=\,(14:25:10.3,-60:05:49) with a distance of $2\farcm5$ to
the \HI\ position which matches the profile well.  However, due to the high
extinction and faintness of the object we cannot say for certain that this
is a galaxy.


{\bf J1435$-$61:} There are several galaxies visible on the VISTA image. Due
to the high extinction ($A_{\rm B}=26\fm6$) only the bulges are clearly
visible and it is not possible to compare sizes and morphological types. We
have chosen the largest as the most likely counterpart which is also red on
WISE.

{\bf J1448$-$54:} The \HI\ detection is a near-by dwarf galaxy ($D=11.5$\,Mpc). The candidate
appears very small but with an extinction of $A_{\rm B}=3\fm3$ a large LSB
halo is possible. 

{\bf J1452$-$56:} The cross match is a very edge-on late-type spiral
galaxy. It seems to have a companion to the North which is less inclined
but of similar morphological type. According to the data cube it is likely
that both contribute to the profile.

{\bf J1501$-$57:} There is a galaxy pair in the field, both very
similar-looking inclined medium-type spirals. The profile, however, implies
a galaxy with little inclination. A closer inspection of the data cube
reveals a broader profile under the detection limit, similar to
J1004$-$58. On the other hand, the extinction in this region is fairly high
($A_{\rm B}=10\fm5$) and varies across the field (which is most prominent
in the \II -band). Therefore, either the cross match is an invisible face-on
/ late-type galaxy or we have detected a lopsided horn of one of the
galaxies in the pair (note that the baseline is very variable here).
%

{\bf J1504$-$55:} There is a possible galaxy visible in the NIR at
(RA,Dec)\,=\,(15:04:24.3,-55:29:09) with a distance of $1\farcm2$ to
the \HI\ position. It lies between stars and this cannot be unambiguously
identified as a galaxy.




{\bf J1509$-$52:} This \HI\ detection is resolved with respect to the
Parkes beam.  

{\bf J1513$-$54:} This is a close galaxy pair, possibly interacting (IRAS
source). The larger component is listed here.

{\bf J1514$-$52:} This \HI\ detection is resolved with respect to the
Parkes beam.  



{\bf J1530$-$54:} There is a possible galaxy visible in the NIR at
(RA,Dec)\,=\,(15:31:00.5,-54:03:37) with a distance of $0\farcm2$ to
the \HI\ position, which matches the profile well.

{\bf J1530$-$59:} The cross match is a close galaxy pair, possibly
interacting (IRAS source). It is not possible to tell whether one or both
contribute to the profile.

{\bf J1532$-$56:} This galaxy is extended and has been observed with the
Australia Telescope Compact Array by Staveley-Smith \etal (1998) with a
positional accuracy of 15\,arcsec (NED). At an extinction of $A_{\rm
B}=57^{\rm m}$ nothing is visible at this position.



{\bf J1549$-$60:} There are three galaxies in the field. The counterpart
seems slightly larger in the \B -band as well as with WISE than WKK\,5337
and the third galaxy at (RA,Dec)\,=\,(15:49:39.6,-60:15:24).

{\bf J1550$-$58:} Schr\"oder \etal (2009) observed WKK 5366 (at $d_{\rm sep}=4\farcm9$)
with almost the same flux. We have excluded WKK 5366 as a counterpart since
it has an optical velocity of $4822\pm82$\kms\ (Woudt \etal 2004), and
the distance is fairly large for such a bright detection.

{\bf J1553$-$50:} Despite the high extinction ($A_B=9\fm2$) several small
galaxies of similar appearance can be seen, possibly forming a group. We
have chosen the largest with obvious bulge and disk as the probable
counterpart.

{\bf J1553$-$49 and J1554$-$50:} This is an area of complex emission which
cannot easily be disentangled. The data cube shows also some fainter
emission below our detection limit. While the cross match for J1554$-$50 is a
nearly edge-on galaxy with no other galaxy visible within a 5-arcminute
radius, for J1553$-$49 there are four galaxies visible in the NIR (two of
which are unknown). The largest, 2MASX J15534458-5000182, seems to be an
early type galaxy and is an unlikely candidate. The galaxy at
(RA,Dec)\,=\,(15:54:01.1,-49:54:06) shows a diffuse halo and a small bulge
and seems to be the most likely candidate. It is also bright in the WISE
bands W3 and W4. A candidate for possible confusion could be 2MASX
J15534427-5002132 at $4\farcm8$ from the \HI\ position.
%

{\bf J1553$-$61:} Schr\"oder \etal (2009) observed WKK\,5430 which is
$5\farcm3$ to the North of our position. Their detection is lopsided and
has less flux than ours which rules out WKK\,5430 as the counterpart. In
addition, the HIPASS position (RA,Dec)\,=\,(15:53:43.5,-61:13:34) favours
WKK\,5447.

{\bf J1557$-$50A and J1557$-$50B:} These two \HI\ detections are $6\farcm5$
apart, and the two profiles are slightly confused. The extinction is fairly
high ($A_{\rm B}\simeq 17^{\rm m}$) and a hidden galaxy cannot be excluded.

{\bf J1603$-$49:} There are two candidates, both nearly edge-on. They are of
similar size, possibly a galaxy pair. The candidate at
(RA,Dec)\,=\,(16:03:34.6,-49:51:16) is slightly larger with a larger bulge
(it is also brighter on WISE).

{\bf J1612$-$56:} The detection of WKK\,6219 by Schr\"oder \etal (2009) seems
to be the low-velocity horn of J1612$-$56.

{\bf J1616$-$48A and J1616$-$48B:} These two \HI\ detections are $2\farcm0$
apart and the profiles are confused.

{\bf J1616$-$55:} There are ATCA observations on this \HI\ detection
(Staveley-Smith \etal 1998) which shows that the \HI\ is extended and has
two bright components about $25\arcmin$ apart. There is no counterpart
visible in the optical or NIR despite a moderate extinction ($A_B = 2\fm7$)
indicating that the counterpart is likely to have a low surface
brightness. Due to its extent, the coordinates given in previous
publications, Henning \etal (2000) and Juraszek \etal (2000), are $11\farcm2$
and $10\farcm3$ east of our position, while the exact position given in
Staveley-Smith \etal (1998) is only $5\farcm0$ (west) from our position.

{\bf J1622$-$44A:} There is a possible galaxy visible in the NIR at
(RA,Dec)\,=\,(16:22:05.2,-44:22:36) with a distance of $1\farcm5$ to
the \HI\ position, which matches the profile well. It lies next to a bright
star which makes the identification uncertain.

{\bf J1625$-$42:} The spectrum also shows J1624$-$42 at the lower
velocities. It appears that the profiles overlap and the linewidths and
flux measurement are uncertain.

{\bf J1625$-$55:} The cross match is a galaxy pair: 2MASX J16253644-5533498
is bright and nearly edge-on (wide profile) while WKK\,6819 is a late-type
spiral (narrow profile).


{\bf J1633$-$44:} There are two similar looking small galaxies visible in the
NIR, possibly a galaxy pair. The \HI\ detection is too faint to tell
whether one or both galaxies contribute to the profile.

{\bf J1648$-$49:} The \HI\ cross match seems to be a galaxy pair of similar
appearance where both are likely to contribute.

{\bf J1651$-$40:} There is a possible galaxy visible in the NIR at
(RA,Dec)\,=\,(16:51:53.2,-40:48:03) with a distance of $1\farcm1$ to
the \HI\ position. WISE shows also a faint reddish patch, though there is a
possibility that it is Galactic.

{\bf J1653$-$44:} This galaxy is only visible with WISE and could be also a
background galaxy.

{\bf J1653$-$48:} There is a possible galaxy visible in the NIR at
(RA,Dec)\,=\,(16:53:21.1,-48:01:53) with a distance of $2\farcm5$ to
the \HI\ position, which matches the profile well.


{\bf J1705$-$40:} WISE shows a bright red extended emission at
(RA,Dec)\,=\,(17:05:04.8,-40:58:08) with a distance of $1\farcm3$ to the \HI\
position. According to the WISE colors it could be either a star burst
galaxy or an \HII\ region (Jarrett \etal 2011).

{\bf J1716$-$42:} The only visible candidate seems too small for this
distance, though a larger LSB halo cannot be excluded.

{\bf J1716$-$35:} There are two galaxies visible in the NIR. The edge-on
galaxy at (RA,Dec)\,=\,(17:16:35.5,-35:57:26) does not match the profile as
well as the less inclined galaxy at (17:16:51.8,-35:55:48).

{\bf J1719$-$37:} This galaxy is only visible with WISE and could also be a
background galaxy.


{\bf J1811$-$21 and J1812$-$21:} Both spectra show emission of the other galaxy
as well. 

{\bf J1821$-$06:} The only candidate appears too small for a velocity of
$v=1788$\kms\ and $\log M_{\rm HI}=8.8$ even taking the extinction of $A_{\rm
B}=6\fm1$ into account, but a very LSB halo cannot be excluded. 

{\bf J1824$-$20:} This galaxy was found on WISE images and is faintly visible
on the IRSF images.

{\bf J1825$-$07:} There are four galaxies visible in the NIR, two of which
are NVSS sources and one is an AGN with an optical velocity of
$v=10887$\kms\ (Burenin \etal 2009). We assume the crossmatch to be the
most inclined. It is possible that all galaxies belong to a group or
cluster.

{\bf J1846$-$07B and J1846$-$07A:} These two \HI\ detections are only
$2\farcm2$ apart but well separated in velocity. There are two candidates
visible in the NIR: both are very inclined, and the larger is assumed to be
the cross match for J1846$-$07A, while the cross match for J1846$-$07B is
smaller and seems to be of later morphological type.

{\bf J1847+04:} This galaxy lies at a Galactic latitude of $l=36\fdg06$ and
belongs technically to the Northern Extension HIZOA (Donley \etal 2005).
%

{\bf J1855$-$03B:} There is a possible galaxy faintly visible in the NIR at
(RA,Dec)\,=\,(18:56:00.5,-03:12:21) with a distance of $0\farcm3$ to
the \HI\ position. It is large and diffuse and, if real, matches the
profile well.

{\bf J1858+00:} The profile could be confused or come from a disturbed \HI\
distribution. The extinction is high ($A_{\rm B}=12\fm2$) and only one
candidate could be found on the images. No halo is visible, and a hidden
galaxy cannot be excluded.

\clearpage

\section{Tables and figures available in electronic form only } \label{online}


\bigskip
\noindent {\bf Table~2:} \HI\ and derived parameters 

\bigskip
\noindent {\bf Table~3:} Crossmatches of the \HI\ detections 

\bigskip
\noindent {\bf Figure~1:} \HI\ spectra of the newly detected galaxies in the HIZOA-S
survey. Low order baselines (indicated by the solid line) are fitted, excluding the
detections themselves (which are bracketed by the dash-dot vertical lines) and excluding the low and high-velocity 
edges to the left and right of the dashed vertical lines, respectively. 20\% and 50\% profile
markers are visible.



\begin{thebibliography}{}

\bibitem[Acker et al.(1991)]{1991A&AS...87..499A} 
  Acker, A., Stenholm, B., \& Veron, P.\ 1991, \aaps, 87, 499 

\bibitem[Abell et al.(1989)]{1989ApJS...70....1A} 
  Abell, G.~O., Corwin, H.~G., Jr., \& Olowin, R.~P.\ 1989, \apjs, 70, 1

\bibitem[Alpaslan et al.(2014)]{2014MNRAS.440L.106A} 
  Alpaslan, M., Robotham, A.~S.~G., Obreschkow, D., et al.\ 2014, \mnras,
  440, L106

\bibitem[Alves et al.(2015)]{2015MNRAS.450.2025A} Alves, M.~I.~R., 
 Calabretta, M., Davies, R.~D., et al.\ 2015, \mnras, 450, 2025 

\bibitem[Barnes et al.(2001)]{2001MNRAS.322..486B} 
 Barnes, D.~G., Staveley-Smith, L., de Blok, W.~J.~G., et al.\
 2001, \mnras, 322, 486  

\bibitem[Begum et al.(2005)]{2005MNRAS.359L..53B} 
  Begum, A., Chengalur, J.~N., Karachentsev, I.~D., \& Sharina, M.~E.\
  2005, \mnras, 359, L53  

\bibitem[Burenin et al.(2009)]{2009AstL...35...71B} 
  Burenin, R.~A., Bikmaev, I.~F., Revnivtsev, M.~G., et al.\ 2009,
  Astronomy Letters, 35, 71  

\bibitem[Calabretta et al.(2014)]{2014PASA...31....7C} Calabretta, M.~R., 
Staveley-Smith, L., \& Barnes, D.~G.\ 2014, \pasa, 31, e007 

\bibitem[Chamaraux et al.(1999)]{1999MNRAS.307..236C} 
  Chamaraux, P., Masnou, J.-L., Kaz{\'e}s, I., et al.\ 1999, \mnras, 307,
  236  

\bibitem[Chamaraux \& Masnou(2004)]{2004MNRAS.347..541C} 
  Chamaraux, P., \& Masnou, J.-L.\ 2004, \mnras, 347, 541

\bibitem[Cluver et al.(2008)]{2008ApJ...686L..17C} Cluver, M.~E., Jarrett, 
T.~H., Appleton, P.~N., et al.\ 2008, \apjl, 686, L17 

\bibitem[Cluver et al.(2010)]{2010ApJ...725.1550C} Cluver, M.~E., Jarrett, 
T.~H., Kraan-Korteweg, R.~C., et al.\ 2010, \apj, 725, 1550 

\bibitem[Corwin et al.(1985)]{1985sgcc.book.....C} 
  Corwin, H.~G., de Vaucouleurs, A., \& de Vaucouleurs, G.\ 1985,
  University of Texas Monographs in Astronomy, Austin: University of Texas,
  1985,   

\bibitem[Courtois et al.(2012)]{2012ApJ...744...43C}
  Courtois, H.~M., Hoffman, Y., Tully, R.~B., \& Gottl\"ober, S.\ 2012  
  \apj, 774, 43

\bibitem[de Lapparent et al.(1986)]{1986ApJ...302L...1D} 
  de Lapparent, V., Geller, M.~J., \& Huchra, J.~P.\ 1986, \apjl, 302, L1 

\bibitem[Donley et al.(2005)]{2005AJ....129..220D} 
  Donley, J.~L. Staveley-Smith, L., Kraan-Korteweg, R.~C. et al.\ 2005,
  \aj, 129, 220

\bibitem[Donley et al.(2006)]{2006MNRAS.369.1741D}  
  Donley, J.~L., Koribalski, B.~S., Staveley-Smith, L., et al.\
  2006, \mnras, 369, 1741  
   
\bibitem[Ebeling et al.(2002)]{2002ApJ...580..774E} 
  Ebeling, H., Mullis, C.~R., \& Tully, R.~B.\ 2002, \apj, 580, 774 

\bibitem[Ebeling et al.(2005)]{2005ASPC..329...83E} 
  Ebeling, H., Kocevski, D., Tully, R.~B., \& Mullis, C.~R.\ 2005, Nearby
  Large-Scale Structures and the Zone of Avoidance, 329, 83

\bibitem[Erdo{\v g}du et al.(2006)]{2006MNRAS.373...45E} 
  Erdo{\v g}du, P., Lahav, O., Huchra, J.~P., et al.\ 2006, \mnras, 373, 45 
  
\bibitem[Fairall(1980)]{1980MNRAS.192..389F} 
  Fairall, A.~P.\ 1980, \mnras, 192, 389 

\bibitem{}
  Fairall, A.~P 1998, Large-scale structures in the universe (Wiley:
  Praxis), 1998

\bibitem[Fairall et al.(1998)]{1998A&AS..127..463F} 
  Fairall, A.~P., Woudt, P.~A., \& Kraan-Korteweg, R.~C.\ 1998, \aaps, 127,
  463  

\bibitem[Gooch(1996)]{1996ASPC..101...80G} Gooch, R.\ 1996, Astronomical 
Data Analysis Software and Systems V, 101, 80 

\bibitem[Hasegawa et al.(2000)]{2000MNRAS.316..326H} 
  Hasegawa, T., Wakamatsu, K.-i., Malkan, M., et al.\ 2000, \mnras, 316,
  326

\bibitem[Helou \& Walker(1988)]{1988iras....7.....H} 
  Helou, G., \& Walker, D.~W.\ 1988, Infrared astronomical satellite (IRAS)
  catalogs and atlases.~Volume 7, p.1-265, 7,   

\bibitem[Henning et al.(2000)]{2000AJ....119.2686H} 
  Henning, P.~A., Staveley-Smith, L., Ekers, R.~D., et al.\ 2000, \aj, 119,
  2686 (HIZSS)

\bibitem[Henning et al.(2005)]{2005ASPC..329..199H} 
  Henning, P.~A., Kraan-Korteweg, R.~C., \& Staveley-Smith, L.\ 2005,
  Nearby Large-Scale Structures and the Zone of Avoidance, 329, 199

\bibitem[Henning et al.(2010)]{2010AJ....139.2130H}  
  Henning, P.~A., Springob, C.~M., Minchin, R.~F., et al.\ 2010, \aj, 139,
  2130  


\bibitem{}
 Hoppmann, L., Staveley-Smith, L., Freudling, W., et al.\ 2015, \mnras,  452, 3726


\bibitem[Huchra et al.(2012)]{2012ApJS..199...26H} 
  Huchra, J.~P., Macri, L.~M., Masters, K.~L., et al.\ 2012, \apjs, 199, 26

\bibitem[Huchtmeier et al.(2001)]{2001A&A...377..801H} 
  Huchtmeier, W.~K., Karachentsev, I.~D., \& Karachentseva, V.~E.\
  2001, \aap, 377, 801  

\bibitem[Jarrett et al.(2000)]{2000AJ....119.2498J} 
  Jarrett, T.~H., Chester, T., Cutri, R., et al.\ 2000, \aj, 119, 2498 

\bibitem[Jarrett et al.(2007)]{2007AJ....133..979J} 
  Jarrett, T.~H., Koribalski, B.~S., Kraan-Korteweg, R.~C., et al.\
  2007, \aj, 133, 979  

\bibitem[Jarrett et al.(2011)]{2011ApJ...735..112J} 
  Jarrett, T.~H., Cohen, M., Masci, F., et al.\ 2011, \apj, 735, 112 

\bibitem[Johnston et al.(2007)]{2007PASA...24..174J} 
  Johnston, S., Bailes, M., Bartel, N., et al.\ 2007, \pasa, 24, 174 

\bibitem[Johnston et al.(2007)]{2007MNRAS.376.1757J} 
  Johnston, R., Teodoro, L., \& Hendry, M.\ 2007, \mnras, 376, 1757 

\bibitem[Juraszek et al.(2000)]{2000AJ....119.1627J} 
  Juraszek, S.~J., Staveley-Smith, L., Kraan-Korteweg, R.~C., et al.\
  2000, \aj, 119, 1627  

\bibitem[Kerr \& Henning (1987)]{1987ApJ...320L..99K}  
  Kerr, F.~J., \& Henning, P.~A.\ 1987, \apjl, 320, L99 

\bibitem[Kerp et al.(2011)]{2011AN....332..637K}
  Kerp, J., Winkel, B., Ben Bekhti, N., Fl{\"o}er, L., \& Kalberla,
  P.~M.~W.\ 2011, Astronomische Nachrichten, 332, 637  

\bibitem[Kilborn et al.(2002)]{2002AJ....124..690K} 
  Kilborn, V.~A., Webster, R.~L., Staveley-Smith, L., et al.\ 2002, \aj,
  124, 690  

\bibitem[Koribalski et al.(2004)]{2004AJ....128...16K} 
  Koribalski, B.~S., Staveley-Smith, L., Kilborn, V.~A., et al.\ 2004, \aj,
  128, 16  

\bibitem[Koribalski(2012)]{2012PASA...29..359K} Koribalski, B.~S.\ 2012, 
\pasa, 29, 359 

\bibitem[Kraan-Korteweg(1989)]{1989AGAb....2...54K} 
  Kraan-Korteweg, R.~C.\ 1989, Astronomische Gesellschaft Abstract Series,
  2, 54  

\bibitem[Kraan-Korteweg \& Huchtmeier(1992)]{1992A&A...266..150K} 
  Kraan-Korteweg, R.~C., \& Huchtmeier, W.~K.\ 1992, \aap, 266, 150 

\bibitem[Kraan-Korteweg et al.(1994)]{1994ASPC...67...99K} 
  Kraan-Korteweg, R.~C., Cayette, V., Balkowski, C., Fairall, A.~P., \&
  Henning, P.~A.\ 1994, Unveiling Large-Scale Structures Behind the Milky
  Way, 67, 99

\bibitem[Kraan-Korteweg \& Woudt(1994)]{1994ASPC...67...89K} 
  Kraan-Korteweg, R.~C., \& Woudt, P.~A.\ 1994, Unveiling Large-Scale
  Structures Behind the Milky Way, 67, 89  

\bibitem[Kraan-Korteweg et al.(1995)]{1995A&A...297..617K} 
  Kraan-Korteweg, R.~C., Fairall, A.~P., \& Balkowski, C.\ 1995, \aap, 297,
  617  

\bibitem[Kraan-Korteweg et al.(1996)]{1996Natur.379..519K} 
  Kraan-Korteweg, R.~C., Woudt, P.~A., Cayatte, V., et al.\ 1996, \nat,
  379, 519  

\bibitem[Kraan-Korteweg \& Woudt(1999)]{1999PASA...16...53K} 
  Kraan-Korteweg, R.~C., \& Woudt, P.~A.\ 1999, \pasa, 16, 53 

\bibitem[Kraan-Korteweg(2000)]{2000A&AS..141..123K} 
  Kraan-Korteweg, R.~C.\ 2000, \aaps, 141, 123 

\bibitem[Kraan-Korteweg \& Lahav(2000)]{2000A&ARv..10..211K}
  Kraan-Korteweg, R.~C., \& Lahav, O.\ 2000, \aapr, 10, 211


\bibitem[Kraan-Korteweg et al.(2002)]{2002A&A...391..887K} 
  Kraan-Korteweg, R.~C., Henning, P.~A., \& Schr{\"o}der, A.~C.\
  2002, \aap, 391, 887  

\bibitem[Kraan-Korteweg(2005)]{2005RvMA...18...48K} 
  Kraan-Korteweg, R.~C.\ 2005, Reviews in Modern Astronomy, 18, 48 

\bibitem[Kraan-Korteweg et al.(2005)]{2005IAUS..216..203K} 
  Kraan-Korteweg, R.~C., Staveley-Smith, L., Donley, J., Koribalski, B., 
  \& Henning, P.~A.\ 2005, Maps of the Cosmos, 216, 203 

\bibitem[Kraan-Korteweg et al.(2008)]{2008glv..book...13K} 
  Kraan-Korteweg, R.~C., Shafi, N., Koribalski, B.~S., et al.\ 2008,
  Galaxies in the Local Volume, 13

\bibitem[Kraan-Korteweg et al.(2011)]{2011arXiv1107.1069K} 
  Kraan-Korteweg, R.~C., Riad, I.~F., Woudt, P.~A., Nagayama, T., \&
  Wakamatsu, K.\ 2011, Ten Years of IRSF - and the Future, Nagoya
  University: Nagoya, 98; arXiv:1107.1069

\bibitem[Lavaux et al.(2010)]{2010ApJ...709..483L}  
  Lavaux, G., Tully, R.~B., Mohayaee, R., \& Colombi, S.\ 2010, \apj, 709,
  483  

\bibitem[Lavaux \& Hudson(2011)]{2011MNRAS.416.2840L}
  Lavaux, G., \& Hudson, M.~J.\ 2011, \mnras, 416, 2840

\bibitem[Loeb \& Narayan(2008)]{2008MNRAS.386.2221L}  
  Loeb, A. \& Narayan, R.\ 2008, \mnras, 386, 2221L

\bibitem[Ma et al.(2012)]{2012MNRAS.425.2880M} 
  Ma, Y.-Z., Branchini, E., \& Scott, D.\ 2012, \mnras, 425, 2880 

\bibitem[Massey et al.(2003)]{2003AJ....126.2362M} Massey, P., Henning, 
P.~A., \& Kraan-Korteweg, R.~C.\ 2003, \aj, 126, 2362 

\bibitem[Matthews \& Gallagher(1996)]{1996AJ....111.1098M} 
  Matthews, L.~D., \& Gallagher, J.~S., III 1996, \aj, 111, 1098 

\bibitem[Matthews et al.(1995)]{1995AJ....110..581M} 
  Matthews, L.~D., Gallagher, J.~S., III, \& Littleton, J.~E.\ 1995, \aj,
  110, 581  
 
\bibitem[McIntyre et al. (2015)]{}
  McIntyre, T.~P., Henning, P.~A., Minchin, R.~F., Momjian, E., and
  Butcher, Z.\ 2015, \aj, 150, 28

\bibitem[Meyer et al.(2004)]{2004MNRAS.350.1195M} 
  Meyer, M.~J., Zwaan, M.~A., Webster, R.~L., et al.\ 2004, \mnras, 350,
  1195

\bibitem[Mullis et al.(2005)]{2005ASPC..329..183M} 
  Mullis, C.~R., Ebeling, H., Kocevski, D.~D., \& Tully, R.~B.\ 2005,
  Nearby Large-Scale Structures and the Zone of Avoidance, 329, 183

\bibitem[Nagayama et al.(2004)]{2004MNRAS.354..980N} 
  Nagayama, T., Woudt, P.~A., Nagashima, C., et al.\ 2004, \mnras, 354, 980

\bibitem[\protect\citeauthoryear{Paturel et al.}{2003}]{2003A&A...412...45P} 
  Paturel G., Petit C., Prugniel P., Theureau G., Rousseau J., Brouty M.,
  Dubois P., Cambr{\'e}sy L., 2003, A\&A, 412, 45

\bibitem[Radburn-Smith et al.(2006)]{2006MNRAS.369.1131R} 
  Radburn-Smith, D.~J., Lucey, J.~R., Woudt, P.~A., Kraan-Korteweg,
  R.~C., \& Watson, F.~G.\ 2006, \mnras, 369, 1131

\bibitem[Ramatsoku et al.(2014)]{2014arXiv1412.5324R}  
  Ramatsoku, M., Kraan-Korteweg, R.~C, Schr\"oder, A.~C., \& van Driel, W.\  2014, arXiv:1412.5324

\bibitem[Rauzy(2001)]{2001MNRAS.324...51R} 
  Rauzy, S.\ 2001, \mnras, 324, 51

\bibitem[Ryan-Weber et al.(2002)]{2002AJ....124.1954R} 
  Ryan-Weber, E., Koribalski, B.~S., Staveley-Smith, L., et al.\ 2002, \aj,
  124, 1954  


\bibitem[Said et al.(2015)]{2015MNRAS.447.1618S} Said, K., Kraan-Korteweg, 
R.~C., \& Jarrett, T.~H.\ 2015, \mnras, 447, 1618 

\bibitem[Sault et al.(1995)]{1995ASPC...77..433S} Sault, R.~J., Teuben, 
P.~J., 
\& Wright, M.~C.~H.\ 1995, Astronomical Data Analysis Software and Systems IV, 77, 433 

\bibitem[Saunders et al.(2000)]{2000MNRAS.317...55S} 
  Saunders, W., Sutherland, W.~J., Maddox, S.~J., et al.\ 2000, \mnras,
  317, 55  

\bibitem[Schlafly \& Finkbeiner(2011)]{2011ApJ...737..103S} 
  Schlafly, E.~F., \& Finkbeiner, D.~P.\ 2011, \apj, 737, 103 

\bibitem[Schlegel et al.(1998)]{1998ApJ...500..525S} 
  Schlegel, D.~J., Finkbeiner, D.~P., \& Davis, M.\ 1998, \apj, 500, 525 

\bibitem[Schr{\"o}der et al.(2007)]{2007A&A...466..481S} 
  Schr{\"o}der, A.~C., Mamon, G.~A., Kraan-Korteweg, R.~C., \& Woudt,
  P.~A.\ 2007, \aap, 466, 481  

\bibitem[Schr{\"o}der et al.(2009)]{2009A&A...505.1049S} 
  Schr{\"o}der, A.~C., Kraan-Korteweg, R.~C., \& Henning, P.~A.\
  2009, \aap, 505, 1049  

\bibitem[Shafi (2008)]{}  
  Shafi, N.\ 2008,  A HI Search for Galaxies Hidden by the Galactic Bulge, 
  M.Sc.~thesis, University of Cape Town, 2008

\bibitem[Springob et al.(2014)]{2014MNRAS.445.2677S} 
  Springob, C.~M., Magoulas, C., Colless, M., et al.\ 2014, \mnras, 445,
  2677
 
\bibitem[Staveley-Smith et al.(1996)]{1996PASA...13..243S} Staveley-Smith, 
L., Wilson, W.~E., Bird, T.~S., et al.\ 1996, \pasa, 13, 243 

\bibitem[Staveley-Smith(1997)]{1997PASA...14..111S} Staveley-Smith, L.\ 
1997, \pasa, 14, 111 

\bibitem[Staveley-Smith et al.(1998)]{1998AJ....116.2717S} 
  Staveley-Smith, L., Juraszek, S., Koribalski, B.~S., et al.\ 1998, \aj,
  116, 2717  

\bibitem[Stein(1997)]{1997A&A...317..670S} 
  Stein, P.\ 1997, \aap, 317, 670 

\bibitem[Tully \& Fisher(1987)]{1987nga..book.....T} 
  Tully, R.~B., \& Fisher, J.~R.\ 1987, Cambridge: University Press, 1987,

\bibitem[Tully et al.(2009)]{2009AJ....138..323T}  
  Tully, R.~B., Rizzi, L., Shaya, E.~J., et al.\ 2009, \aj, 138, 323 

\bibitem[van Driel et al.(2009)]{}  
  van Driel, W., Schneider, S.~E., Kraan-Korteweg, R.~C., \& Monnier
  Ragaigne, D.\ 2009, \aap, 505, 29  

\bibitem[Verheijen et al.(2009)]{2009pra..confE..10V} 
  Verheijen, M., Oosterloo, T., Heald, G., \& van Cappellen, W.\ 2009,
  Panoramic Radio Astronomy: Wide-field 1-2 GHz Research on Galaxy
  Evolution, 10

\bibitem[Visvanathan \& Yamada(1996)]{1996ApJS..107..521V} 
  Visvanathan, N., \& Yamada, T.\ 1996, \apjs, 107, 521 

\bibitem[Wakamatsu et al.(2005)]{2005ASPC..329..189W} 
  Wakamatsu, K., Malkan, M.~A., Nishida, M.~T., et al.\ 2005, Nearby
  Large-Scale Structures and the Zone of Avoidance, 329, 189

\bibitem[Watkins et al.(2009)]{2009MNRAS.392..743W}
  Watkins, R., Feldman, H.~A., \& Hudson, M.~J.\ 2009, \mnras, 392, 743 

\bibitem[Williams et al.(2014)]{2014MNRAS.443...41W} 
  Williams, W.~L., Kraan-Korteweg, R.~C., \& Woudt, P.~A.\ 2014, \mnras,
  443, 41  

\bibitem[Wong et al.(2006)]{2006MNRAS.371.1855W} 
  Wong, O.~I., Ryan-Weber, E.~V., Garcia-Appadoo, D.~A., et al.\
  2006, \mnras, 371, 1855  

\bibitem[Woudt et al.(1999)]{1999A&A...352...39W} 
  Woudt, P.~A., Kraan-Korteweg, R.~C., \& Fairall, A.~P.\ 1999, \aap, 352,
  39 

\bibitem[Woudt \& Kraan-Korteweg(2000)]{2000ASPC..218..193W} 
  Woudt, P.~A., \& Kraan-Korteweg, R.~C.\ 2000, Mapping the Hidden
  Universe: The Universe behind the Milky Way - The Universe in HI, 218, 193

\bibitem[Woudt et al.(2000)]{2000ASPC..218..203W} 
  Woudt, P.~A., Kraan-Korteweg, R.~C., \& Fairall, A.~P.\ 2000, Mapping the
  Hidden Universe: The Universe behind the Milky Way - The Universe in HI,
  218, 203

\bibitem{} 
  Woudt, P.A., \& Kraan-Korteweg, R.C., 2001, A\&{A} 380, 441

\bibitem[Woudt et al.(2004)]{2004A&A...415....9W} 
  Woudt, P.~A., Kraan-Korteweg, R.~C., Cayatte, V., Balkowski, C., \&
  Felenbok, P.\ 2004, \aap, 415, 9  

\bibitem[Woudt et al.(2008)]{2008MNRAS.383..445W} 
  Woudt, P.~A., Kraan-Korteweg, R.~C., Lucey, J., Fairall, A.~P., \& Moore,
  S.~A.~W.\ 2008, \mnras, 383, 445

\bibitem[Wright \& Otrupcek(1990)]{1990PKS...C......0W} 
  Wright, A., \& Otrupcek, R.\ 1990, PKS Catalog (1990), 0 

\bibitem[Wright et al.(2010)]{2010AJ....140.1868W} 
  Wright, E.~L., Eisenhardt, P.~R.~M., Mainzer, A.~K., et al.\ 2010, \aj,
  140, 1868  

\bibitem[Zwaan et al.(2004)]{2004MNRAS.350.1210Z} Zwaan, M.~A., Meyer, 
M.~J., Webster, R.~L., et al.\ 2004, \mnras, 350, 1210 




\end{thebibliography}
\end{document}